\documentstyle[12pt]{article}
\begin{document}

\pagestyle{empty}

\rule{0in}{2in}

\Large
\centerline{EVIDENCES FOR A NEW FUNDAMENTAL INTERACTION} 
\normalsize
\vskip .5in
\large
\centerline{M\'{A}RIO EVERALDO DE SOUZA}
\vskip .2in
\centerline{DEPARTAMENTO DE F\'{I}SICA}
\centerline{UNIVERSIDADE FEDERAL DE SERGIPE}
\centerline{Aracaju, Sergipe, Brazil}

\newpage

\rule{0in}{1.5in}
\Large

\begin{center}
\parbox{4.2in}
{\bf{To the memory of Sir Isaac Newton for his genius
and for his humility.}}
\end{center}

\vskip 2in

\begin{center}
\parbox{5in}
{\it{\bf{I do not know what I may appear to the world; but to myself I seem to 
have been only like a boy playing on the seashore, and diverting myself in now 
and then finding a smoother pebble or prettier shell than ordinary, while the 
great ocean of truth lay all undiscovered before me.}}}
\end{center}

\vskip .2in
\hskip 3.5in {\it{Sir Isaac Newton}}

\normalsize
\newpage

\setcounter{page}{1}
\Large
\centerline{EVIDENCES FOR A NEW FUNDAMENTAL INTERACTION} 
\normalsize
\vskip .5in
\large
\centerline{M\'{A}RIO EVERALDO DE SOUZA}
\vskip .2in
\centerline{DEPARTAMENTO DE F\'{I}SICA}
\centerline{UNIVERSIDADE FEDERAL DE SERGIPE}
\centerline{Aracaju, Sergipe, Brazil}
\centerline{Report No. 02.07.1999.UFS}
\centerline{March 24, 1999}
\normalsize
\vskip .5in
\noindent
\large
\centerline{CONTENTS}
\normalsize
\vskip .25in
\begin{enumerate}
\item Preliminary Ideas on Prequarks and the Number of Quarks  
\item Distribution of the Electric Charge in Baryons, The Pion Cloud in 
Nucleons, The Size of Quarks, Proton's Stability and Deep Inelastic Scattering
\item Calculation of the Light Antiquark Flavor Asymmetry in the Nucleons
\item The Quark Sea Content of the Other Baryons
\item Quark Confinement, The Sizes of Quarks, Primon Mass and the Generation 
of Quark Mass
\item The True Potential of a Quark Pair and the 'Usual' QCD Potential
\item Asymptotic Freedom
\item The Success of QCD, the Nature of Gluons and the Number of Leptons
\item The Supernovae Evidence for the Superstrong Interaction
\item The Planetary Evidence for the Superstrong Interaction
\item Evidences for the Superstrong Interaction from Galactic Formation and
Evolution and from the Formation of Structure
\item The Rotation of Spiral Galaxies
\item General Classification of Matter
\item The Superweak Force
\item The Galactic Medium
\item The Structured State
\item The Energies of Baryons
\item Generalization of the Gell-Mann-Okubo Mass Formula
\item The Excited States of Quarks
\item Hadronic Molecules
\item The Energies of Mesons
\item The Nuclear Potential and the Stability of the Deuteron
\item The Absence of Nuclides with $A=5$ and the Instability of 
$Be^{8}$
\item The Desired Unity
\end{enumerate}

\newpage
\noindent
1) PRELIMINARY IDEAS ON PREQUARKS AND THE NUMBER OF QUARKS 
\vskip .1in
\par It has been proposed by De Souza$^{(1,2,3,4,5,6,7,8,9,10,11,12)}$ that 
nature has six fundamental forces. One of the new forces, called superstrong 
force, acts between any two quarks and between quarks constituents. Actually, 
quark composition is an old idea, although it has been proposed on different 
grounds$^{(13,14,15,16)}$. A major distinction is that in this work leptons 
are supposed to be elementary particles. This is actually consistent with the 
smallness of the electron mass which is already too small for a particle with 
a very small radius$^{(17)}$. 
\par In order to distinguish the model proposed in this work from other models
of the literature we will name these prequarks with a different name.   We may 
call them {\it{primons}}, a word derived from the latin word {\it{primus}} 
which means first. 
\par Let us develop some preliminary ideas which will help us towards the 
understanding of the superstrong interaction. Since a baryon is composed of
three quarks it is reasonable to consider that a quark is composed of two
primons. The new interaction between them exists by means of the exchange of
new particles. 
\par In order to reproduce the spectrum of 6 quarks and their colors we need 
4 primons in 3 supercolor states. As to the charge, one has charge (+5/6)e and 
any other one has charge (-1/6)e.  And what about spin? How can we have 
prequarks with spins equal to 1/2 and also have quarks with spins equal
to 1/2? There are two solutions to this question. One is to consider that
at the prequark level Planck's constant is redefined as 1/2. I adopted
this solution in another version of this work and in other publications on
prequarks. It is too drastic and leads to some problems. One of them is that
in the end we will have to deal with anyons. But anyons violate {\bf{P}} and 
{\bf{T}} while the strong interaction does not. The other solution I believe 
is more plausibe although it depends on a postulate which may be expressed in 
the following way: {\bf{Primons are fermions with spins equal to 1/2 but 
each spin makes an angle of ${\pi}/3$ with the direction of the hadron spin, 
so that the total spin of the quark is 1/2.}} This means that the system of 
primons in a baryon is a very cooperative system in the sense that primons 
interact in such a way as to maintain their spins making either ${\pi}/3$ or 
$2{\pi}/3$ with the baryon's spin direction as is shown below in Fig. 1 in the 
case of a proton. The total angle between the two spins of the two primons of 
a quark should always be $2{\pi}/3$. Of course, we are saying that primons are 
special fermions because their spins can not be aligned by a magnetic field 
due to their mutual interaction that couples each pair making a rigid angle 
between their spins, and only the total spin of each pair is aligned by the 
field. Thus, with respect to spin such a system is highly ordered. Maybe 
this spin picture will shed some light onto the proton spin puzzle for, as we
see, each primon contributes with $1/4$, and if we make the confusion of
considering them as quarks, then we obtain that they contribute with half of
the proton's spin.
\par Let us choose the +Z direction as the direction of the proton's 
spin. Each primon spin contributes with $(\hbar/2)\cos{(\pi/3)}=(\hbar/4)$ 
along the Z direction. Thus, each quark has a spin equal to $\hbar/2$. We also 
see that the spins of the two primons in a quark can rotate freely around the 
Z axis but they have to rotate at the same time so that the two components in 
the XY plane cancel out.  We easily see in Fig. 01 that it is possible to 
have the exchange of scalar and vector bosons between primons of different 
quarks. We also see that between the primons of a quark scalar bosons are 
exchanged. This is quite in line with the known properties of the nuclear 
potential which may be described with terms due to the exchange of pions as 
well as exchange of vector mesons such as $\rho$ and $\omega$. Let us consider 
that the superstrong field is mediated by the exchange of vector bosons.
\par It will be shown later on in section 2 that the primons that form 
a quark do not rotate much about their centers of mass. In order to have an 
effective potential well(sort of molecular potential) between the two primons 
%\newpage
%\rule{0in}{1in}
%\newpage
%\noindent
forming a quark the superstrong force between them should be repulsive, and 
the strong force must be attractive. Therefore, the superstrong charges of 
primons $p_{1}$, $p_{2}$, $p_{3}$ and $p_{4}$ have the same sign and there is, 
therefore, always repulsion between $p_{i}$ and $p_{j}$ for any $i$ and $j$. 
Moreover, for $i=j$ the repulsion should be stronger. Since we need to have 
attraction between primons we have to postulate that at the level of primons 
the strong interaction manifests itself by means of supercolors in such a way 
that two different primons with different supercolors attract each other and 
primons with the same supercolors repel each other. Quark colors are formed 
from the mixing of different supercolors as shown in Table 1. 
\par Taking into account the above considerations on spin and charge we have 
the following table for primons(Table 2).  With this table we are able to form 
all quarks as shown in Table 3 and to affirm that the maximum number of quarks
is six. There should exist similar tables for the corresponding antiparticles.
\par Many different bosons may mediate the strong interaction among primons. 
They are the combinations $p_{i}\bar{p_{j}}$. For example, between $p_{1}$ and 
$p_{3}$ the bosons $\bar{p_{1}}p_{3}$ and $\bar{p_{3}}p_{1}$ are exchanged. 
Considering that $p_{i}\bar{p_{i}}$ are also formed inside hadrons, the bosons 
$\bar{p_{1}}p_{3}$ and $\bar{p_{3}}p_{1}$ may be, of course, the pions 
$\pi^{-}=\overline{p_{1}p_{2}}p_{2}p_{3}$ and 
$\pi^{+}=p_{1}p_{2}\overline{p_{2}p_{3}}$. With $p_{1}$, $p_{2}$, $p_{3}$ and 
$p_{4}$(and $\bar{p_{1}}$, $\bar{p_{2}}$, $\bar{p_{3}}$ and $\bar{p_{4}}$), 
we form the mesons $\pi^{+}$, $\pi^{0}$, $\pi^{-}$, $d\bar{c}=D^{-}$, 
$\bar{d}c=D^{+}$,  $u\bar{c}={\bar{D}}^{0}$, 
$\bar{u}c=D^{0}$, $b\bar{s}={\bar{B}}_{s}^{0}$, $\bar{b}s=B_{s}^{0}$, 
$t\bar{s}$, $\bar{t}s$, $t\bar{b}$, $\bar{t}b$, $c\bar{c}$, $b\bar{b}$, 
$t\bar{t}$, $\rho$, $\omega$, $\eta_{8}$, $\eta_{0}$, $\phi$. In sections 3 
and 4 we will calculate the constitution of the quark sea in the nucleons and 
will indicate how to calculate it in the other baryons. 
\par  The superstrong field needs three bosons, which, due to the very 
small range of the interaction, should be very heavy. Let us call them 
${\aleph}^{-}$, $\aleph^{+}$ and $\aleph^{0}$. Between $p_{1}$ and 
$p_{i}$($i{\neq}1$) ${\aleph}^{-}$ and $\aleph^{+}$ are exchanged; 
$\aleph^{0}$ is exchanged between any two equal primons and between $p_{i}$ 
and $p_{j}$ with $i,j{\neq}1$, and, thus, the symmetry group should is SU(2).
\par We may have an estimation of the strength of the superstrong interaction
using the similarities and differences that exist among the different kinds of
structured states as defined in reference 8 in the following way. In ordinary 
matter(liquids, solids and gasses) which is formed by the gravitational and 
electromagnetic forces  the energy levels are in the eV region. In nuclei which
are formed by the electromagnetic and strong forces the energy levels are in 
the MeV region, and in quarks which are formed by means of the action of the 
strong and superstrong forces the energy leves are in the GeV region. 
Therefore, {\it{the superstrong interaction bosons are in the GeV region}}. We
will see later on that their masses are about (1-3)GeV.
\par We can construct the eight gluons with the combinations of $\alpha$,
$\beta$ and $\gamma$ with $\bar{\alpha}$, $\bar{\beta}$ and $\bar{\gamma}$.
Let us analyse, for example, the gluon interaction 
$u_{b} + \bar{b}r {\rightarrow} u_{r}$ which may be described in terms of the 
two interactions $p_{1}^{\alpha} + 
\bar{\alpha}\beta {\rightarrow} p_{2}^{\beta}$
and $p_{2}^{\beta} + \bar{\beta}\gamma {\rightarrow} p_{1}^{\gamma}$, omitting 
the exchange of ${\aleph}$'s. The combination of $\bar{\alpha}\beta$ with
$\bar{\beta}\gamma$ forms the gluon $\bar{b}r$. \newline
\vskip .2in
\noindent
2) DISTRIBUTION OF THE ELECTRIC CHARGE IN BARYONS, THE PION CLOUD IN NUCLEONS,
 THE SIZE OF QUARKS, PROTON'S STABILITY AND DEEP INELASTIC SCATTERING
\vskip .15in
\noindent
a) Distribution of Electric Charge in the Nucleon$^{(12)}$
\par Deep inelastic electron scattering$^{(18,19)}$ has shown that the 
distributions of electric charge

\rule{0in}{.2in}

\begin{center}
\begin{tabular}{||ccc||ccc|ccc|ccc||} \hline\hline 
& & & & & & & & & & & \\ 
&          & & &  $\alpha$ & & & $\beta$ & & & $\gamma$ & \\
& & & & & & & & & & & \\
\hline\hline
& & & & & & & & & & & \\
& $\alpha$ & & &           & & & blue    & & & green    & \\
& & & & & & & & & & & \\
\hline
& & & & & & & & & & & \\
& $\beta$  & &  & blue    & & &       & & & red      & \\
& & & & & & & & & & & \\
\hline
& & & & & & & & & & & \\
& $\gamma$ & & &  green    & & & red   & & &          & \\
& & & & & & & & & & & \\
\hline\hline 
\end{tabular}
\end{center}
\vskip .15in

\begin{center}
\parbox{3.8in}
{Table 1. Generation of colors
out of the supercolors.}
\end{center}

\rule{0in}{.4in}

\begin{center}
\begin{tabular}{||cc||ccccc|} \hline\hline 
& & & & & & \\ 
superflavor & & & charge & & spin & \\
& & & & & & \\
\hline \hline 
& & & & & & \\
$p_{1}$ & & & $\frac{5}{6}$  & & $\frac{1}{2}$ & \\
& & & & & & \\
\hline 
& & & & & & \\
$p_{2}$ & & & $-\frac{1}{6}$ & & $\frac{1}{2}$ & \\
& & & & & & \\
\hline 
& & & & & & \\
$p_{3}$ & & & $-\frac{1}{6}$ & & $\frac{1}{2}$ & \\
& & & & & & \\
\hline 
& & & & & & \\
$p_{4}$ & & & $-\frac{1}{6}$ & & $\frac{1}{2}$ & \\
& & & & & & \\
\hline\hline
\end{tabular}
\end{center}
\vskip .15in

\begin{center}
\parbox{3.5in}
{Table 2. Electric charges and spins of primons.}
\end{center}

\newpage

\begin{center}
\begin{tabular}{||cc||ccc|ccc|ccc|ccc||} \hline\hline 
& & & & & & & & & & & & &  \\ 
 & & & $p_{1}$ & & & $p_{2}$ & & & $p_{3}$ & & & $p_{4}$ &  \\
& & & & & & & & & & & & &  \\
\hline\hline 
& & & & & & & & & & & & &  \\
$p_{1}$ & & &   & & & u & & & c & & & t &  \\
& & & & & & & & & & & & &  \\
\hline 
& & & & & & & & & & & & &  \\
$p_{2}$ & & & u & & &   & & & d & & & s &  \\
& & & & & & & & & & & & &  \\
\hline 
& & & & & & & & & & & & &  \\
$p_{3}$ & & & c & & & d & & &   & & & b &  \\
& & & & & & & & & & & & &  \\
\hline 
& & & & & & & & & & & & &  \\
$p_{4}$ & & & t & & & s & & & b & & &   &  \\
& & & & & & & & & & & & &  \\
\hline\hline
\end{tabular}
\end{center}
\vskip .15in

\begin{center}
\parbox{2.8in}
{Table 3. Composition of quark flavors.}
\end{center}
\vskip .3in

\rule{0in}{.2in}

%\newpage
%\rule{0in}{.2in}
%\newpage
%\rule{0in}{1in}
%\newpage

\noindent
in the nucleons are represented by the two
graphs below(Figs. 2a and 2b). These distributions have inspired the pion cloud
model of the nucleon which has been quite sucessful at explaining many of its 
properties.
\par Analyzing these two figures one easily sees that shells of electric 
charge exist in both nucleons. The proton has two shells with mean radii of 
about 0.2fm and 0.7fm and the neutron has three shells with radii of about 
0.15fm, 0.65fm, and 1.8fm. Let us disregard the 
outermost shell of the neutron. Therefore, each nucleon has two shells of 
primons located at about 0.16fm and 0.67fm from the center. We can only  
explain the existence of these shells if we admit that quarks are composite 
and formed of prequarks. The two shells are, then, prequark shells, showing 
that a quark is composed of two prequarks. Considering what was presented above
primons with the same supercolors tend to stay away from each other and primons
with different supercolors attract each other. Therefore, primons are arranged
inside the proton as is shown in Fig. 3. The charge of each one of the two 
shells(inner and outer shells) is +1/2. In
terms of primon shells we can say that the proton has the configuration
\begin{eqnarray*} (p_{1}^{\alpha}p_{2}^{\beta}p_{3}^{\gamma})^{1}
(p_{2}^{\beta}p_{1}^{\gamma}p_{2}^{\alpha})^2.\end{eqnarray*} 
\noindent
The superscripts 1 and 2 mean the inner and outer layers, respectively. Let us
dispose the primons of the inner layer clockwise. A primon of one layer with 
the closest primon of the other layer forms a quark. Each layer forms a plane 
of primons. The two planes are linked by the three strong bonds, that is, by 
the three quarks. A primon of a quark with a primon of another quark 
forms a weak bond when they are different and have different supercolors. The 
three quarks of the inner layer of the proton, for example, are linked by weak 
bonds. Due to the exchange of gluons the colors change, and therefore the weak 
bonds change all the time, but the lowest potential energy of the inner layer 
should happen when it has three different supercolors since equal colors repel 
each other. Thus, all possible configurations of the proton are: \newline
$(p_{1}^{\alpha}p_{2}^{\beta}p_{3}^{\gamma})^{1}
(p_{2}^{\beta}p_{1}^{\gamma}p_{2}^{\alpha})^{2}$;  
$(p_{1}^{\alpha}p_{2}^{\beta}p_{3}^{\gamma})^{1}
(p_{2}^{\gamma}p_{1}^{\alpha}p_{2}^{\beta})^{2}$;  
$(p_{1}^{\alpha}p_{2}^{\gamma}p_{3}^{\beta})^{1}
(p_{2}^{\beta}p_{1}^{\alpha}p_{2}^{\gamma})^{2}$; 
$(p_{1}^{\alpha}p_{2}^{\gamma}p_{3}^{\beta})^{1}
(p_{2}^{\gamma}p_{1}^{\beta}p_{2}^{\alpha})^{2}$;\newline
$(p_{1}^{\beta}p_{2}^{\alpha}p_{3}^{\gamma})^{1}
(p_{2}^{\gamma}p_{1}^{\beta}p_{2}^{\alpha})^{2}$;
$(p_{1}^{\beta}p_{2}^{\alpha}p_{3}^{\gamma})^{1}
(p_{2}^{\alpha}p_{1}^{\gamma}p_{2}^{\beta})^{2}$;
$(p_{1}^{\beta}p_{2}^{\gamma}p_{3}^{\alpha})^{1}
(p_{2}^{\alpha}p_{1}^{\beta}p_{2}^{\gamma})^{2}$;
$(p_{1}^{\beta}p_{2}^{\gamma}p_{3}^{\alpha})^{1}
(p_{2}^{\gamma}p_{1}^{\alpha}p_{2}^{\beta})^{2}$;\newline
$(p_{1}^{\gamma}p_{2}^{\alpha}p_{3}^{\beta})^{1}
(p_{2}^{\beta}p_{1}^{\gamma}p_{2}^{\alpha})^{2}$;
$(p_{1}^{\gamma}p_{2}^{\alpha}p_{3}^{\beta})^{1}
(p_{2}^{\alpha}p_{1}^{\beta}p_{2}^{\gamma})^{2}$;
$(p_{1}^{\gamma}p_{2}^{\beta}p_{3}^{\alpha})^{1}
(p_{2}^{\alpha}p_{1}^{\gamma}p_{2}^{\beta})^{2}$ and
$(p_{1}^{\gamma}p_{2}^{\beta}p_{3}^{\alpha})^{1}
(p_{2}^{\beta}p_{1}^{\alpha}p_{2}^{\gamma})^{2}$. 
\par Since the $u$ quark does not decay $p_{1}$ and $p_{2}$ have to be stable 
and since $d$ decays $p_{3}$ has to decay according to the reaction
$p_{3}{\rightarrow}p_{1}e^{-}\bar{\nu_{e}}$. Why then does not the proton 
decay since it contains a $d$ quark? The outer layer of the proton contains 
the primons $p_{1}$ and $p_{2}$ which are stable. Why does not the $p_{3}$ 
primon of the inner layer decay? It does not decay 
%\newpage
%\rule{0in}{1in}
%\newpage
%\rule{0in}{5.5in}
%\noindent
because if it decayed the inner layer would have two $p_{1}$ primons and, 
then, it would have a higher potential energy(because two ${p_{1}}'s$ repel 
each other), that is, it would be less bound. Therefore, it does not decay 
because the configuration of the inner layer which is $p_{1}p_{2}p_{3}$ is 
already the most stable one. The neutron, on the other hand, has a $p_{3}$ 
primon in the outer layer, and therefore, may decay. That is, if the 
$p_{3}$ of the outer layer is in a well it must be so shallow that this 
primon may not be bound(that is, there is no bound state). 
\par Following the same reasoning the configuration of primons in 
the neutron should be as shown in Fig. 4. The charge of the inner layer is
+1/2 and the charge of the outer layer is -1/2. As we saw above
the inner layer of the neutron should be equal to that of the proton. The
configuration of the neutron is  
$(p_{1}p_{2}p_{3})^{1}(p_{2}p_{3}p_{2})^2$ which differs from the proton{'}s 
in the outer layer. It is implied that the supercolors{'} configurations are 
as displayed above for the proton. From now on we will not write them out 
explicitly. 
\par In atoms or in molecules the probability density of electrons form the
orbitals. We may extend the same ideas to baryons and talk about primon 
orbitals. Their existence may help us understand, for example, some 
properties of the deuteron. Its size, for instance, of about 2.1fm can not be 
explained if we consider that quarks are pointlike. On the other hand, if we 
consider that there is a repulsive force(the superstrong force) among 
primons(and quarks) we see that its size is a consequence of the arrangement 
of primons in the whole pn system(Fig. 5).  There are two cores with equal 
positive charges, +1/2 each. Due to the exchange of pions between $p_{1}$ and 
$p_{3}$ there is a net positive charge of about +2/3 in the  
%\newpage
%\rule{0in}{1in}
%\newpage
%\rule{0in}{4.5in}
%\noindent
middle, between the two cores. And there is a negative 
charge cloud of -1/3 on each side around each core. Such a distribution 
yields two opposite electric dipoles moments of about $(1/3){\times}0.6e(fm)$ 
and, therefore a quadrupole moment of about 
$0.2e(fm){\times}1fm=2{\times}10^{-3}$e(barn), which is close to the 
experimental value of $2.82{\times}10^{-3}$e(barn)$^{(20)}$. 
{\it{Pointlike quarks moving randomly can not produce 
such moments.}} A quark is not, then, a usual 
particle such as a lepton is. It is an extended object with an average size 
of about 0.5fm and a size that varies up to a about 1fm. 
\par Since the neutron decays via the weak interaction into 
$n{\rightarrow}p^{+}e^{-}\bar{\nu_{e}}$ the primon $p_{3}$ should decay 
according to the reaction $p_{3}{\rightarrow}p_{1}e^{-}\bar{\nu_{e}}$ as we
saw above. Other weak decays of primons are: \newline
$p_{4}{\rightarrow}p_{1}e^{-}\bar{\nu_{e}}$, 
$p_{4}{\rightarrow}p_{1}\mu^{-}\bar{\nu_{\mu}}$, 
$p_{4}{\rightarrow}p_{1}\tau^{-}\bar{\nu_{\tau}}$,
$p_{3}{\rightarrow}p_{1}\mu^{-}\bar{\nu_{\mu}}$,
$p_{4}{\rightarrow}p_{3}\pi^{0}$,
$p_{3}{\rightarrow}p_{1}\pi^{-}$,
$p_{3}{\rightarrow}p_{2}\pi^{0}$.
\par The exchange of pions between two different nucleons happens due to the 
exchange of pions between primons. In order to explain some decays of hadrons 
the electromagnetic decay $p_{4}{\rightarrow}p_{3}\gamma$ should also exist. 
Since pions may be expressed in terms of primons the pion exchange between
$p_{1}$ and $p_{3}$, $p_{1} + \pi^{-}{\rightarrow}p_{3}$, is, actually
described by $p_{1} +  p_{2}p_{3}\overline{p_{1}p_{2}} {\rightarrow}  p_{3}$
and by $p_{1} +  p_{4}p_{3}\overline{p_{1}p_{4}} {\rightarrow}  p_{3}$. 
In order to explain some decays of hadrons the electromagnetic decay 
$p_{4} {\rightarrow}  p_{3}\gamma$ should also exist.
\par Figures 3 and 4 are planar displays of three-dimensional 
spatial configurations. In this way we reconcile the pion cloud vision of the 
nucleon with the quark model. We easily see that a bare nucleon is a nucleon 
without its outer layer since it is this layer that makes the difference 
between nucleons. 
\par We may idenfify primons as partons$^{11}$ which are supposed to be 
pointlike and with spin equal to 1/2. According to the above 
considerations we have just to admit that a baryon has at least six partons 
that are combined in three quarks. At intermediate $Q^{2}$(around $1GeV^2$)
$F_{2}(x)$ peaks around $x=1/3$, showing thus that a proton has three quarks.
At  higher $Q^{2}$ primons should appear. In this region we have to 
rewrite the structure functions, $F_{2}^{ep}(x)$ and $F_{2}^{en}(x)$, 
accordingly. Considering that primons are partons we have 
\begin{eqnarray}\frac{1}{x}F_{2}^{ep}(x)&=
&\left(\frac{5}{6}\right)^{2}p_{1}^{p}(x) + 
\left(\frac{1}{6}\right)^{2}p_{2}^{p}(x) + 
\left(\frac{1}{6}\right)^{2}p_{3}^{p}(x)\end{eqnarray}                
\noindent
and
\begin{eqnarray}\frac{1}{x}F_{2}^{en}(x)&=
&\left(\frac{5}{6}\right)^{2}p_{1}^{n}(x) +
\left(\frac{1}{6}\right)^{2}p_{2}^{n}(x) + 
\left(\frac{1}{6}\right)^{2}p_{3}^{n}(x)\end{eqnarray}			
\noindent
where $p_{i}(x)$ are the probability distributions of primons in each nucleon.
There are as many $p_{3}$'s in a proton as $p_{1}$ in a neutron and both 
nucleons have the same number of $p_{2}$'s. Therefore, 
$p_{3}^{p}(x)=p_{1}^{n}(x)$, $p_{3}^{n}(x)=p_{1}^{p}(x)$, and 
$p_{2}^{p}(x)=p_{2}^{n}(x)$. Thus, $F_{2}^{ep}$ and $F_{2}^{en}$ become
\begin{eqnarray}F_{2}^{ep}(x)&=&x\left(\frac{25}{36}p_{1}(x) +
\frac{1}{36}p_{2}(x) + \frac{1}{36}p_{3}(x)\right)\end{eqnarray}	 
\noindent
and
\begin{eqnarray}F_{2}^{en}(x)&=&x\left(\frac{1}{36}p_{1}(x) +
\frac{1}{36}p_{2}(x) + \frac{25}{36}p_{3}(x)\right)\end{eqnarray}	 
\noindent
where $p_{i}(x)$ refer only to the proton. The difference 
$F_{2}^{ep}(x) - F_{2}^{en}(x)$ is then
\begin{eqnarray}F_{2}^{ep}(x) - F_{2}^{en}(x)&=&
x\frac{24}{36}(p_{1}(x) - p_{3}(x)).\end{eqnarray}			 
\noindent
Of course, $F_{2}^{ep}(x) - F_{2}^{en}(x){\rightarrow}0$ as 
$x{\rightarrow}0$ tends to zero which is in line with the SLAC data presented 
in Fig. 6a. From this result we also obtain that 
$\frac{F_{2}^{ep}(x)}{F_{2}^{en}(x)}{\rightarrow}1$, which agrees well with 
the experimental data. For higher $Q^{2}$, $F_{2}^{ep}(x) - F_{2}^{en}(x)$ 
should have a peak at $x{\approx}1/6$ which is quite hard to be seen due to
the smearing of $F_{2}^{ep}(x) - F_{2}^{en}(x)$ about 1/3 and the influence of
the quark sea. In this way we
can understand why $F_{2}(x,Q^{2})$ increases as $x{\rightarrow}0$ even 
without the contribution of the quark sea. Actually, the smearing of 
$F_{2}(x,Q^{2})$ around $x=1/3$ for $Q^{2}=1GeV^{2}$ may be an indication 
of the size of a quark, that is, an indication of the motion of the two primons
that form a quark. Summarizing the discussion on deep inelastic scattering
we can say that for low values of $Q^{2}(<0.5GeV^{2})$ the virtual photon
sees the whole nucleon and the scattering is mainly elastic. For higher
$Q^{2}$(few $GeV^{2}$), the virtual photon probes the internal structure of the
nucleons and the scattering from the three quarks occurs. At higher values of
$Q^{2}$ the photons are mainly scattered by primons which makes $F_{2}(x)$ to
peak at $x{\approx}1/6$. At still higher values of $Q^{2}$ the quark sea should
contribute. According to the above picture of the nucleons both of them have
a common core made of $p_{1}p_{2}p_{3}$ with a radius of about 0.15fm. Thus,
in a certain range of high $Q^{2}$ the virtual photon will probe this core, 
and therefore, $\frac{F_{2}^{ep}(x)}{F_{2}^{en}(x)}{\rightarrow}1$. It
represents a plateau(around x=0.1) which is shown in SLAC's data(Fig. 6b). 
As will be seen later on primons should have equal masses and must each carry 
the same momentum fraction.  Summing over the momenta of all primons we obtain 
the total momentum of the proton. The integrals of $F_{2}^{ep}(x)$ and 
$F_{2}^{en}(x)$ over x from 0 to 1 are equal to$^{(21)}$ 0.18 and 0.15. Using 
the above expressions for $F_{2}^{ep}(x)$ and $F_{2}^{en}(x)$ we obtain 
$\varepsilon_{1}=\varepsilon_{2}=\varepsilon_{3}=0.24$. Hence, 
the total momentum fraction carried by primons is about 0.72 and only about 
28\% is carried by the quark sea and gluons. 
\par Therefore, it looks like that nature has been fooling us since a long time
ago, at least since the sixties:{\bf{The pointlike particles that we have 
observed in the nucleon are not quarks, they are prequarks.}}
%\newpage
%\rule{0in}{5.2in}
\par A very important quantity that corroborates the arrangement of primons in
the nucleons is the value of the electric dipole moment(EDM) in each nucleon.
The values of the EDM for the proton and the neutron are$^{(22)}$ 
$d=(-4\pm6){\times}10^{-23}$ecm and $d<1.1{\times}10^{-25}$ecm, respectively. 
According to the above picture we expect that the neutron EDM should be 
smaller than the proton{'}s because the outer layer of the neutron is 
$(p_{2}p_{3}p_{2})$ while the proton{'}s is $(p_{2}p_{1}p_{2})$. Therefore, 
since the primons $p_{2}$ and $p_{3}$ have the same charge(-1/6), while the 
primons $p_{1}$ and $p_{2}$ have quite different charges(+5/6 and -1/6), the 
outer layer of the neutron should be more spherical than the proton{'}s. And 
since the inner layer is the same for both nucleons, the neutron EDM should be 
smaller than the proton{'}s.
\par It is very interesting to notice that there may exist the exchange of
primons from the inner to the outer layer. For example, in the proton if the
primons $p_{1}$ and $p_{2}$ of the inner layer get too close they may form the
$u$ quark. But this makes the other $p_{1}$ and one $p_{2}$ of the outer layer
to be free until they combine and form another $u$ quark. This means that
the $c$ quark($p_{1}p_{3}$) may exist for short a time inside the 
proton(please, see the sections on the constitution of the nucleon quark seas).
It can not exist for long times because two $p_{2}${'}s do not combine.
\par This picture also means that each quark(i.e., each pair of primons) does 
not rotate much about its center of mass. 
\vskip .2in
\noindent
b) Distribution of Electric Charge in the Other Baryons$^{(12)}$
\vskip .1in
\par If we assume that the other quarks behave in the same way we may extend 
the same reasoning to them and predict the shape of the electric charge 
distribution in the other baryons.  
\par Let us begin with the $\Delta^{++}$(uuu). It should have the primon 
configurations \newline
$(p_{1}p_{2}p_{2})^{1}(p_{2}p_{1}p_{1})^2$ and 
$(p_{1}p_{2}p_{1})^{1}(p_{2}p_{1}p_{2})^2$ which are represented 
in Figs. 7 and 8, respectively, and produce the charge 
distributions shown in Figs. 9 and 10, respectively. In both configurations 
the inner layer has only two weak bonds and are, therefore, less bound than 
that of each nucleon. We should investigate if there is an alternation between 
the two possible configurations above mentioned. In this case the effective 
charge distribution would be an average of the charge distributions displayed 
in Figs. 9 and 10. Due to the repulsion between the two inner primons the 
inner layer may be larger than that of the nucleons. This is what happens with 
the outer layer of the nucleons. That is, the mean radius of the inner layer 
is larger than 0.15fm. Thus, the size of $\Delta^{++}$ is larger than that  
of the nucleons. From now on the figures of the configurations will be omitted 
but they may be easily constructed from the ones above. 
\par The configurations of $\Delta^{-}$(ddd) are 
$(p_{2}p_{3}p_{2})^{1}(p_{3}p_{2}p_{3})^2$ and
$(p_{3}p_{2}p_{3})^{1}(p_{2}p_{3}p_{2})^2$ which yield the same charge 
distribution curve displayed in Fig. 11. The configurations and charge 
distributions of $\Delta^{+}$ and $\Delta^{0}$ are the same as those of the 
proton and neutron, respectively. The configuration of $\Sigma^{+}(uus)$ is 
$(p_{1}p_{2}p_{4})^{1}(p_{2}p_{1}p_{2})^2$  which yields a charge 
distribution as that of the proton.
\par The $\Sigma^{0}$(dus) and $\Lambda$(dus) have the configurations
$(p_{1}p_{2}p_{3})^{1}(p_{2}p_{4}p_{2})^2$, 
$(p_{1}p_{3}p_{4})^{1}(p_{2}p_{2}p_{2})^2$, \newline
$(p_{1}p_{2}p_{4})^{1}(p_{2}p_{3}p_{2})^2$ and
$(p_{2}p_{3}p_{4})^{1}(p_{1}p_{2}p_{2})^2$.
\noindent
The first three give the charge distribution of the neutron and the fourth
produces the charge distribution shown in Fig. 12. But, probably, the first
configuration above is the most stable one, and may be the preferred 
configuration due to the stability of $p_{1}$ and $p_{2}$. The 
$\Sigma_{c}^{++}$(uuc), which has the configuration 
$(p_{1}p_{2}p_{3})^{1}(p_{2}p_{1}p_{1})^{2}$, has the charge distribution 
given by Fig. 13.
\par The particles $\Sigma^{-}$(dds) and $\Xi^{-}$(dss)have the respective 
configurations $(p_{2}p_{3}p_{4})^{1}(p_{3}p_{2}p_{2})^2$ and 
$(p_{2}p_{3}p_{4})^{1}(p_{4}p_{2}p_{2})^2$, and $\Omega$(sss) has the two 
configurations $(p_{2}p_{2}p_{4})^{1}(p_{4}p_{4}p_{2})^2$ and\newline 
$(p_{2}p_{4}p_{4})^{1}(p_{4}p_{2}p_{2})^2$. These three configurations yield 
the same charge distribution shown in Fig. 11($\Delta^{-}$).
\par The $\Xi^{0}$(uss), which has the primon configuration
$(p_{1}p_{2}p_{4})^{1}(p_{2}p_{4}p_{2})^{2}$ has the same charge 
distribution of the neutron.  The particles $\Sigma_{c}^{0}$(ddc) and 
$\Sigma_{c}^{+}$(udc), which have the respective configurations 
$(p_{1}p_{2}p_{3})^{1}(p_{3}p_{3}p_{2})^2$ and 
$(p_{1}p_{2}p_{3})^{1}(p_{2}p_{3}p_{1})^2$ produce the configurations of the 
neutron and proton, respectively.
\par The particle $\Xi_{c}^{0}$(dsc) has the four possible configurations
$(p_{1}p_{2}p_{3})^{1}(p_{3}p_{4}p_{2})^2$, \newline
$(p_{1}p_{2}p_{4})^{1}(p_{3}p_{3}p_{2})^2$, 
$(p_{1}p_{3}p_{4})^{1}(p_{3}p_{2}p_{2})^2$ and
$(p_{2}p_{3}p_{4})^{1}(p_{3}p_{1}p_{2})^2$. The first three give the 
configuration of the neutron and the fourth
produces the configuration shown in Fig. 12. As in the case of $\Sigma^{0}$,
probably, the first configuration above is the most stable one, and may be 
the preferred configuration.
\par Doing in the same way we may find the charge configurations of all the
other baryons.\newline
%\newpage
%\rule{0in}{1in}
%\newpage
%\rule{0in}{5.7in}
%\noindent
%\end{document}
\vskip .15in
\noindent
3) CALCULATION OF THE LIGHT ANTIQUARK FLAVOR ASYMMETRY IN THE NUCLEONS
\vskip .1in
\noindent
a) The Proton Sea Content
\par In order to find all the interactions let us divide in three parts: 
interactions in the outer shell, interactions in the inner shell and 
interactions between the two shells, and let us first calculate the 
transitions in Fig. 3.\newline
\noindent
i) Interactions in the outer shell 
\begin{eqnarray*}
p_{1}^{\alpha}  & + &  
\overline{p_{1}^{\alpha}p_{3}^{\gamma}}p_{3}^{\gamma}p_{2}^{\beta}
{\rightarrow}  p_{2}^{\beta} \\
p_{1}^{\alpha}  & + &  
\overline{p_{1}^{\alpha}p_{4}^{\gamma}}p_{4}^{\gamma}p_{2}^{\beta}
{\rightarrow}  p_{2}^{\beta} \\
p_{2}^{\beta}  & + & 
\overline{p_{2}^{\beta}p_{3}^{\gamma}}p_{3}^{\gamma}p_{1}^{\alpha}
{\rightarrow}  p_{1}^{\alpha} \\
p_{2}^{\beta}  & + & 
\overline{p_{2}^{\beta}p_{4}^{\gamma}}p_{4}^{\gamma}p_{1}^{\alpha}
{\rightarrow}  p_{1}^{\alpha} \\
p_{1}^{\alpha}  & + &  
\overline{p_{1}^{\alpha}p_{3}^{\beta}}p_{3}^{\beta}p_{2}^{\gamma}
{\rightarrow}  p_{2}^{\gamma} \\
p_{1}^{\alpha}  & + &  
\overline{p_{1}^{\alpha}p_{4}^{\beta}}p_{4}^{\beta}p_{2}^{\gamma}
{\rightarrow}  p_{2}^{\gamma} \\
p_{2}^{\gamma}  & + & 
\overline{p_{2}^{\gamma}p_{3}^{\beta}}p_{3}^{\beta}p_{1}^{\alpha}
{\rightarrow}  p_{1}^{\alpha} \\
p_{2}^{\gamma}  & + & 
\overline{p_{2}^{\gamma}p_{4}^{\beta}}p_{4}^{\beta}p_{1}^{\alpha}
{\rightarrow}  p_{1}^{\alpha} \\
p_{2}^{\gamma}  & + & 
\overline{p_{2}^{\gamma}p_{3}^{\alpha}}p_{3}^{\alpha}p_{1}^{\beta}
{\rightarrow}  p_{2}^{\beta} \\
p_{2}^{\gamma}  & + & 
\overline{p_{2}^{\gamma}p_{4}^{\alpha}}p_{4}^{\alpha}p_{1}^{\beta}
{\rightarrow}  p_{2}^{\beta} \\
p_{2}^{\beta}  & + & 
\overline{p_{2}^{\beta}p_{3}^{\alpha}}p_{3}^{\alpha}p_{1}^{\gamma}
{\rightarrow}  p_{2}^{\gamma} \\
p_{2}^{\beta}  & + & 
\overline{p_{2}^{\beta}p_{4}^{\alpha}}p_{4}^{\alpha}p_{1}^{\gamma}
{\rightarrow}  p_{2}^{\gamma} 
\end{eqnarray*}

\noindent
ii) Interactions in the inner shell

\begin{eqnarray*}
p_{1}^{\gamma}  & + &  
\overline{p_{1}^{\gamma}p_{3}^{\alpha}}p_{3}^{\alpha}p_{2}^{\beta}
{\rightarrow}  p_{2}^{\beta} \\
p_{1}^{\gamma}  & + &  
\overline{p_{1}^{\gamma}p_{4}^{\alpha}}p_{4}^{\alpha}p_{2}^{\beta}
{\rightarrow}  p_{2}^{\beta} \\
p_{2}^{\beta}  & + & 
\overline{p_{2}^{\beta}p_{3}^{\alpha}}p_{3}^{\alpha}p_{1}^{\gamma}
{\rightarrow}  p_{1}^{\gamma} \\
p_{2}^{\beta}  & + & 
\overline{p_{2}^{\beta}p_{4}^{\alpha}}p_{4}^{\alpha}p_{1}^{\gamma}
{\rightarrow}  p_{1}^{\gamma} \\
p_{1}^{\gamma}  & + &  
\overline{p_{1}^{\gamma}p_{2}^{\beta}}p_{2}^{\beta}p_{3}^{\alpha}
{\rightarrow}  p_{3}^{\alpha} \\
p_{1}^{\gamma}  & + &  
\overline{p_{1}^{\gamma}p_{4}^{\beta}}p_{4}^{\beta}p_{2}^{\alpha}
{\rightarrow}  p_{2}^{\alpha} \\
p_{3}^{\alpha}  & + & 
\overline{p_{3}^{\alpha}p_{2}^{\beta}}p_{2}^{\beta}p_{1}^{\gamma}
{\rightarrow}  p_{1}^{\gamma} \\
p_{3}^{\alpha}  & + & 
\overline{p_{3}^{\alpha}p_{4}^{\beta}}p_{4}^{\beta}p_{1}^{\gamma}
{\rightarrow}  p_{1}^{\gamma} \\
p_{2}^{\beta}  & + & 
\overline{p_{2}^{\beta}p_{1}^{\gamma}}p_{1}^{\gamma}p_{3}^{\alpha}
{\rightarrow}  p_{3}^{\alpha} \\
p_{2}^{\beta}  & + & 
\overline{p_{2}^{\beta}p_{4}^{\gamma}}p_{4}^{\gamma}p_{3}^{\alpha}
{\rightarrow}  p_{3}^{\alpha} \\
p_{3}^{\alpha}  & + &  
\overline{p_{3}^{\alpha}p_{1}^{\gamma}}p_{1}^{\gamma}p_{2}^{\beta}
{\rightarrow}  p_{2}^{\beta} \\
p_{3}^{\alpha}  & + &  
\overline{p_{3}^{\alpha}p_{4}^{\gamma}}p_{4}^{\gamma}p_{2}^{\beta}
{\rightarrow}  p_{2}^{\beta} 
\end{eqnarray*}

\noindent
iii) Interactions between the two shells

\begin{eqnarray*}
p_{1}^{\alpha}  & + &  
\overline{p_{1}^{\alpha}p_{3}^{\gamma}}p_{3}^{\gamma}p_{2}^{\beta}
{\rightarrow}  p_{2}^{\beta} \\
p_{1}^{\alpha}  & + &  
\overline{p_{1}^{\alpha}p_{4}^{\gamma}}p_{4}^{\gamma}p_{2}^{\beta}
{\rightarrow}  p_{2}^{\beta} \\
p_{2}^{\beta}  & + & 
\overline{p_{2}^{\beta}p_{3}^{\gamma}}p_{3}^{\gamma}p_{1}^{\alpha}
{\rightarrow}  p_{1}^{\alpha} \\
p_{2}^{\beta}  & + & 
\overline{p_{2}^{\beta}p_{4}^{\gamma}}p_{4}^{\gamma}p_{1}^{\alpha}
{\rightarrow}  p_{1}^{\alpha} \\
p_{2}^{\gamma}  & + &  
\overline{p_{2}^{\gamma}p_{1}^{\beta}}p_{1}^{\beta}p_{3}^{\alpha}
{\rightarrow}  p_{3}^{\alpha} \\
p_{2}^{\gamma}  & + &  
\overline{p_{2}^{\gamma}p_{4}^{\beta}}p_{4}^{\beta}p_{3}^{\alpha}
{\rightarrow}  p_{3}^{\alpha} \\
p_{3}^{\alpha}  & + & 
\overline{p_{3}^{\alpha}p_{1}^{\beta}}p_{1}^{\beta}p_{2}^{\gamma}
{\rightarrow}  p_{2}^{\gamma} \\
p_{3}^{\alpha}  & + & 
\overline{p_{3}^{\alpha}p_{4}^{\beta}}p_{4}^{\beta}p_{2}^{\gamma}
{\rightarrow}  p_{2}^{\gamma} \\
p_{1}^{\gamma}  & + &  
\overline{p_{1}^{\gamma}p_{3}^{\alpha}}p_{3}^{\alpha}p_{2}^{\beta}
{\rightarrow}  p_{2}^{\beta} \\
p_{1}^{\gamma}  & + &  
\overline{p_{1}^{\gamma}p_{4}^{\alpha}}p_{4}^{\alpha}p_{2}^{\beta}
{\rightarrow}  p_{2}^{\beta} \\
p_{2}^{\beta}  & + & 
\overline{p_{2}^{\beta}p_{3}^{\alpha}}p_{3}^{\alpha}p_{1}^{\gamma}
{\rightarrow}  p_{1}^{\gamma} \\
p_{2}^{\beta}  & + & 
\overline{p_{2}^{\beta}p_{4}^{\alpha}}p_{4}^{\alpha}p_{1}^{\gamma}
{\rightarrow}  p_{1}^{\gamma} \\
p_{1}^{\alpha}  & + & 
\overline{p_{1}^{\alpha}p_{2}^{\beta}}p_{2}^{\beta}p_{1}^{\gamma}
{\rightarrow}  p_{1}^{\gamma} \\
p_{1}^{\alpha}  & + & 
\overline{p_{1}^{\alpha}p_{3}^{\beta}}p_{3}^{\beta}p_{1}^{\gamma}
{\rightarrow}  p_{1}^{\gamma} \\
p_{1}^{\alpha}  & + & 
\overline{p_{1}^{\alpha}p_{4}^{\beta}}p_{4}^{\beta}p_{1}^{\gamma}
{\rightarrow}  p_{1}^{\gamma} \\
p_{1}^{\gamma}  & + &  
\overline{p_{1}^{\gamma}p_{2}^{\beta}}p_{2}^{\beta}p_{1}^{\alpha}
{\rightarrow}  p_{1}^{\alpha} \\
p_{1}^{\gamma}  & + &  
\overline{p_{1}^{\gamma}p_{3}^{\beta}}p_{3}^{\beta}p_{1}^{\alpha}
{\rightarrow}  p_{1}^{\alpha} \\
p_{1}^{\gamma}  & + &  
\overline{p_{1}^{\gamma}p_{4}^{\beta}}p_{4}^{\beta}p_{1}^{\alpha}
{\rightarrow}  p_{1}^{\alpha} \\
p_{2}^{\gamma}  & + &  
\overline{p_{2}^{\gamma}p_{1}^{\alpha}}p_{1}^{\alpha}p_{2}^{\beta}
{\rightarrow}  p_{2}^{\beta} \\
p_{2}^{\gamma}  & + &  
\overline{p_{2}^{\gamma}p_{3}^{\alpha}}p_{3}^{\alpha}p_{2}^{\beta}
{\rightarrow}  p_{2}^{\beta} \\
p_{2}^{\gamma}  & + &  
\overline{p_{2}^{\gamma}p_{4}^{\alpha}}p_{4}^{\alpha}p_{2}^{\beta}
{\rightarrow}  p_{2}^{\beta} \\
p_{2}^{\beta}  & + & 
\overline{p_{2}^{\beta}p_{1}^{\alpha}}p_{1}^{\alpha}p_{2}^{\gamma}
{\rightarrow}  p_{2}^{\gamma} \\
p_{2}^{\beta}  & + & 
\overline{p_{2}^{\beta}p_{3}^{\alpha}}p_{3}^{\alpha}p_{2}^{\gamma}
{\rightarrow}  p_{2}^{\gamma} \\
p_{2}^{\beta}  & + & 
\overline{p_{2}^{\beta}p_{4}^{\alpha}}p_{4}^{\alpha}p_{2}^{\gamma}
{\rightarrow}  p_{2}^{\gamma} \\
p_{3}^{\alpha}  & + &  
\overline{p_{3}^{\alpha}p_{1}^{\gamma}}p_{1}^{\gamma}p_{2}^{\beta}
{\rightarrow}  p_{2}^{\beta} \\
p_{3}^{\alpha}  & + &  
\overline{p_{3}^{\alpha}p_{4}^{\gamma}}p_{4}^{\gamma}p_{2}^{\beta}
{\rightarrow}  p_{2}^{\beta} \\
p_{2}^{\beta}  & + & 
\overline{p_{2}^{\beta}p_{1}^{\gamma}}p_{1}^{\gamma}p_{3}^{\alpha}
{\rightarrow}  p_{1}^{\alpha} \\
p_{2}^{\beta}  & + & 
\overline{p_{2}^{\beta}p_{4}^{\gamma}}p_{4}^{\gamma}p_{3}^{\alpha}
{\rightarrow}  p_{1}^{\alpha} \\
p_{1}^{\alpha}  & + &  
\overline{p_{1}^{\alpha}p_{2}^{\beta}}p_{2}^{\beta}p_{3}^{\alpha}
{\rightarrow}  p_{3}^{\alpha} \\
p_{1}^{\alpha}  & + &  
\overline{p_{1}^{\alpha}p_{2}^{\gamma}}p_{2}^{\gamma}p_{3}^{\alpha}
{\rightarrow}  p_{3}^{\alpha} \\
p_{1}^{\alpha}  & + &  
\overline{p_{1}^{\alpha}p_{4}^{\beta}}p_{4}^{\beta}p_{3}^{\alpha}
{\rightarrow}  p_{3}^{\alpha} \\
p_{1}^{\alpha}  & + &  
\overline{p_{1}^{\alpha}p_{4}^{\gamma}}p_{4}^{\gamma}p_{3}^{\alpha}
{\rightarrow}  p_{3}^{\alpha} \\
p_{3}^{\alpha}  & + &  
\overline{p_{3}^{\alpha}p_{2}^{\beta}}p_{2}^{\beta}p_{1}^{\alpha}
{\rightarrow}  p_{1}^{\alpha} \\
p_{3}^{\alpha}  & + &  
\overline{p_{3}^{\alpha}p_{2}^{\gamma}}p_{2}^{\gamma}p_{1}^{\alpha}
{\rightarrow}  p_{1}^{\alpha} \\
p_{3}^{\alpha}  & + &  
\overline{p_{3}^{\alpha}p_{4}^{\beta}}p_{4}^{\beta}p_{1}^{\alpha}
{\rightarrow}  p_{1}^{\alpha} \\
p_{3}^{\alpha}  & + &  
\overline{p_{3}^{\alpha}p_{4}^{\gamma}}p_{4}^{\gamma}p_{1}^{\alpha}
{\rightarrow}  p_{1}^{\alpha} \\
p_{1}^{\gamma}  & + &  
\overline{p_{1}^{\gamma}p_{3}^{\alpha}}p_{3}^{\alpha}p_{2}^{\gamma}
{\rightarrow}  p_{2}^{\gamma} \\
p_{1}^{\gamma}  & + &  
\overline{p_{1}^{\gamma}p_{3}^{\beta}}p_{3}^{\beta}p_{2}^{\gamma}
{\rightarrow}  p_{2}^{\gamma} \\
p_{1}^{\gamma}  & + &  
\overline{p_{1}^{\gamma}p_{4}^{\alpha}}p_{4}^{\alpha}p_{2}^{\gamma}
{\rightarrow}  p_{2}^{\gamma} \\
p_{1}^{\gamma}  & + &  
\overline{p_{1}^{\gamma}p_{4}^{\beta}}p_{4}^{\beta}p_{2}^{\gamma}
{\rightarrow}  p_{2}^{\gamma} \\
p_{2}^{\gamma}  & + &  
\overline{p_{2}^{\gamma}p_{3}^{\alpha}}p_{3}^{\alpha}p_{1}^{\gamma}
{\rightarrow}  p_{1}^{\gamma} \\
p_{2}^{\gamma}  & + &  
\overline{p_{2}^{\gamma}p_{3}^{\beta}}p_{3}^{\beta}p_{1}^{\gamma}
{\rightarrow}  p_{1}^{\gamma} \\
p_{2}^{\gamma}  & + &  
\overline{p_{2}^{\gamma}p_{4}^{\alpha}}p_{4}^{\alpha}p_{1}^{\gamma}
{\rightarrow}  p_{1}^{\gamma} \\
p_{2}^{\gamma}  & + &  
\overline{p_{2}^{\gamma}p_{4}^{\beta}}p_{4}^{\beta}p_{1}^{\gamma}
{\rightarrow}  p_{1}^{\gamma} \\
p_{2}^{\beta}  & + &  
\overline{p_{2}^{\beta}p_{1}^{\alpha}}p_{1}^{\alpha}p_{2}^{\beta}
{\rightarrow}  p_{2}^{\beta} \\
p_{2}^{\beta}  & + &  
\overline{p_{2}^{\beta}p_{1}^{\gamma}}p_{1}^{\gamma}p_{2}^{\beta}
{\rightarrow}  p_{2}^{\beta} \\
p_{2}^{\beta}  & + &  
\overline{p_{2}^{\beta}p_{3}^{\alpha}}p_{3}^{\alpha}p_{2}^{\beta}
{\rightarrow}  p_{2}^{\beta} \\
p_{2}^{\beta}  & + &  
\overline{p_{2}^{\beta}p_{3}^{\gamma}}p_{3}^{\gamma}p_{2}^{\beta}
{\rightarrow}  p_{2}^{\beta} \\
p_{2}^{\beta}  & + &  
\overline{p_{2}^{\beta}p_{4}^{\alpha}}p_{4}^{\alpha}p_{2}^{\beta}
{\rightarrow}  p_{2}^{\beta} \\
p_{2}^{\beta}  & + &  
\overline{p_{2}^{\beta}p_{4}^{\gamma}}p_{4}^{\gamma}p_{2}^{\beta}
{\rightarrow}  p_{2}^{\beta}.
\end{eqnarray*}

\noindent
We count a total of 74 interactions. Identifying the $q\bar{q}$'s we have:
06 $u\bar{u}$, 06 $d\bar{d}$, 06 $s\bar{s}$, 02 $c\bar{c}$, 02 $t\bar{t}$,
07 $c\bar{d}$, 07 $d\bar{c}$, 07 $t\bar{s}$, 07 $s\bar{t}$, 03 $u\bar{c}$,
03 $c\bar{u}$, 03 $b\bar{s}$, 03 $s\bar{b}$, 03 $u\bar{d}$, 03 $d\bar{u}$,
03 $t\bar{b}$, 03 $b\bar{t}$. Now if we rotate counterclockwise the 
supercolors of the inner shell and count the transitions and the $q\bar{q}$'s
we obtain 70 transitions with the following pairs: 
06 $u\bar{u}$, 06 $d\bar{d}$, 06 $s\bar{s}$, 02 $c\bar{c}$, 02 $t\bar{t}$,
06 $c\bar{d}$, 06 $d\bar{c}$, 06 $t\bar{s}$, 06 $s\bar{t}$, 04 $u\bar{c}$,
04 $c\bar{u}$, 04 $b\bar{s}$, 04 $s\bar{b}$, 02 $u\bar{d}$, 02 $d\bar{u}$,
02 $t\bar{b}$, 02 $b\bar{t}$.
\par After averaging over the two supercolor configurations we obtain 
that the proton 
sea is constituted of 12 $u$, 12 $\bar{u}$, 15 $d$, 15 $\bar{d}$, 16 $s$, 
16 $\bar{s}$, 12 $c$, 12 $\bar{c}$, 6 $b$, 6 $\bar{b}$, 11 $t$, 11 $\bar{t}$. 
Hence, there is an asymmetry between $\bar{u}$ and $\bar{d}$. There is an
excess of 11.11\% of $\bar{d}$ with respect to $\bar{u}$. This result agrees 
quite well with the NMC result$^{23}$ of about $0.147 {\pm} 0.039$. Also, the 
ratio $\bar{u}/\bar{d}=0.787$ is quite close to the Cern NA51 result$^{24}$ 
of about $0.51{\pm}0.04{\pm}0.05$. Moreover, it also agrees with the 
Fermilab E86NuSea 
results$^{25}$ that found the peak of $\bar{d}/\bar{u}$ around $x=0.15$. 
According to the ideas of this work the peak should be around $1/6 {\approx} 
0.17$ since there are six primons in the proton. The ratio 
$\bar{d}/\bar{u}= 1.25$ agrees very well with Fermilab E86NuSea results. It 
is interesting to notice that there is a complete symmetry between each quark 
and its corresponding antiquark. In E86NuSea$^{25}$ one notices that as $x$
increases the ratio $\bar{d}/\bar{u}$ diminishes and tends to 1. The above
results agrees well with these data for as $x$ increases we probe more and 
more the inner shell and its $q\bar{q}$ transitions are 01 $c\bar{d}$, 01 
$d\bar{c}$, 01 $t\bar{s}$, 01 $s\bar{t}$, 01 $u\bar{c}$,
01 $c\bar{u}$, 01 $b\bar{s}$, 01 $s\bar{b}$, 01 $u\bar{d}$, 01 $d\bar{u}$,
01 $t\bar{b}$ and 01 $b\bar{t}$. Hence we obtain 02 $u$, 02 $\bar{u}$, 02 $d$, 
02 $\bar{d}$, 02 $s$, 02 $\bar{s}$, 02 $c$, 02 $\bar{c}$, 02 $b$, 
02 $\bar{b}$, 02 $t$, and 02 $\bar{t}$. Thus, the ratio 
$\bar{d}/\bar{u}{\rightarrow}1$. 
\par We may find all the other asymmetries: ${\bar{s}}/{\bar{c}}=1.333$, 
${\bar{b}}/{\bar{t}}=0.546$, ${\bar{b}}/{\bar{s}}=0.375$, 
${\bar{t}}/{\bar{c}}=0.917$, ${\bar{s}}/{\bar{u}}=1.333$, 
${\bar{s}}/{\bar{d}}=1.067$, and ${\bar{b}}/{\bar{c}}=0.5$. And, with 
respect to  $\bar{u}$ we have the following ratios: 
${\bar{d}}/{\bar{u}}=1.250$,
${\bar{s}}/{\bar{u}}=1.333$, ${\bar{c}}/{\bar{u}}=1$, 
${\bar{b}}/{\bar{u}}=0.5$, ${\bar{t}}/{\bar{u}}=0.917$.
\newline
\vskip .15in
\noindent
b) The Neutron Sea Content
\par Doing the same for the neutron (Fig 4) we obtain the same number of 
transitions. After averaging over the two configurations we obtain the 
following $q\bar{q}$'s: 3.5 $d\bar{c}$, 3.5 $s\bar{t}$, 3.5 $t\bar{s}$, 
3.5 $c\bar{d}$, 6.5 $b\bar{s}$, 6.5 $c\bar{u}$, 6.5 $u\bar{c}$, 
6.5 $s\bar{b}$, 02 $d\bar{u}$, 02 $b\bar{t}$, 02 $u\bar{d}$, 02 $t\bar{b}$, 
06 $u\bar{u}$, 08 $d\bar{d}$, 06 $s\bar{s}$, and 02 $c\bar{c}$. Therefore, 
we find that the neutron sea is constituted of: 
14.5 $u$, 14.5 $\bar{u}$, 13.5 $d$, 13.5 $\bar{d}$, 16 $s$, 16 $\bar{s}$, 
12 $c$, 12 $\bar{c}$, 10.5 $b$, 10.5 $\bar{b}$, 5.5 $t$, 5.5 $\bar{t}$. The 
asymmetry in the neutron is the other way around, that is, there is an 
excess of 3.58\% of $\bar{u}$ with respect to $\bar{d}$. The other 
asymmetries change accordingly to: ${\bar{s}}/{\bar{c}}=1.333$, 
${\bar{b}}/{\bar{t}}=1.909$, ${\bar{b}}/{\bar{s}}=1.524$, 
${\bar{t}}/{\bar{c}}=2.182$, ${\bar{s}}/{\bar{u}}=1.103$, 
${\bar{s}}/{\bar{d}}=1.185$, and ${\bar{b}}/{\bar{c}}=0.875$. And, with 
respect to  $\bar{u}$  we have the following ratios: 
${\bar{d}}/{\bar{u}}=0.931$, ${\bar{s}}/{\bar{u}}=1.103$, 
${\bar{c}}/{\bar{u}}=0.828$, ${\bar{b}}/{\bar{u}}=0.724$, 
${\bar{t}}/{\bar{u}}=0.379$. 
\newline
\vskip .2in
\noindent
4) THE QUARK SEA CONTENT OF THE OTHER BARYONS
\vskip .1in
\par Calculating in the same way we may find the sea content of all the other
baryons. It is quite a lengthy job. Just to have the taste of it, let us 
calculate the sea content of $\Delta^{++}$. We obtaint the  $q\bar{q}$'s:
08 $d\bar{c}$, 08 $c\bar{d}$, 08 $s\bar{t}$, 08 $t\bar{s}$, 12 $u\bar{u}$,
06 $d\bar{d}$, 06 $c\bar{c}$, 06 $s\bar{s}$ and 06 $t\bar{t}$. Thus, the 
$\Delta^{++}$ sea is constituted of: 14 $d$, 14 $\bar{d}$, 12 $u$, 
12 $\bar{u}$, 14 $s$, 14 $\bar{s}$, 14 $c$, 14 $\bar{c}$, 14 $t$ and 
14 $\bar{t}$. Then, there is also an asymmetry of $\bar{d}$ with respect to 
$\bar{u}$ which is an excess of about 7.6\%. And in this baryon there is no 
$b\bar{b}$. 
We obtain a very important conclusion concerning the quark sea in baryons:
a certain $q\bar{q}$ may be absent in a baryon containing three equal quarks.
In this case it happened because the $\Delta^{++}$ has only $p_{1}$'s and
$p_{2}$'s, and so we can not combine none of them to form either $b$ or 
$\bar{b}$. We suspect, then, that the $\Delta^{-}$, for example, has no
$t\bar{t}$.
\vskip .2in
\noindent
5) QUARK CONFINEMENT, THE SIZES OF QUARKS, PRIMON MASS AND THE GENERATION OF 
QUARK MASS
\vskip .1in
\par It is reasonable to consider that the range of the superstrong 
interaction is smaller than that of the strong interaction. Taking into account 
the above considerations it is reasonable to assume that the effective molecular 
type potential, $V'(r)$(Fig.14), between two primons has the form
\begin{eqnarray} V'(r) &=& \frac{{\alpha}e^{-\mu_{1}r}}{r} - 
\frac{{\beta}e^{-\mu_{2}r}}{r}\end{eqnarray}			
\noindent
where $\alpha = \mho_{1}\mho_{2}>0$ and $\beta = g_{1}g_{2}>0$. The strong 
charges are $g_{1}$ and $g_{2}$ and the superstrong charges are given 
by $\mho_{i}$. Let us try to find the conditions for having a minimum in the
potential energy. At the minimum we have 
\begin{eqnarray} \left(\frac{\alpha}{\beta}\right) e^{-(\mu_{1} - 
\mu_{2})r_{0}} &=& \frac{\mu_{2} + \frac{1}{r_{0}}}{\mu_{1} + 
\frac{1}{r_{0}}}.\end{eqnarray}					
\noindent
This transcendental equation may only be solved numerically, but considering
the graphs of both sides of it(Fig. 15) we are able to see that if 
$\alpha>\beta$, and $\mu_{1}>\mu_{2}$, then there is one value of $r_{0}$ which 
satisfies the equation. The position $r'$(in Fig.14) is 
\begin{eqnarray} r' &=& \frac{1}{\mu_{1} - 
\mu_{2}}\ln{\left(\frac{\alpha}{\beta}\right)}.\end{eqnarray}	
\noindent
The relation between $r_{0}$ and $r'$ is given by
\begin{eqnarray} r_{0} &=& r' + \frac{1}{\mu_{1} - 
\mu_{2}}\ln{\left(\frac{\mu_{1} + \frac{1}{r_{0}}}{\mu_{2} + 
\frac{1}{r_{0}}}\right)}.\end{eqnarray}					
\noindent
The minimum is easily found to be 
\begin{eqnarray} V'(r_{0}) &=& - \frac{\alpha(\mu_{1} - 
\mu_{2})e^{-\mu_{1}r_{0}}}{1 + \mu_{2}r_{0}} < 0.\end{eqnarray}		
\noindent
It is easy to see that $V'(r_{0})$ and $r_{0}$ are very influenced by $\alpha$ 
and $\mu_{1}$. 
\par The heavier a quark is the deeper should be the well generated by the
two primons. Also, the well should be narrower because the heavier a quark is 
the more it must be bound. For a given quantum number, $n$, the energy of a 
well increases as it narrows. The potential of the top quark is extremely deep 
since it is much more massive than the other quarks are. We are able, then, to 
understand the decays of quarks. The lowest level is, of course, the ground 
state of the $u$ quark. The ground state of the $d$ quark is slightly above 
that of the $u$ quark, and the ground state of the $s$ quark is above the 
ground state of the $d$ quark. This should also happen for the other heavier 
quarks. Therefore, we expect that the potentials of all quarks should be as 
shown in Fig.(16)(the well of the $d$ quark is not shown). {\it{Quark masses 
are just the levels in the wells}}. These potentials are in line with the 
observed decay chain $b{\rightarrow}c{\rightarrow}s{\rightarrow}u$ and with 
the decays $d{\rightarrow}ue{\bar{\nu}}_{e}$, $b{\rightarrow}s\gamma$.
What about the masses of primons? Since the quarks $u$ and $d$ have about the
same mass of 0.3GeV, we expect that $p_{1}$, $p_{2}$ and $p_{3}$ have the
same mass. But, since the combination $p_{1}p_{3}$ generates the $c$ quark
which is much heavier(about 1.5GeV)than $u$, we can infer that the different 
masses of quarks comes from the strong and superstrong interactions. Thus, we 
may suspect that all primons have the same mass which is a sort of primitive, 
inherent mass, which may be of the same kind of the mass that leptons have.
\par Let us now see what is behind quark masses. Several researchers have 
tried to relate their masses to something more fundamental. In order to do 
this let us approximate each well of Fig. 16 by an infinite potential well. 
Since each mass corresponds to a single level in each well, and considering 
that primons have the same mass, we obtain that each mass should be related 
to the average distance between each pair of primons, that is, to the width 
of each well. Therefore, we should have the approximate relations:
\begin{eqnarray} 0.3 &=& \frac{C}{R_{u}^{2}};\end{eqnarray}	
\begin{eqnarray} 0.5 &=& \frac{C}{R_{s}^{2}};\end{eqnarray}	
\begin{eqnarray} 1.5 &=& \frac{C}{R_{c}^{2}};\end{eqnarray}	
\begin{eqnarray} 5.0 &=& \frac{C}{R_{b}^{2}};\end{eqnarray}	
\begin{eqnarray} 150 &=& \frac{C}{R_{t}^{2}}.\end{eqnarray}	

\noindent
where $C$ is a constant and $R_{u}$, $R_{s}$,$R_{c}$,$R_{b}$ and $R_{t}$ are
the widths of the wells. As we discussed in section 2, $R_{u}{\approx}0.5$F. We
may assume that $R_{u}=R_{d}$. We arrive at the very important relations about
quark sizes:
\begin{eqnarray} R_{s}^{2} &=& \frac{3}{5}R_{u}^{2} = 
0.6R_{u}^{2}{\approx}0.15F^{2},\end{eqnarray}			
\begin{eqnarray} R_{c}^{2} &=& \frac{5}{15}R_{s}^{2} = 
\frac{5}{15}\frac{3}{5}R_{u}^{2} = 0.2R_{u}^{2}{\approx}0.05F^{2},
\end{eqnarray}
\begin{eqnarray} R_{b}^{2} &=& \frac{15}{50}R_{c}^{2} = 
\frac{15}{50}\frac{5}{15}\frac{3}{5}R_{u}^{2} = 
0.06R_{u}^{2}{\approx}0.015^{2}F ,\end{eqnarray}			
\begin{eqnarray} R_{t}^{2} &=& \frac{5}{150}R_{b}^{2} = 
\frac{5}{150}\frac{15}{50}\frac{5}{15}\frac{3}{5}R_{u}^{2} {\approx} 
0.002R_{u}^{2}{\approx}0.0005F^{2}.\end{eqnarray} 			

\par It is quite interesting that there are some very fascinating relations.
A very important one is:
\begin{eqnarray} \frac{R_{u}^{2}}{R_{d}^{2}} &=& 1 \end{eqnarray}
\begin{eqnarray} \frac{R_{s}^{2}}{R_{c}^{2}} &=& 3 \end{eqnarray}
\begin{eqnarray} \frac{R_{b}^{2}}{R_{t}^{2}} &=& 30, \end{eqnarray}	

\noindent
and, thus, there is a factor of 10 between the last two relations. Other 
quite important relations are:
\begin{eqnarray} \frac{R_{s}^{2}}{R_{b}^{2}} &=& 10 \end{eqnarray}	
\begin{eqnarray} \frac{R_{c}^{2}}{R_{t}^{2}} &=& 100, \end{eqnarray}	

\noindent
which has the same factor of 10. Therefore, the heavier a quark is the smaller
it is. The top quark is extremely small, 
with a radius around $0.02F$. 
\par We may explain quark (and prequark) confinement in the following way. Let 
us consider the three quarks that compose a certain baryon, a proton, for 
example. When it is hit by a very energetic proton, for example,  due to the 
transformation of kinetic energy into potential energy at the moment of the 
interaction, higher levels of the potential wells are reached, that is, the 
excited states of the proton(s) are reached. It is reasonable to assume that 
quarks may also have excited states, that is, beyond the ground state mass a 
quark may have excited state masses. Thus, the larger the energy used the more 
primons and quarks interact in the potential wells, and then higher excited 
states(that is, larger masses) of the quarks and of baryons are reached. 
Therefore, the more we try to free them, the more they get bound. Moreover,
the harder nucleons are hit the more $q\bar{q}$(that is, $p\bar{p}$) pairs are 
generated, and thus, quarks never get free.
\par In the light of the previous section we may consider that 
$r_{0}{\approx}0.5$fm and $\mu_{2}{\approx}1fm^{-1}$. Considering that the 
superstrong interaction is about 500 times stronger than the strong 
interaction, we have $\beta{\approx}1$ and $\alpha{\approx}500$. With these 
numbers we obtain  $r' {\approx} 0.385$fm and $V(r_{0}){\approx}0.3GeV$ which 
is of the same order of the $u$ mass and is a quite reasonable figure. Of 
course, $r'$ is not very 
influenced by $\alpha$ because $\ln{\alpha}$ changes slowly. For example, if 
we use $\alpha{\approx}100$ and $\beta{\approx}1$, we obtain 
$r'{\approx}0.369$fm. In the following sections we will discuss the range of 
the superstrong interaction and, therefore, we will be able again to estimate 
a lower bound of the energies of its bosons. 
\vskip .2in
\noindent
6) THE TRUE POTENTIAL OF A QUARK PAIR AND THE `USUAL' QCD POTENTIAL
\vskip .1in
\par As two quarks($Q\bar{Q}$) are brought to a very close distance(below 
0.5 fm, presumably) from each other they should experience the strong force
and also the superstrong force. The latter will be experienced to a lesser
degree as compared to its action between primons because the interacting
distance between quarks is supposed to be larger than in the case of the 
distance between two primons. Because $Q\bar{Q}$ form bound states there should
exist a net molecular potential well between them. At large distances it should
be dominated by the strong force(Yukawa) potential 
\begin{eqnarray} V_{Q\bar{Q}}(r) &=& 
- \frac{(g_{s}^{Q})^{2}e^{-\mu_{s}{r}}}{r}.\end{eqnarray}              

\noindent
Let us analyze how  this potential is compared to the usual QCD potential
\begin{eqnarray} V_{QCD}(r) &=& -\frac{4}{3}\frac{\alpha_{s}}{r} + {\beta}r.
\end{eqnarray}                                                  

\noindent
We do not know the value of $g_{s}^{Q}$, but we may assume that 
$(g_{s}^{Q})^{2}$ is of the order of $\alpha_{s}$. Then, it is easy to see 
that for $\mu{r}{\ll}1$ the two potentials may have the same order of 
magnitude. When $r$ increases $V_{Q\bar{Q}}$ will be above the first term of 
$V_{QCD}$, which decreases slowly to zero. The term ${\beta}r$ raises the 
potential and makes it get closer to $V_{Q\bar{Q}}$, as is shown in Fig.(17). 
\par It has been said in most textbooks on elementary particles that the data
of the experiments UA1 (Arnison et al.)$^{(26)}$ and UA2 (Bagnaia et 
al.)$^{(27)}$ at the CERN $p\bar{(p)}$ collider provide the best 
direct evidence that the QCD potential at small $r$ is proportional to $1/r$. 
But, the data show much more than this simplistic conclusion. The data is shown
in Fig.(18). Parameterizing the data in the form 
$(sin{\frac{\theta}{2}})^{-n}$ one obtains $n=4.16$ for the slope up to
$\sin^{4}{\theta/2} {\approx} 0.1$. Notice that the center of the first point 
at the top is off the straight line somewhat. This deviation may indicate that
the differential cross section tends to saturate as we go to small angles.
A better fitting to the data may be provided by a differential cross section
of the form 
\begin{eqnarray} \frac{d\sigma}{d{\Omega}} &{\propto}& \frac{1}
{(1 + 4(\frac{k}{\mu_{ss}})^{2}{sin}^{2}\frac{\theta}{2})^{2}}\end{eqnarray}

\noindent
which means that the interacting potential for very short distances is of the 
Yukawa type.  Since $q^{2}=2000 GeV^{2}$, $q$ is about 45 GeV, and so, 
$k=1.56{\times}10^{3} fm^{-1}$. For $sin^{4}{\theta/2} {\approx} 0.01$, 
$\theta {\approx} 37^{o}$. This is not a small angle, and if the 
saturation is already beginning for such angles, then 
\begin{eqnarray} 4(\frac{k}{\mu_{ss}})^{2}{sin}^{2}{\theta/2} &{\sim}& 1.
\end{eqnarray}                                                

\noindent
This means that $\mu_{ss} {\sim} (10^{2} - 10^{3}) fm^{-1}$, and thus the order 
of magnitude of the range of the superstrong interaction is  
$(10^{-2} - 10^{-3}) fm$. Therefore, its bosons have masses in the range 
(10 - 100) GeV.  Several experiments have, indeed, shown that the `strong
force' becomes repulsive at distances less than about 0.45 fm. Of course, it is
not the strong force. It is the action of the superstrong force.
\par The relative success of the usual QCD potential is due to the use of
several adjustable parameters in the models and due to the existence of the
two primon shells described in section 2. As we saw the inner shell is
quite close to the center(mean radius of about $r_{1}=0.15$fm) while the outer 
layer has quite a large mean radius $r_{2}$ of about 0.65fm, that is, 
$r_{2}{\approx}4r_{1}$. Therefore, it is almost a central potential for the
primons of the outer layer. That is, we can say that there is an approximate
central potential due to the existence of a strong charge $g_{1}$ at the center
and another strong charge in the outer layer. Let us now estimate the value of
the coupling constant $\alpha_{s}{\approx}g_{1}g_{2}$ between the two layers. 
Each quark has a strong charge of about 1/3. Thus, a primon has a strong charge 
of about 1/6. But each layer has three primons, and therefore, each layer has a 
strong charge of about $3{\times}\frac{1}{6}=\frac{1}{2}$. Then, the product 
$g_{1}g_{2}$, that is, $\alpha_{s}$, is about 0.25 which is the experimental 
value of $\alpha_{s}$ at $Q=3$GeV. As discussed above, at very high Q the 
effective coupling should diminish due to the action of the superstrong 
interaction. The lowest value of $\alpha_{s}$(around 0.1) at $Q=100$GeV does 
include the effect of the superstrong force. Please, find below a very 
important discussion on $\alpha_{s}$.\newline
\vskip .2in
\noindent
7) ASYMPTOTIC FREEDOM
\vskip .1in
\par ``Asymptotic'' freedom is, actually, another evidence of the existence of 
the superstrong interaction. As we go to higher energies(i.e., to smaller $r$) 
there is more and more the influence of the superstrong force, which being 
repulsive, diminishes the strength of the strong force.  As has been shown 
there should exist a molecular type of potential between two quarks whose 
mathematical expression may be very complicated. The effective force may even 
become zero at the bottom of each well. Just to show that the effective 
coupling constant diminishes with $r$, let us approximate the effective 
potential by 
\begin{eqnarray} V_{eff} &=& -\beta_{s}\frac{e^{-\mu_{s}r}}{r} + 
\beta_{ss}\frac{e^{-\mu_{ss}r}}{r}\end{eqnarray}              

\noindent
where $\beta_{s}=(g_{s}^{Q})^{2}$($Q$ for quark) , and 
$\beta_{ss}=(g_{ss}^{Q})^{2}$ are the 
strong and superstrong couplings, respectively. But according to QCD the 
effective potential for small $r$ should be given by
\begin{eqnarray} V_{eff} &=& - \frac{\alpha_{s}}{r}.\end{eqnarray}     

\noindent
We expect that $\beta_{ss}{\gg}\beta_{s}$, and $\mu_{ss}{\gg}\mu_{s}$. Just to 
have a practical example, let us make $\beta_{ss} = 500\beta_{s}$ 
and $\mu_{ss} = 20\mu_{s}$. This means a boson with a mass of about 3GeV. 
Making $\beta_{s} = 0.3$ GeVfm, and considering that 
$\mu_{s}{\approx}1$ fm$^{-1}$, we obtain that $\alpha_{s}$ should be 
\begin{eqnarray} \alpha_{s} &=& 0.3(e^{-r} - 500e^{-20r}).\end{eqnarray}
                                                                 
\noindent
The values of $\alpha_{s}$ for different values of $r$ are shown in Table 4.
We will see later on that the above values for $\beta_{ss}$ and $\mu_{ss}$ 
agree quite well with supernovae data. Since a baryon has 
6 primons, $\beta_{ss}$ of each primon, $\beta_{ss}^{p}$ is about
$\frac{1}{6}$500X0.3(GeV)(fermi)=100 GeVfm, and because of this the net 
repulsion between two baryons is the result of 36 repulsive terms. 
Therefore the effective superstrong coupling between two baryons is about 
36x100GeVfm=3600GeVfm.
\par Looking at Fig. 17 we see that the higher the energy the better QCD gets.
\newline
\vskip .2in
\noindent
8) THE SUCCESS OF QCD, THE NATURE OF GLUONS AND THE NUMBER OF LEPTONS
\vskip .1in
\par Since we cannot ignore the great success of QCD, gluons have to exist. But
as we saw in the last three sections primons and quarks have to carry strong
charges $g_{s}$ and $g_{ss}$, respectively. Therefore, pions(and their 
resonances) are exchanged inside hadrons. And this must be the reason for the 
size of a hadron which is about $m_{\pi}^{-1}$. Thus, the further two primons
can go apart is about 1fm. Due to the existence of the superstrong interaction 
among primons and quarks, its bosons also act among them. 
\par What are gluons, then? As we saw above there should exist a sort of 
effective molecular potential between two quarks. Then, gluons have 
to be the quanta of the vibrations in such a well, that is, they are 
collective excitations of the effective potentials among quarks.
\par QCD is, then, a mean field theory, that is, it is a theory of an 
effective field which is an admixture of the strong and superstrong fields. 
And it describes Nature so well exactly because it already takes into account 
the two fields. Although it is great theory it has inconsistencies. For 
example, as we know the strong coupling constant considered in QCD is energy 
dependent. Also, sometimes QCD considers that
quarks do not have masses and sometimes it treats them as constituent quarks.
Sometimes it treats them as being free, and then sometimes it considers that 
there is a Coulombian potential acting between any two quarks. Moreover, 
we also treat them as being the so-called partons and these, also being 
$q\bar{q}$ pairs. We may suspect that when we consider them as being partons
we may disregard their masses and when we consider them as being constituent
quarks we need to take into \newpage

\rule{0in}{.5in}
\begin{center}
\begin{tabular}{||l|c||} \hline
& \\
$r$(fm) & $\alpha_{s}$(GeV$\cdot$fm) \\
& \\
\hline\hline 
0.500 & 0.175 \\ 
\hline 
0.400 & 0.151 \\
\hline
0.390 & 0.142 \\
\hline
0.380 & 0.130 \\
\hline
0.370 & 0.116 \\
\hline
0.360 & 0.097 \\
\hline
0.350 & 0.075 \\
\hline
0.340 & 0.047 \\
\hline
0.330 & 0.012 \\
\hline
0.329 & 0.008 \\
\hline
0.328 & 0.004 \\
\hline
0.327 & -0.0003 \\
\hline
... & ... \\
\hline\hline
\end{tabular}
\end{center}
\vskip .3in
\begin{center}
\parbox{4.5in}
{Table 4. An example of how the effective coupling constant(which is the 
result of the strong and superstrong interactions) varies with $r$. In this
case the superstrong coupling constant is chosen to be 5000 times the strong 
coupling constant(of a quark) and the ranges of the strong and supertrong
interactions are 1 fm and 0.02 fm, respectively.}
\end{center}

\rule{0in}{.2in}

\noindent
account their masses. The worst confusion is, of
course, to assume that quarks are partons since partons are primons. It is, 
therefore, more than obvious that we need to put some order in all this 
picture. In order to construct a definitive theory of elementary particles we 
need to take some important steps, among which we may include:
\begin{enumerate}
\item A quark is a system of two primons(prequarks) which has
an average size ranging from 0.5F($u$ and $d$) down to 0.02F(top quark).
\item There are four primons which combined in pairs form the six quarks.
Therefore, there are only six quarks.
\item There is a superstrong interaction that acts between two primons.
\item The strong interaction also acts between two primons. 
\item Each color is generated by two different supercolors which are
manifestations of the strong field at the prequark level. 
\item The basic bosons of the strong interaction are the three pions but other
mesons also act between any two primons. 
\item The superstrong interaction is carried out by three heavy vector
bosons with masses in the GeV range.
\item There is a residual superstrong interaction between two quarks. The
strong force also acts between two quarks.
\item Primons and leptons have intrinsic, inherent, inborn masses. 
\item Gluons are effective bosons of a collective interaction.
\item The masses of quarks and hadrons are the energy levels of effective 
molecular potential wells which are the result of the superstrong and strong 
interactions combined.
\end{enumerate}
\par A possible consequence of the existence of primons is the number of 
leptons which is such that the weak decays among primons take place correctly.
As we saw above $p_{4}$ has the following decays \newline
$p_{4}{\rightarrow}p_{1}e^{-}\bar{\nu_{e}}$, 
$p_{4}{\rightarrow}p_{1}\mu^{-}\bar{\nu_{\mu}}$, and
$p_{4}{\rightarrow}p_{1}\tau^{-}\bar{\nu_{\tau}}$.

\noindent
Therefore, we do not need to have other leptons to cover all the weak decays of
$p_{4}$, that is, to cover all the weak decays of hadrons.
\vskip .15in
\noindent
9) THE SUPERNOVAE EVIDENCE FOR THE SUPERSTRONG INTERACTION
\vskip .1in
\par Type II supernovae release the enormous energy$^{28}$ of about $10^{51}$ 
ergs, which corresponds to initial shock velocities of $5{\times}10^7$m/s. 
Several different models have attempted without success$^{28}$ to explain how 
the gravitational energy released during collapse could reach the outer layers 
of the star. Moreover it is not clear at all why the core explodes in the 
first place. No known theoretical model to date has been able to make the core 
collapse and to produce both a gas remnant and a neutron star$^{28}$. We may
shed some light onto this issue by simply proposing that the 
explosion is caused the superstrong interaction. The mechanism of the 
explosion may be as follows: Due to gravitational collapse all 
nucleons(and electrons) of the star get more and more squeezed up to the point 
when the repulsion among the nucleons begin to play an important role
because of the very small range of this new interaction. With further collapse 
a point is reached when the repulsion overcomes the gravitational attraction 
and a rapid expansion takes place in the star as a whole so that the expansion 
of the outer layers is not caused by the energy released by the core. While 
the core is collapsing the outer layers are also 
doing the same and when the core begins to expand the gravitational collapse 
of the outer layers diminishes so that the overall repulsion between the 
nucleons in the outer layers halt the collapse and these layers also expand. 
\par The energy of $10^{51}$ergs corresponds to an energy of about $0.6$MeV per
baryon. A nucleon has a radius of about 1.4fm and the superstrong interaction
will be dominant only if the nucleons are very close to each other. Taking
the distance from their centers we may consider that when they are very 
squeezed they are separated by about 2fm(from center to center). This is 
consistent with nuclear physics data. Therefore, we can write
\begin{eqnarray} 3600(GeV) \frac{e^{-2\mu_{ss}}}{2} &=& 0.6(MeV)
\end{eqnarray}							

\noindent 
where and $\mu_{ss}$ is given in $fm^{-1}$.  We obtain 
$\mu_{ss}\approx{7}$ $fm^{-1}$, which 
means that the mediator of the new interaction has a mass of about 1GeV. This
is quite in line with what we calculated in section 7. It is important to say
that the search for quark composition has been aimed at too high energies, in
the TeV region. But, as we see, the superstrong bosons are not as massive.
\par These considerations are also in line with the repulsion which is one of 
the features attributed to the strong force at very small distances. 
Walecka$^{29}$ 
has developed a theory of highly condensed matter, in the domain of the strong
force, assuming that the strong repulsion is due to the exchange of $\omega$. 
He constructed a relativistic Lagrangian that allows nucleons to interact
attractively by means of scalar pion exchange and repulsively by means of the 
more massive vector meson $\omega$. At very high densities he finds that
the vector meson field dominates and one recovers Zel'dovich result
\begin{eqnarray} \displaystyle P \rightarrow {\rho}c^{2};v_{s}{\rightarrow}c.
\end{eqnarray}							

\noindent 
where $v_s$ is the thermodynamic speed of sound in the medium$^4$, $P$ is the
pressure, and $\rho$ is the density. In his article he defines the two 
dimensionless coupling constants 
\begin{eqnarray} \displaystyle {c_s}^{2} &=& \frac{{g_s}^2}{{\hbar}c^3}
\frac{M^2}{\mu^2},
\end{eqnarray}							
\begin{eqnarray} \displaystyle {c_v}^{2} &=& \frac{{g_v}^2}{{\hbar}c}
\frac{M^2}{m^2},
\end{eqnarray}								

\noindent
in which $\frac{{g_s}^2}{{\hbar}c^3}$ and $\frac{{g_v}^2}{{\hbar}c}$
are, respectively, the coupling constants of the strong(pionic) and vectorial
fields, and $M$, $\mu$ and $m$ are the inverse Compton wavelengths
\begin{eqnarray} M &{\equiv}& \frac{m_{b}c}{{\hbar}} \\
\mu &{\equiv}& \frac{m_{s}c}{{\hbar}} \\
m &{\equiv}& \frac{m_{v}c}{{\hbar}}					
\end{eqnarray}

\noindent
where $m_b$ is the proton mass. Using data of nuclear matter he obtains 
${c_s}^{2}=266.9$ and ${c_v}^{2}=195.7$. Considering that the vector field is 
actually caused by the superstrong interaction we observe that 
the value of $c_v$ is consistent with the figures that we obtained for 
$\mu_{ss}$ and $\beta_{ss}^2$.
\par Claims of the experimental discovery of a new interaction have been 
made by Baurov and Kopajev$^{30}$(and references therein). According to them 
the new interaction is manifested by the magnetic activity of solar flares 
on the surface of the Sun. As they say ``The new interaction must be very 
strong in that case because the vector potential $\rightarrow{A}$ of the 
tubes is of the order of ${\sim}10^{11}$Gs.cm ..." It may actually be the 
same superstrong
interaction we have discussed above. Although it has been manifested on the
surface of the Sun, its origin may be traced to its center where the 
density is of the order of the nuclear density. Moreover we may explain why
it happens in flares in the following way: Due to gravitational contraction
the density may increase momentaneously to such a point that the superstrong 
interaction becomes important. This is especially true right at the center of 
the Sun. Then, the very squeezed baryons(nucleons) are
expelled to the outer layers of the Sun due to their mutual repulsion, and, 
in such a process, we expect that the magnetic activity will increase and, 
thus, the solar flares are generated.\newline
\vskip .15in
\noindent
\noindent
10) THE PLANETARY EVIDENCE FOR THE SUPERSTRONG INTERACTION
\vskip .1in
\par As McCaughrean and Mac Low$^{31}$ say ``Mass outflow is known to be a 
common and perhaps inevitable part of star formation". Edwards et al.$^{32}$
also states that observations of young low-mass stars at optical, 
near-infrared, and milimeter wavelengths often reveal highly collimated bipolar
jets and molecular outflows. And jets carry large amounts of energy and 
momentum from the central regions of young stellar objects$^{33}$(YSOs). 
Moreover, between 25\% to 75\% of YSOs in the Orion nebula appear to have 
disks$^{34}$. 
\par It is very important to point out that no theory of planet formation is
able to offer a reasonable explanation for the origin of the large amount of 
iron that is found in the cores of all planets, and the heavy elements(such as
uranium) found on Earth and for sure in other planets. Of course, the jets and 
outflows mentioned above contain the planetary iron. How was it formed and 
expelled? We may explain it as follows: Due to inhomogeneities when the solar 
nebular collapsed some parts of it got so squeezed that all heavy elements
were formed and expelled due to the action of the superstrong
interaction. That is, a small part of the sun suffered a supernova-like 
explosion. We easily observe that it was just a small ejection since the mass
of all planets is only about one thousandth of the mass of the Sun. We expect
that more massive stars eject more mass from their center.\newline
\vskip .2in
\noindent
11) EVIDENCES FOR THE SUPERSTRONG INTERACTION FROM GALACTIC\newline
FORMATION AND EVOLUTION AND FROM THE FORMATION OF STRUCTURE
\newline
\vskip .1in
\noindent
I) The Formation of Galaxies
\newline
\par The superstrong force explains the formation of galaxies in a quite easy
way. As M\'{a}rquez et al.$^{35}$ and Yahil$^{36}$ have shown, high redshift
galaxies are very small, having diameters smaller than 1kpc. This is so
because they are born as quasars which have sizes much smaller than 1kpc.
In the beginning of the Universe, because of the repulsion caused by the
superstrong interaction, the nucleons had high velocities in the range of 
$10^4$km/s, but due to the action of the strong force this velocity 
diminished, and probably went down to $10^3$km/s. Gravity diminished further
this velocity to the range $10^{2}$-$10^{3}$km/s,  which is the range of the
peculiar velocities of galaxies. Assuming $v{\approx}10^{3}$km/s, and using
the virial theorem we obtain that a new born quasar had a radius
\begin{eqnarray} R &{\approx}& \frac{GM}{2v^2}\end{eqnarray}

\noindent
which is about 500 light years, where $M$ is the typical mass of a 
galaxy($10^{11}$ suns). This is quite consistent with the
data of M\'{a}rquez et al.$^{35}$ and Yahil$^{36}$. In the data of M\'{a}rquez
et al. we see that some quasars have companion galaxies just some kpc away from
them. For example, there is a galaxy only 28kpc away from the quasar 3C 215 and
it is surrounded by 14 galaxies within ${\pm}30$". If we assume that 
they are going away from each other with velocities around $10^{3}$km/s, 
going backwards to the time when they were were formed(touching each other), 
they had sizes of approximately 100pc, which is very consistent with the above
calculation. This means that quasars(galaxies) were formed when the temperature
was about $10^{9}$K, just after the formation of the first light nuclei. The 
Universe was very 
young, less than a second old. Considering the typical number of baryons of
a galaxy in a sphere with a radius of about 500 light years we obtain that
the average distance among nucleons was only $10^{-4}$m and corresponds to
an average density $\rho{\approx}10^{-16}$kg/m$^3$. For 
$v{\approx}10^3$km/s we find that the Universe was about $10^{-11}$s old. 
The size of an atom is about 1\AA, so that, atoms were formed  $10^{-17}$s 
after the Big Bang. The Big Bang was then a sort of supernova explosion. If
we form an enormous squeezed nucleus with all baryons we obtain a radius of
about 50000km, which is approximately equal to Jupiter radius. It was not a 
black hole because it expanded due to the action of the superstrong 
interaction.
\par Let us now use Jeans criterion. The mass contained in a sphere of radius
$\lambda_{J}$  is the Jeans mass
\begin{eqnarray} M_{J} &=& \frac{4}{3}{\pi}{\lambda_J}^{3}\rho\end{eqnarray}

\noindent
where $\rho$ is the density and $\lambda_J$ is given by
\begin{eqnarray} \lambda_{J} &=& \sqrt{\frac{\pi}{G\bar{\rho}}}c_{S}
\end{eqnarray}

\noindent
where $c_S$ is the sound velocity. Using $c_{S}=10^3$km/s and the
above figure for $\rho$  we obtain $\lambda_{J}{\approx}10^{19}$m and 
$M_{J}=10^{41}$kg which is quite consistent with the virial theorem 
calculation. This $M_J$ is the mass of a typical galaxy like the Milky Way, 
and $10^{19}$m is about 300kpc. 
\par Very close to the beginning of the Universe, when the density was just
above nuclear density ($10^{18}$kg/m$^3$) and the nucleons were still with 
supernova velocities of about 50000km/s we obtain 
$\lambda_{J}{\approx}10^4$m and $M_{J}{\approx}10^{30}$kg. This $M_J$ is the
typical mass of a star like our Sun. This calculation means that quasars and 
stars were formed almost at the same time. These were the primordial stars. 
This is in line with the arguments and data of Shaver et al.$^{37}$ 
and Pettini et al.$^{38}$ Shaver et al. show in their work
that the formation rate of stars and the space density of quasars peak at
the same redshift($z{\approx}2.5$) and have the same redshift dependence.
This fact links the formation of quasars to the formation of stars and rules
out the existence of black holes. It is the superstrong interaction that
avoids the formation of these hypothetical objects. Therefore, we should 
expect to have old stars and also neutron stars in the bulges of galaxies. 
\par This picture of stars and quasars formation raises the possibility of
having neutron quasars which we may call 'quasons'. They could provide the
energy for the very energetic cosmic rays and would provide quite a lot of 
dark mass.
\newline
\vskip .1in
\noindent
II) The Formation of Structure(Voids and Clusters)
\newline
\par Only clumps of matter exceeding the Jeans mass stabilize and virialize.
Therefore, clumps of different sizes going apart from each other populated the
young Universe. These clumps were quasars, young galaxies and groups of them
with different numbers. The sizes of the biggest clumps can be calculated using
Jean's criterion. With an average density of about $10^{-29}$g/cm$^3$ and a 
sound velocity around $10^6$m/s we obtain 
$\lambda_{J}{\approx}10^{24}m{\approx}67$Mpc which is of the order of 
magnitude of the largest voids. Solving for the ratio
$v=d/t=67Mpc/10^{17}$ we find $v=2{\times}10^7$m/s, which is quite reasonable
with the above arguments and figures. As the density diminishes
larger and larger structures are formed. When the density was higher the
structures were smaller. For simplicity let us consider that the progenitor 
of a void was a spherical volume with a radius $r$. Since the volume of the 
void increases the density decreases and, due to gravity, galaxies are 
decelerated as they go apart. This will generate underdense and 
overdense regions and, considering two or more voids, we generate the clusters
and superclusters. Let us consider a galaxy on the surface of a sphere of
radius $r$. As $r$ increases $v$ diminishes as we can see below. In order to
make the Universe to close we need to have
\begin{eqnarray} E &=& \frac{1}{2}mv^{2} - {\frac{4\pi}{3}}G{\rho}r^2 = 0.
\end{eqnarray}

\noindent
Thus, we obtain 
\begin{eqnarray} v^{2} &=& \frac{2GM}{r}\end{eqnarray}

\noindent
which is quite consistent with the data and with our reasoning above. As
we know the peculiar velocities of relatively close galaxies are in the
range 150-450km/s. That is, the velocity of a galaxy depends on the mean 
redshift of the bubble where it is. In very far bubbles the peculiar 
velocities of galaxies should be higher than these figures. It should also 
depend on the size of each bubble. Galaxies close to the surface of large  
bubbles should have small peculiar velocities. Therefore, in each void the
law between velocity and distance should be given by the above equation.
\par In the light of what we are saying above we have to reinterpret 
Hubble's law
\begin{eqnarray} v = Hr. \end{eqnarray}

\noindent
This equation should be right for the centers of voids(bubbles), but it 
cannot be right for any two galaxies. Let us consider, for example, the 
velocities of two galaxies of a void, one close to the center and the other 
close to its
surface. In such a case we would find two different values for H. In other
words, we should always discount local effects due to the expansion of each
bubble and to the position of a particular galaxy in a void.
\par It is quite remarkable the similarity between a supernova explosion and
the Big Bang. We can observe it visually. In supernova debris we find sheets 
and filaments of gas, and underdense and overdense regions. We find the same in
the large scale structure of the Universe: sheets, filaments and voids.
There are more similarities. In supernova debris we find shells of gas 
expanding at speeds in the range $10^{3}-10^{4}$km/s. There are also shells
in the Universe. As di Nella and Paturel$^{39}$ shows "The distribution of
galaxies up to a distance of 200 Mpc (650 million light-years is flat and shows
a structure like a shell roughly centered on the Local Supercluster (Virgo
Cluster). This result clearly confirms the existence of the hypergalactic
large scale structure noted in 1988. this is presently the largest structure
ever seen." This is so because both explosions, either in the supernova or in
the Universe, are caused by the same force: THE SUPERSTRONG FORCE. It is
worth mentioning that the above picture of the universal expansion maintains
nucleosynthesis untouched. We only have to reinterpret the cosmic background
radiation(CBM). 
\par The recent data on CBM indicate a temperature $T_{0}=2.7$K. As we konw
the frequency at the peak of the spectrum, $\nu_{max}$, is related to $T$ by 
$\nu_{max}/T = 5.88{\times}10^{10}$Hzdeg$^{-1}$.  On the other hand, during 
collapse, the temperature and the density of a collapsing 
mass(supernova) obeys the equation$^{28}$

\begin{eqnarray} T &=& T_{c,i}\left(frac{\rho}{\rho_{c,i}}\right)^{\frac{1}{3}}
\end{eqnarray}

\noindent
where $T_{c,i}=8.0{\times}10^{9}$K, and 
$\rho_{c,i}=3.7{\times}10^{9}$gcm$^{-3}$, and are the temperature and the
density at the onset of collapse. Using the above equation for a density 
slightly higher than nuclear density around $10^{15}$gcm$^{-3}$ we obtain 
$T=5.2{\times}10^{11}$K, and $\nu_{max}=3{\times}10^{22}$Hertz which
corresponds to an energy of 124 MeV. This is quite close to the mass of
pions. For example, the annihilation of $\pi^{+}$ with $\pi^{-}$ produce
photons with energies of about 140 MeV. Therefore, the primordial photons
that produced the CMB may have been created by pion annihilation.
\newline
\vskip .1in
\noindent
III) The Evolution of Galaxies
\newline
\par The remarkable work of M\'{a}rquez et al.$^{35}$ has shown that very high
redshift elliptical galaxies harbor quasars. They have also shown that such
galaxies are very small(diameters smaller than 1kpc) and all of them are 
ellipticals. All the studied objects(about 15 quasars) have extended structures
of ionized gas around them(this fact had already been presented by other
researchers). They have found other galaxies in the fields of the
studied objects only a few kpc away from them. Some of the quasars present 
asymmetric radio sources with collimated one-sided jets of extended ionized 
gas. This means that galaxies are born as quasars which become galaxies by 
means of the shedding of matter(ionized gas) from their cores as a result of
the strong repulsion among their baryons caused by the superstrong interaction.
The authors have found that the quasar 3C 281 is a double radio source. They
also confirm the results of Miley and Hartsuijker$^{40}$ that found that the 
quasar 3C 206 is also a classical double radio source. At such high redshifts
it is very unlikely that this double source was caused by merging. It 
probably was caused by the breaking of the core of these quasars into two 
cores, separated by a very small distance. This breaking was caused by
repulsion due to the superstrong interaction. The same kind of phenomenon has 
been observed in galaxies. For example, our well behaved normal galaxy M31
has two nuclei separated by just 5 light years$^{41}$. Very recent data$^{42}$
of NGC 6240, which is considered a typical protogalaxy, show that 
``approximately 70\% of the total radio power at 20cm originates from the
nuclear region ($\leq$1.5kpc), of which half is emitted by two unresolved 
(R$\leq$30pc) cores and half by a diffuse component. Nearly all of the other 
30\% of the total radio power comes from an arm-like region extending
westward from the nuclear region". A very important property of many quasars is
their brightness which can vary from night to night. This flickering may 
have its origin in the outward motion of large quantities of matter from their 
cores. This brightness variability is also present in seyfert galaxies which
are powerful sources of infrared radiation.  Many of them are also strong 
radio emitters.  For example, over a period of a few months, the nucleus of 
the Seyfert galaxy M77(or NGC1068) switches on and off a power output 
equivalent to the total luminosity of our galaxy$^{43}$. It is also worth 
noting that the nuclei of Seyfert galaxies are very bright and have a general 
starlike appearance. Researchers have found that some Seyfert galaxies exhibit 
explosive phenomena$^{43}$. For example, M77 and NGC4151 expel huge amounts 
of gas from their nuclei.  The spectra of both galaxies show strong emission 
lines, just as quasars'. Shaver et al.$^{44}$ have found that there is a 
redshift cutoff in the number of quasars around z=2, none at 5<z<7, and almost
no quasar for z<0.5. This clearly shows that quasars change into
galaxies.
\par We can show a long list of similar phenomena that evidences the 
superstrong interaction. Let us mention just some of them. NGC 2992 presents a 
jet-like structure and a circum-nuclear ring$^{45}$. Falcke and Biermann$^{46}$
report that there is a large scale emission-like jet going outward from the 
core of NGC 4258 with a mass of about 4${\times}10^{35}$kg and with a kinetic 
power of approximately $10^{42}$ergs/s and expansion velocity of 2000km/s. 
This velocity implies an energy per baryon of about 0.02MeV, which according to
Eq. 32 means that before being expanded the distances among the centeres of 
baryons were only about 2.6 fermi, which are of the same order of the 
distances among  baryons in the core of a supernova(before explosion). 
It is well known that BL Lacertae objects are powerful sources of radio
waves and infrared radiation.  They share with quasars the fact of exhibiting
a starlike appearance and of showing short-term brightness fluctuations.  As
some quasars do, they also have a nebulosity around the bright nucleus. 
Researchers$^{43}$ have managed to obtain the spectrum of their nebulosity.
{\it{The spectrum of the nebulosity is strikingly similar to the spectrum of 
an elliptical galaxy}}(M32's spectrum, in this case). In terms of 
the evolution above described they are simply an evolutionary stage of a 
quasar towards becoming a normal galaxy.
\par Radio galaxies share with BL Lacertae objects many of the properties of
quasars. As Heckman et al.$^{47}$ have shown, in the middle and far 
infrared(MFIR) quasars are more powerful sources of MFIR radiation than radio 
galaxies. Also, there have been investigations showing that the emission from
the narrow-line region(NLR) in radio-loud quasars is stronger than in radio 
galaxies of the same radio power$^{48,49,50}$. Goodrich and 
Cohen$^{51}$ have studied the polarization in the broad-line radio galaxy
3C 109. After the intervening dust is taken into account the absolute 
V-magnitude of this galaxy becomes $-26.6$ or brighter, which puts it in the
quasar luminosity range. The investigators suggest that "many radio galaxies
may be quasars with their jets pointed away from our direct line of sight".
It has also been established that radio galaxies are found at intermediate
or high redshifts and that they are clearly related to galactic evolution 
because as the redshift increases cluster galaxies become bluer on average, 
and contain more young stars in their nuclei. This is also valid for radio
galaxies: the higher the redshift, the higher their activity. All these data
show that a radio galaxy is just an evolutionary stage of a galaxy towards
becoming a normal galaxy, i.e., it is just a stage of the slow transformation 
by means of an overall expansion of a quasar into a normal galaxy. 
\par In the light of the above considerations the nuclei of old spirals must 
exhibit a moderate activity.  This is actually the case.  The activity must
be inversely proportional to the galaxy's age, i.e., it must be a function of
luminosity.  The bluer they are, the more active their nuclei must be.  As
discussed above there must also exist a relation between this activity and
the size of the nucleus(as compared to the disk) in spiral galaxies.  Our 
galaxy has a mild activity at its center.  Most of the activity is 
concentrated in a region called Sagittarius A, which includes the galactic 
center.  It emits synchroton and infrared radiations.  Despite its large 
energy output Sagittarius A is quite small, being only about 40 light years in 
diameter.  Besides Sagittarius A our galaxy exposes other evidences showing 
that in the past it was a much more compact object: 1) Close to the center, 
{\it{on opposite sides of it}}, there are two enormous expanding arms of 
hydrogen going away from the center at speeds of 53km/s and 153km/s; b) Even 
closer to the center there is the ring called Sagittarius B2 which is 
expanding at a speed of 110km/s$^{(43,52)}$. It is worth noting that the 
speeds are  low(as compared to the velocities of relativistic electrons 
from possible black holes).  This phenomenon is not restricted to 
our galaxy. Recent high-resolution molecular-line observations of external 
galaxies have revealed that galactic nuclei are often associated with similar 
expanding rings$^{53}$.   
\par A new born quasar, as discussed above, must have most of 
its mass as hydrogen, the rest being the primordial helium.  Therefore, it 
is mainly constituted of protons. We expect that different parts of it will be
subjected to the superstrong force, especially close to its center  where
the gravitational field is small. The expansion of the quasar has to be, thus, 
from within, that is, from its center to its outer layers. The repulsion makes 
the quasar increase in size and go through the intermediate stages which may 
include radio galaxies and BL Lacertae objects.  Far from the center big 
clumps of hydrogen and helium gasses form stars. Considering what is exposed 
above we may propose the following evolutionary scenario: 
\vskip .15in
\noindent
a) Elliptical galaxy. 
\newline
\par A quasar may become an elliptical galaxy by expanding slowly as a whole.  
Because of rotation we may have several types of ellipticals according to 
their oblateness. As is well known ellipticals do not exhibit much rotation(as
compared to spiral galaxies). This is explained as follows: As a quasar 
expands its angular velocity decreases because of angular momentum 
conservation. For example, the angular velocity of an EO must be given 
by(disregarding mass loss)
\begin{eqnarray} {\omega}_{EO} &=& 
{\omega}_{Q}\left({\frac{R_{Q}}{R_{EO}}}\right)^{2}
\end{eqnarray}                                                  

\noindent
where $\omega_Q$ is the angular velocity of the quasar which gave origin to the
galaxy; $R_{EO}$ and $R_{Q}$ are the radii of the elliptical galaxy and the 
quasar, respectively. Because $R_{EO}{\gg}R_{Q}$, 
${\omega}_{EO}{\ll}{\omega}_{Q}$. This is consistent with the slow rotation
of ellipticals.  There is also the following consistency to be considered. 
Most galaxies in the Universe are ellipticals(about 60\%) and as was shown 
above this means that most quasars expand slowly. Therefore, most quasars must 
not show rapid variability and must also be radio quite.  This is exactly what 
has been reported$^{54}$. Another evidence to be taken into account is the 
reported nebulosity around some quasars.  Boroson and Person$^{55}$ have 
studied this nebulosity spectroscopically.  The emission lines they found are 
of the same type as the emissions from a plasma. 
\vskip .15in
\noindent
b) Spiral galaxy.  
\newline
\par There are two possibilities in this case: normal spiral and barred 
spiral. This happens when, at some point in its expansion towards becoming a 
galaxy, a quasar expands rapidly by pouring matter outwards from its center, 
mainly in opposite sides accross a diameter. This pouring will give origin to 
two jets which will wind up around the central bulge because of rotation and 
will create the two spiral arms. A possible mechanism is the following: Due 
to rotation we expect to have some bulging in the spherical shape, and 
because of angular 
momentum conservation  the outpouring of matter may only happen in a plane 
perpendicular to the axis of rotation.  Because of rotation the core of the 
quasar becomes also an ellipsoid.  This core(which has a higher concentration 
of baryons) may be broken into two parts, going to a state of lower potential 
energy(of the superstrong interaction). These two parts repel each other and 
form two centers(lobes) in the equator of the quasar(or young galaxy). The
quasar 3C 281, for example, is a double radio source and has extended ionized
gas around it$^{35}$. These two lobes are also seen in many radio galaxies. As 
a consequence of the outpouring 
of matter from each center there must exist all kinds of radiations covering 
the whole electromagnetic spectrum, especially in the form of synchroton 
radiation caused by collisions among atoms.  Because of these collisions we 
expect to have electrons stripped from  hydrogen and helium atoms.  These 
electrons create the observed synchroton radiation which is associated with 
jets in very active galaxies.  These collisions provide also the enormous 
energy output observed in quasars.
\par  As a galaxy ages its bulge diminishes and leaves the globular
clusters alone without the embedding gas since the gas has either escaped 
or has been transformed into globular clusters.  Also, the activity at the 
galactic center diminishes as the age increases due to decrease of mass
in the nucleus. 
\par  The barred spirals are galaxies that have expel matter more 
vigorously. That is exactly why their arms are not tightly wound.
As the galaxy ages the arms will curl up more and more and the bar will 
disappear because of the ejection of material outwards.  It is worth 
noting that the more spirals(including barred ones) are wound up the smaller 
are their nuclei and, conversely, the larger are their bulges, the less they 
are wound up.  This happens because of the shedding of matter outwards from 
their  nuclei throughout the galaxy's life due to the action of the superstrong
force. The bar can be explained in terms of a more vigorous shedding 
of matter outwards as compared to the shedding that takes place in normal
spirals. Therefore, as a spiral evolves its nucleus diminishes and the two
arms become more and more tightly wound up. In summary, the evolution of
galaxies probably follows one of the nine branches:
\newline
i) Quasar(without jets) $\rightarrow$ BL Lacertae or radio galaxy 
  $\rightarrow \left\{ \begin{array}{l}
                          \mbox{Seyfert Galaxies} \\
			  \mbox{Elliptical Galaxies$\rightarrow$Spiral Galaxies} 
			  \end{array}
	       \right. $
\newline
ii) Quasar(without jets) $\rightarrow$ BL Lacertae or radio galaxy 
 $\rightarrow \left\{ \begin{array}{l}
                          \mbox{Seyfert Galaxies} \\
			  \mbox{Elliptical Galaxies} 
			  \end{array}
	       \right. $
\newline
iii) Quasar(without jets) ${\rightarrow} \left\{ \begin{array}{l}
                          \mbox{Seyfert Galaxies} \\
			  \mbox{Elliptical Galaxies$\rightarrow$Spiral Galaxies} 
			  \end{array}
	       \right. $
\newline
iv) Quasar(without jets) ${\rightarrow} \left\{ \begin{array}{l}
                          \mbox{Seyfert Galaxies} \\
			  \mbox{Elliptical Galaxies} 
			  \end{array}
	       \right. $
\newline
v) Quasar(without jets) ${\rightarrow}$ Elliptical Galaxies$\rightarrow$Spiral Galaxies
\newline
vi) Quasar(without jets) ${\rightarrow}$ Elliptical Galaxies
\newline
vii) Quasar(with jets) $ \rightarrow$ radio galaxy 
 $\rightarrow \left\{ \begin{array}{l}
		       \mbox{Seyfert Galaxies}\\
		       \mbox{Spiral Galaxies}
		       \end{array}
	                	      \right. $		
\newline
viii) Quasar(with jets) $ \rightarrow \left\{\begin{array}{l}
		       \mbox{Seyfert Galaxies}\\
		       \mbox{Spiral Galaxies}
		       \end{array}
	                	      \right. $		
\newline
ix) Quasar(with jets) $\rightarrow$ Spiral Galaxies.
\vskip .2in

\par Let us, now, make a general analysis including all kinds of galaxies.  
Considering the evolution above proposed we do not expect to have very small
spiral galaxies because spirals must come from strong expulsion of matter from
quasars nuclei, and this must happen only when the number of baryons is 
sufficiently large.  This is the case, indeed, because  dwarf galaxies are
either irregular or elliptical galaxies.  Ellipticals have masses in the range
between $10^{5}$ and $10^{13}$ solar masses while spirals' masses are 
comprised between $10^{9}$ and $10^{11}$ solar masses.  Also, we expect that
spirals have faster rotations than ellipticals and, indeed, they do.  This
is just because the nuclei of spirals are smaller than the nuclei of 
ellipticals(for the same mass, of course). Therefore, according to Eq.(38) 
above, spirals should have faster rotational velocities.  It is 
also expected that, since spirals shed gas to their disks throughout their 
lifetimes, their disks must have young stars.  This is an established fact.  
Our galaxy's disk, for example, has very hot, young(O-,B-, and A-type) stars, 
type-I Cepheids, supergiants, open clusters, and interstellar gas and dust.  
Each of these types represent young stars or the material from which they are 
formed. Conversely, the globular clusters and the nucleus contain older stars, 
such as RR Lyrae, type-II Cepheids, and long-period variables.  This, of 
course, is a general characteristic of all spirals.  For example, Young O- and 
B-type stars are the stars which outline the beautiful spiral arms of the 
Whirlpool galaxy. Because of the lack of gas(i.e., because of the lack of a 
disk) ellipticals also have primarily very old stars.
\par A very important support to the above evolution scheme is provided by 
the number-luminosity relation $N(>l)$. When expanded in terms of the apparent 
luminosity, $l$, the first term(Euclidean term) is  given by$^{56}$
\begin{eqnarray} N(>l) &=& \frac{4{\pi}n(t_{o})}{3}
{\left(\frac{L}{4{\pi}l}\right)}^{1.5}\end{eqnarray}      

\noindent
where $n(t_{o})$ is the present density of galaxies and $L$ is the absolute 
magnitude. The correction term is always negative, so that the number of 
faint objects($l$ small) should always be less than the number that the 
$l^{-1.5}$ predicts. This conclusion is strongly contradicted by observations 
on radio sources: many surveys of radio sources agree that there are more 
faint sources than the $l^{-1.5}$ law predicts.  The fitting of the 
experimental data provides a law of the form$^{53}$
\begin{eqnarray} N(>l) {\approx} \frac{constan}{l^{1.8}}.\end{eqnarray}  

\noindent
Since the formula breaks down for small $l$(i.e., faint distant sources), we
must conclude that in the past radio sources were brighter and/or more 
numerous than they are today. This, of course, lends support to the above
evolutionary scheme.
\par It is worth mentioning that there is a very important 
drawback against the traditional view of explaining the formation of arms in 
spirals by the bulging effect of rotation. If this were the case we would find
a higher proportion of pulsars off the galactic equator of our galaxy. But the
real distribution reveals that these sources are mostly concentrated in the
galactic equator.  The traditional view does not explain either why all spirals
have large amounts of gas in their disks. Besides, within the traditional
framework quasars are just exotic objects. Evolution is clearly out of question
without a repulsive short range force. 
\newline
\vskip .2in
\noindent
12) THE ROTATION OF SPIRAL GALAXIES$^{6}$
\vskip .1in
\par The rotational curve of spiral galaxies is one of the biggest puzzles of
nature. It is possible to give a reasonable explanation for this puzzle 
in terms of the action of the superstrong force.  In the process we will also
explain the formation of the spiral structure of the arms. 
\par First, let us consider the central nucleus(or bulge). The whole bulge
expands slowly throughout the lifetime of a galaxy. For simplification let us 
consider a uniform density for the bulge.  Because mass varies as $r^3$ and 
the gravitational force varies as $r^{-2}$ we expect the tangential velocity 
to be proportional to $r$.  
\par Now, let us consider the tangential velocities of stars in the disk.
As was shown above the disk was formed by the shedding of matter from the
center of the galaxy where a denser core existed.  The mass is expelled with 
speeds in the range $10^{2}-10^{3}$km/s. Let us consider that the
bulge has a radius $R_{B}$ and also that, because of the action of
the superstrong force, a certain mass of gas $m$ is expelled from the center.
Because of its radial velocity, the mass $m$ will continue to 
distance itself from the bulge, but its tangential velocity is kept fixed 
because of the action of repulsion and because of the transfer of angular 
momentum from the bulge to the mass.  This may be shown in the following 
way: As the mass goes away from the center it increases its angular 
momentum. At a distance $r$ the angular momentum is given by 
\begin{eqnarray} J &=& mrv_{t}\end{eqnarray}                          

\noindent
where $v_t$ is the tangential velocity. Because $J$(of the mass $m$) increases 
with time(and with $r$) we have
\begin{eqnarray} \frac{dv_{t}}{v_{t}} &>& -\frac{dr}{r}.\end{eqnarray} 

\noindent
Integrating, we obtain
\begin{eqnarray} ln{\frac{v_{t}}{v_{to}}} &>& ln{\frac{r_{o}}{r}}\end{eqnarray}
%\newpage
%\rule{0in}{1in}
%\newpage
%\rule{0in}{1in}
%\newpage

\noindent                                                                  
where $r_o$ is the position of the mass at a time $t_o$ and $r$ is its position
at a later time. Both positions are measured from the center. Because the
logarithm is an increasing function of the argument, we must have
\begin{eqnarray} \frac{v_t}{v_{to}} &>& \frac{r_o}{r}.\end{eqnarray}  

\noindent                                                                   
We clearly see that $v_{t} = v_{t_{o}}$ is a solution of the above inequality 
because $r$ is always larger than $r_o$. Thus the mass $m$ gains angular 
momentum. Because of conservation of angular momentum the galactic nucleus 
must decrease its angular momentum by the same amount. A recent study shows 
that the arms of spirals ``transport angular momentum radially within 
galactic disks''$^{57}$. If 
we consider  that the angular velocity of the nucleus does not 
diminish(which is more plausible than otherwise), then its mass must diminish, 
i.e., the nucleus needs to shed more matter outwards.  Since $v_t$ remains the 
same the angular velocity must decrease as the mass goes away from the center. 
This generates the differential rotation observed in all spiral galaxies. The 
formation of the spiral structure is, therefore, directly connected with the 
evolution of the galaxy.  
\par We may easily show that the beautiful spiral arms are described by a 
logarithmic spiral(in an inertial frame). The angle $\theta$ measures the
angular position of $m$ with respect to the center of the bulge and $\phi$
measures the angle in the bulge at position $R$ where the mass left it. The
angular velocity of the bulge is $\Omega$. Let us consider that the tangential
velocity of the mass $m$ is a constant. Therefore, we obtain
\begin{eqnarray} r{\omega} &=& r\frac{d{\theta}}{dt} = R\frac{d{\phi}}{dt} = 
R{\Omega} = v_{t} = constant \end{eqnarray}                           

\noindent
where $R$ is the radius of the galactic bulge and $v_t$ is the tangential 
velocity of the mass $m$. We have that 
\begin{eqnarray} d{\theta} &=& {\omega}dt =  \frac{R{\Omega}}{r}dt = 
\frac{R{\Omega}}{rv_{r}}dr\end{eqnarray}                               

\noindent
where we have used the fact that $v_{r} = \frac{dr}{dt}$. Considering that
$v_r$ varies slowly with $r$(or $t$) we may integrate $d{\theta}$ and obtain
\begin{eqnarray} r &{\approx}& Re^{\frac{v_{r}}{v_{t}}\theta}.\end{eqnarray}

\noindent
{\it{This is the equation of the logarithmic spiral}}. We imediately obtain
that
\begin{eqnarray} \omega &{\approx}& {\Omega}e^{-\frac{v_r}{v_t}\theta}.
\end{eqnarray}                                                        

\noindent
We may also calculate $\phi$. It is given by
\begin{eqnarray} \phi &{\approx}& \kappa\left(e^{\frac{\theta}{k}} - 1\right)
\end{eqnarray}                                                      

\noindent
where $\kappa$ is given by $v_{t}/v_{r}$.
\par The ratio $\kappa=v_{t}/v_{r}$  distinguishes between the two types of 
spiral galaxies. If $\kappa{\ll}1$, then $\omega$ diminishes rapidly with
$\theta$. This corresponds to spirals with 
bars. Conversely, if $\kappa{\gg}1$, then $\omega$ diminishes slowly and only
reaches a very low value for large $\theta$. This is consistent with the
data on spiral galaxies. The middle ground $\kappa{\approx}1$ must
correspond to intermediate cases. A typical spiral without bars should have
$\kappa{\gg}1$. 
\par Let us now consider the problem from the point of view of a frame fixed 
in the galactic bulge and rotating with it. We may define an angle $\psi$ 
related to $\theta$ and $\phi$ by $\psi = \phi - \theta$. Therefore, $\psi$
is given by 
\begin{eqnarray} \psi &=& \kappa\left(e^{\frac{\theta}{\kappa}} - 
1\right) - \theta.\end{eqnarray}                                            

\noindent
For small $\theta$ one has $\psi {\approx} {\theta}^{2}/{2\kappa}$ and 
\begin{eqnarray} r {\approx} Re^{\sqrt{\frac{2}{\kappa}\psi}}\end{eqnarray} 

\noindent
and for large $\theta$ we have $\psi {\approx} {\kappa}
e^{\frac{\theta}{\kappa}}$ and
\begin{eqnarray} r {\approx} R\frac{\psi}{\kappa}.\end{eqnarray}            

\noindent
Therefore, in this rotating frame the mass $m$ also describes a spiral curve as
it moves away from the center.
\par Let us now estimate the order of magnitude of the radial velocity of 
gas(and stars) in the galactic disk.  The radius of our galaxy is $50000$ light
years and the age of our galaxy is of the order of magnitude of the age of the
Universe, $10^{17}$s. The gas which is at the edge of the disk must have
moved from the center with a mean velocity of about 5km/s. Since the gas
was expelled with much higher velocities the mean velocity of stars far from
the center are very small because the gravitational attraction slowed down the
mass to very small velocities. The above figure  is just a rough
estimation.  It is very important to obtain the mean radial velocities of 
stars in the spiral arms of the Milky Way to compare with it. 
\newline
\vskip .2in
\noindent
13) GENERAL CLASSIFICATION OF MATTER
\vskip .1in
\par Science has utilized specific empirical classifications of matter 
which have revealed hidden laws and symmetries.  Two of the most known
classifications are the Periodic Table of the Elements and Gell-Mann's
classification of particles(which paved the way towards the quark model).
\par Let us go on the footsteps of Mendeleev and let us attempt to achieve a 
general classification of matter, including all kinds of matter formed along
the universal expansion, and by doing so we may find the links between the 
elementary particles and the large bodies of the universe. 
\par It is well known that the different kinds of matter appeared 
at different epochs of the universal expansion and that they are imprints of 
the different sizes of the universe along the expansion. 
Taking a closer look at the different kinds of matter  we may classify them  
as belonging to two distinct general states. One state is characterized by a 
single unit with  angular momentum. The angular momentum may either be the 
intrinsic angular momentum, spin, or the 
orbital angular momentum. The other state is characterized by some degree of 
correlation among the interacting particles and may be called 
the structured state. The angular momentum may(or may not) be present in this 
state.  The fundamental units of matter make the structured states, that is, 
they are the building blocks of everything, {\it{stepwise}}. 
In what follows we will not talk about the weak force since it does not form 
any stable matter and is rather related to instability in matter. Along the
universal expansion nature made different building blocks and different
media to fill space.  The weak force did not form any 
building block and is out of our discussion. As is well known this 
force is special in many other ways.  For example, it violates parity and has 
no ``effective potential''(or static potential) as the other interactions do.  
Besides, the weak force is known to be left-handed, that is, particles 
experience this force only when their spin direction is anti-aligned with 
their momentum.  Right-handed particles appear to experience no weak 
interaction, although, if they have electric charge, they may still interact 
electromagnetically. Later on we will include the weak force into the 
discussion. Each structured state is mainly formed by two types of 
fundamental forces. Due to the interactions among the units one  
expects other kinds of forces in the structured state. In this fashion we can 
form a chain from the quarks to the galactic superstructures and extrapolate 
at the two ends towards the constituents of quarks and towards the whole 
Universe. 
\par The units of matter are the nucleons, the atom, the galaxies, etc. 
The `et cetera' will become clearer later on
in this work. In the structured state one finds the quarks, the nuclei, the 
gasses, liquids and solids, and  the galactic liquid. Let us, for example, 
examine the sequence nucleon-nucleus-atom. As is well known a
nucleon is made out of quarks and held together by means of the strong
force. The atom is made out of the nucleus and the electron(we will talk about 
the electron later), and is held together by means of the
electromagnetic force. The nucleus, which is in the middle of the sequence, is 
held together by the strong force(attraction among nucleons) and by the 
electromagnetic force(repulsion among protons). In other words, we may say 
that the nucleus is the result of a compromise between these two forces. 
Let us, now, turn to the sequence atom-(gas,liquid,solid)-galaxy. The gasses, 
liquids and solids are also formed by two forces, namely, the electromagnetic
and the gravitational forces. Because the gravitational force is $10^{39}$ 
weaker than the electromagnetic force the polarization in gasses, liquids and
solids is achieved by the sole action of the electromagnetic force because
it has two signs. But it is well known that large masses of gasses, liquids 
and solids are unstable configurations of matter in the absence of gravity.  
Therefore, they are formed by the electromagnetic and gravitational forces.  
Large amounts of nucleons {\it{gas}} at some time in the history of the 
universe gave origin to galaxies which are the biggest individual units of 
creation. We arrive again at a single fundamental force that holds a galaxy 
together, which is the gravitational force. There is always the same pattern: 
one goes from one fundamental force which holds a single unit(nucleon, 
atom, galaxy) together to two fundamental forces which coexist in a medium. The 
interactions in the medium form a new unit in which the action of another 
fundamental force appears. We are not talking any more about the previous unit 
which exists inside the new unit(such as the nucleons in the nucleus of an 
atom). 
\par By placing all kinds of matter together in a table in the order of the
{\it{universal expansion}}  we can construct the two tables below,
one for the states and another for the fundamental forces.
\par In order to make an atom we need the electron besides the nucleus.
Therefore, just the clumping of nucleons is not enough in this case.
Let us just borrow the electron for now.  Therefore, it looks like that the 
electron belongs to a separate class and is an elementary particle.
The above considerations may be summarized by the following: {\it{the different
kinds of building blocks of the Universe(at different times of the expansion)
are intimately related to the idea of filling space.}}  That is, depending on
its size, the Universe is filled with different units.
\par Following the same reasoning we can say that there should exist a 
force, other  than the strong force, acting between any two prequarks. We call
it superstrong force. Also, for the `galactic liquid', there must be another 
fundamental force at 
play. Because it must be much weaker than the gravitational force(otherwise, 
it would already have been found on Earth) we expect it to be a very weak 
force.  Let us call it the superweak force. 
\par Summing up all fundamental forces  we arrive at {\it{six forces for 
nature: the superstrong, the strong, the electromagnetic, the gravitational, 
the superweak and the weak  forces}}.  \newline
\vskip .3in
\noindent
14) THE SUPERWEAK FORCE
\vskip .2in
\par Let us now try to find a possible mathematical expression for the 
superweak force. There have been reports of a fifth force inferred from the 
reanalysis of the E\"{o}tv\"{o}s experiment and from the mine-gravity 
data$^{58}$.  The discrepancies suggest the existence of a composition 
dependent intermediate-range force. 
\par The potential energy of such hypothetical force is usually represented by 
a Yukawa potential which, when added to the standard Newtonian potential 
energy, becomes$^{58}$
\begin{eqnarray}  V(r)&=& - \frac{Gm_{1}m_{2}}{r}\left(1 + 
\alpha\exp{(-r/\lambda)}\right),\end{eqnarray}                             

\noindent
where $\alpha$ is the new coupling in units of gravity and $\lambda$ is its
range.  The dependence on composition can be made explicit by writing 
${\alpha}=q_{i}q_{j}\zeta$ with 
\begin{eqnarray} \displaystyle q_{i}& =& cos{\theta}(N+Z)_{i}/\mu_{i} + 
sin{\theta}(N-Z)_{i}/\mu_{i},\end{eqnarray}                               

\noindent
where the new effective charge has been written as a linear combination of the
baryon number and the nuclear isospin per atomic mass unit, and $\zeta$ is the 
coupling constant in terms of $G$. 
\par Until now most experimental results have not confirmed the existence of 
this force$^{59}$, although they do not rule it out because its coupling 
constant(s) may be smaller than previously thought. Adelberger et al.$^{59}$ 
have found the upper limit of $10^{-13}$ms${-2}$ in the acceleration which 
means that the fifth force is at least $10^{10}$ smaller than the gravitational
force. Of course, this force may only exist if there is a violation in 
the weak equivalence principle, which has been proven to hold$^{59}$ to a 
precision of one part in $10^{12}$. But it may be violated if the precision 
goes just one more order of magnitude. They may, then, reveal the existence 
of the fifth force. 
\par The superweak force proposed in this paper has the same character of that
 of the fifth force, but has an infinite range.  This means that the mass of
the mediating boson is zero. From the above expression for the fifth force 
potential energy we may express the potential \newpage

\rule{0in}{.7in}

\begin{center}
\begin{tabular}{c c c c c c c} \hline\hline\\ 
& ? & & quark & & nucleon & \\
\\
& nucleon & & nucleus & & atom & \\
\\
& atom & & gas & & galaxy & \\
& & & liquid & & & \\
& & & solid & & & \\
\\
& galaxy & & galactic liquid & & ? \\
\\
\hline\hline\\
\end{tabular}
\end{center}
\vskip .2in

\begin{center}
\parbox{4in}
{Table 5. The two general states  which make everything in the 
Universe, stepwise. The table is arranged in such a way to show the links 
between the structured states and the single states. The interrogation
marks above imply the existence of prequarks and of the Universe as units of
creation}
\end{center}

\vskip .4in

\begin{center}
\begin{tabular}{c c c} \hline\hline\\ ? & ? & strong force \\
& strong force & \\
\\
strong force & strong force & electromagnetic force \\
& electromagnetic force & \\
\\
electromagnetic force & electromagnetic force & gravitational force \\
& gravitational force & \\
\\
gravitational force & gravitational force & ?\\
& ? & \\
\\
\hline\hline\\
\end{tabular}
\end{center}
\vskip .2in

\begin{center}
\parbox{5in}
{Table 6. Three of the fundamental forces of nature. Each force 
appears twice and is linked to another force by means of a structured 
state. The interrogation marks suggest the existence of two other
fundamental forces. Compare with Table 1.}
\end{center}

\newpage

\noindent
energy of the superweak force in 
terms of the baryon numbers and isospins of two very large bodies i and j as
\begin{eqnarray}
 V(r,N,Z)& = & \left(A_{B}(N+Z)_{i} + A_{I}(N-Z)_{i}\right){\times}
\left(A_{B}(N+Z)_{j} + A_{I}(N-Z)_{j}\right){g^2}\frac{1}{r}
\end{eqnarray} 

\noindent
where $A_{B}$ and $A_{I}$ are the force coupling constants of the baryon number
and isospin  terms, respectively, and $g$ is the strong force charge.  
\par Let us assume that the constants $A_{B}$ and $A_I$ are positive. 
The number of neutrons and protons vary from galaxy to galaxy for a given 
time(after $t=0$), but considering that they were formed about the same time 
$N_{i}$ and $Z_{i}$ must be close to mean values $N$ and $Z$, respectively. 
Considering the mean values $N$ and $Z$, $V$ becomes
\begin{eqnarray}
 V(r,\eta)& = & \frac{B^{2}g^{2}}{r}\left({A_{B}}^{2}
 + 2A_{B}A_{I}(2\eta-1)+ {A_{I}}^{2}(2\eta-1)^{2}\right)
\end{eqnarray} 

\noindent
where $\eta=N/B$ and $B=N+Z$ is the baryon number and is a constant,
approximately, since the mass of a galaxy does not vary much with time.
This $V(r)$ is always positive. If we make $N=Z$ in the beginning of the 
Universe the potential becomes
\begin{eqnarray}
V(r)& = & \frac{B^{2}g^{2}}{r}
\end{eqnarray} 

\noindent
which is proportional to the potential of the strong force for very small 
distances. This condition unifies the strong force to the superweak force 
in the beginning of the Universe.
\par We assume that at $t=0$, in the `beginning' of the universe,  $N$ 
was equal to $Z$ and decreased via the weak interaction(neutron decay) 
up to a time called $t_p$.  At this time $N$ reached its minimum value, being  
only 13\% of all baryons, the remaining 87\% being protons.  The halt in 
neutron decay happened due to the formation of atoms which imprisoned  
the remaining neutrons. These neutrons became bound in the nuclei of 
helium and deuterium. From this point on fusion took place in the core of 
stars of all galaxies(or quasars) and the number of 
protons began to decrease slowly. This is explained as follows: As the universe 
ages the stars become white dwarfs, neutron stars and black holes(not observed 
yet). During the aging process the core density of a star increases and the 
high electron Fermi energy drives electron capture onto nuclei and free 
protons. This last process, called neutronization$^{28}$, happens via the 
weak interaction. The most significant neutronization reactions are:
\newline
\begin{itemize}
\item {Electron capture by nuclei,}
\end{itemize}
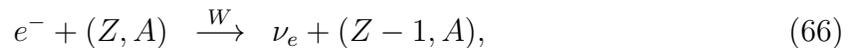
\begin{eqnarray}
& & e^{-} + (Z,A)\;\;\stackrel{W}{\longrightarrow} \;\;\nu_{e} + (Z-1,A),
\end{eqnarray}                                                            
\vskip .3in
\begin{itemize}
\item {Electron capture by free protons,}
\end{itemize}
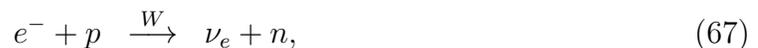
\begin{eqnarray}
& & e^{-} + p \;\;\stackrel{W}{\longrightarrow} \;\;\nu_{e} + n,
\end{eqnarray}                                                            
\vskip .3in
\noindent
where $W$ means that both reactions proceed via  charged currents of the 
electroweak interaction.
\par Of course, neutronization takes place in the
stars of all galaxies, and thus, the number of neutrons increases relative to
the number of protons as the universe ages. For example, a white dwarf 
in the slow cooling stage(for $T{\leq}10^7$K) reaches a steady proton to 
neutron density of about 1/8, and takes about $10^9$ years to  
cool off completely, which is a time close to the present age of the universe. 
By then, most stars have become white dwarfs(or neutron stars).  
This steady increase is expected to be very slow.  Because of the dense and
underdense regions we may have $N$ and $Z$ such that 
$Q(N)=A_{B}B - A_{I}(2N - B)$ can be either negative or positive. Therefore,
there may exist attraction between dense and underdense regions because the
potential energy may become negative
\begin{eqnarray}
 V(r,N,Z)& = & - \left(A_{B}B_{i} - A_{I}(2N-B)_{i}\right){\times}
\left(A_{B}B_{j} - A_{I}(2N-B)_{j}\right){g^2}\frac{1}{r}
\end{eqnarray} 

\noindent
if one of the two $Q(N)$ is negative. Taking into account this interaction let 
us consider again a galaxy on the surface of a sphere of radius $r$. In order 
to make the Universe to close we need to have
\begin{eqnarray} E &=& \frac{1}{2}mv^{2} - \frac{4\pi}{3}G{\rho}r^2 + 
V_{SW} = 0.
\end{eqnarray}

\noindent
where $V_{SW}$ is the potential of the superweak interaction and $\rho$ is
the mean density in the sphere. If $V_{SW}<0$ then this term helps to close 
the Universe.\newline
\vskip .2in
\noindent
15) THE GALACTIC MEDIUM
\vskip .2in
\par Observations have shown that the medium formed by galaxies has quite 
a large degree of order since voids have an average size of about
40$h{-1}$Mpc$^{60}$. This regularity indicates that this medium is a sort of 
liquid because galaxies are not randomly distributed. Because of its very 
small range the superstrong force is only effective when the distances among 
particles are very small. As is argued above and in reference 61, there 
should exist a very weak force acting among voids(bubbles) and superclusters. 
This force together with the gravitational force yield effective potential 
wells where galaxies wander about. Because the universal expansion is very 
slow we may still attempt to use statistical mechanics. Also, because the 
degree of order of the Universe on small scales is small we may treat the 
Universe as a dilute gas. A dilute gas is a very complicated system.  
What is usually done treating such systems is to consider effective potentials 
and make some approximations such as pairwise additivity and cluster expansion. 
The pair potential energy most often used is the well-known Lennard-Jones 
potential energy given by$^{62,63}$ 
\begin{eqnarray} u(r) &=& 
4{\varepsilon_o}\left[\left(\frac{\sigma}{r}\right)^{m} - 
\left(\frac{\sigma}{r}\right)^{n}\right].\end{eqnarray}                    

\noindent
In our case the particles that we are considering are bubbles of galaxies. The
distance r is taken between the centers of mass of two such bubbles. The 
constants in the above equation should be functions of the masses of the 
bubbles and of the relative numbers of protons and neutrons of the clusters. 
The distance $R$ corresponds to the maximum distance between the two bubbles,
and $r_{0}$ to the equilibrium distance, when the net force is zero. During the
contraction $R$ will diminish, of course. The minimum of the potential, 
$u_{0}=u(r_{0})$, is 
\begin{eqnarray} u_{0} &=& 
- 4{\varepsilon_o}\left(\frac{n}{m}\right)^{\frac{n}{m-n}}
\left(\frac{m-n}{m}\right)\end{eqnarray}

\noindent
and the expansion of the potential about $r_{0}$ is
\begin{eqnarray} u(r) &=& u_{0} + \frac{2\varepsilon_o}{{r_0}^{2}}
\left(\frac{n}{m}\right)^{\frac{n}{m-n}}n(m-n)(r - r_{0})^{2}.\end{eqnarray}

\noindent
All the parameters in these equation may depend on the masses of the bubbles
and on their relative numbers of protons and neutrons. Of course, it is an
extremely difficult problem due to the enormous distances(and times) involved
and due to the our extremely small observational times. We may find the
ranges of some parameters by comparing similar bubbles at different
mean redshifts(it must be the redshift of the center of each bubble). If a
particular bubble of the local Universe has not arrived yet at $r=r_{0}$, then
we will never prove the validity of the above equation, simply because of
the times involved. Suppose, for example, that this condition will happen a
billion years from now. 
\par But let us not give up. With the two very important steps mentioned below
and with a lot of fitting we may find out if we are right:

\begin{itemize}
\item Determine the size of bubbles versus the mean redshift(of the center of 
each bubble);
\item Find the velocities of the centers of the bubbles(versus the mean
redshift)and, from them determine the accelerations and the forces and
then the potential.
\end{itemize}
\vskip .3in
\noindent
16) THE STRUCTURED STATE
\vskip .2in
\par The structured state is made by opposing forces, i.e., it represents a
compromise between an attractive force and a repulsive force.  Small bodies of 
ordinary matter(gasses, liquids and solids) is formed by action of the
electromagnetic force.  In order to hold together the mass of a large 
body(an asteroid, for example) we need the action of the gravitational force.
\par In the case of nuclear matter the two main forces are the strong and
electromagnetic forces. For very short distances we also need the action of
the superstrong force(for the so-called hard core). As is well known we may 
describe nuclear matter in terms of an equation of state in a quite similar 
way to what we do with solids and liquids. Nuclear matter exhibits some 
properties of a semiconductor or a liquid such as band gap, phonons, equation
of state, etc. $^{64}$.  It is not by chance that the liquid
drop model provides quite some satisfactory results in nuclear physics. In 
order to keep the same pattern we should expect to have a sort of compromise 
between the superstrong force and the strong force. This compromise forms the 
quark. 
\par In order to have the `galactic liquid' it is also necessary to have 
some sort of compromise. Because of this compromise we find regularity in the
distribution of voids and clusters. Since the gravitational force is always 
attractive the superweak force must be repulsive during the universal
expansion.   
\par The bodies which form any structured state exhibit some degree of 
correlation among them. This degree of correlation is shown by the correlation
function which, in turn, is related to the interacting net potential energy 
among the particles.  The potential energy which is a molecular potential has 
three general features: i) it has a minimum which is related to the mean 
equilibrium positions of the interacting particles; ii) it tends to zero as 
the separation among the particles tends to infinity; and iii) it becomes 
repulsive at close distances. Because of our ignorance in treating the 
many-body problem it has become quite commom the use of the so-called 
semi-empirical interatomic potential energies. Mathematically, there are a few 
of them. The most commonly used in moleuclar physics is the Lennard-Jones 
potential energy $^{62,63}$ but in nuclear physics other similar potentials  
are used. As proposed above the potential energy of the `galactic  
liquid' has the general features of the molecular potential.
\par As is well known the general motion of the particles of a liquid is quite
complex and that is exactly what we are dealing with in the case of the
galactic superstructures.  \newline
\vskip .2in
\noindent
17) THE ENERGIES OF BARYONS
\vskip .2in
\par We will see below another evidence for the existence of the superstrong
interaction, which is the calculation of the energies of hadrons. 
It has been shown above that a sort of effective molecular potential well 
acts among quarks. The expansion of such a potential about its minimum  
yields a harmonic oscillator potential.  Thus, if we consider that in their 
lowest state of energy quarks are separated by a distance $r_{q}$, then for 
small departures from equilibrium the potential must be of the form 

\begin{eqnarray} V(r) &=& V_{o} + K(r - r_{q})^{2}\end{eqnarray}         

\noindent
where $K$ is a constant and $V_{o}$ is a negative constant representing the
depth of the potential well. By doing so we are able to calculate the 
energies of all baryons. The present treatment is very different from other 
calculations of hadron levels found in the literature. In those calculations
ad hoc central harmonic potentials or other ad hoc central potentials have 
been used. {\it{It is a wrong idea because 
the force between any two quarks of a hadron acts along the line 
that joins them, and not between an imaginary point(in which there is no 
particle) and each quark}}. Actually, as deep inelastic scattering has shown, 
the proton, for example, does not have any quark fixed at its center. That is 
the reason why these models predict levels that are never found. The harmonic
potential between the two quarks comes from the combination of the strong and
the superstrong interactions. 
\par I should mention that A. Hosaka, H. Toki and M. Tokayama published a 
work$^{65}$ in 1998 using the same ideas of this section and including 
rotation. It is important to say that my ideas had already been published 
elsewhere since 1991$^{1,2,8}$. Moreover, my ideas 
were also widespread via xxx.lanl.gov with the preprint hep-ph/9311273. I am
mentioning this in name of fairness and in name of intelectual authorship.
\par Let us consider a system composed of three quarks which interact in pairs 
by means of a harmonic potential. Let us disregard the electromagnetic 
interaction which must be considered as a perturbation. Also, let us disregard 
any rotational contribution which may enter as a perturbation too. This is
reasonable because the strong and superwtrong interactions must be much larger 
than the ``centrifugal'' potential. If we consider that quarks do not move
at relativistic speeds, and disregarding  the spin interaction 
among quarks, we may just use Schr\"{o}dinger equation in terms of normal
coordinates$^{66}$

\begin{eqnarray} \sum_{i=1}^{6} 
\frac{{\partial}^{2}\psi}{{\partial}{\xi}_{i}^{2}} +                       
\frac{2}{{\hbar}^2}\left(E - 
\frac{1}{2}\sum_{i=1}^{6}{\omega}_{i}{{\xi_{i}}^2}\right)\psi &=& 
0\end{eqnarray}                                                      
                      
\noindent
where we have used the fact that the three quarks are always in a plane. The 
above equation may be resolved into a sum of 6 equations

\begin{eqnarray} \frac{{\partial}^{2}\psi}{{\partial}{\xi}_{i}^{2}} +
\frac{2}{{\hbar}^2}\left(E_{i} - 
\frac{1}{2}\omega_{i}{\xi_{i}}^{2}\right)\psi &=& 0,\end{eqnarray}    
\noindent
which is the equation of a single harmonic oscillator of potential energy
$\frac{1}{2}\omega_{i}{\xi_{i}}^{2}$ and unitary mass with
\begin{eqnarray} E &=& \sum_{i=1}^{6} E_{i}.\end{eqnarray}            

\noindent
The general solution is a superposition of 6 harmonic motions in the 6 normal
coordinates.
\par The eigenfunctions $\psi_{i}(\xi_{i})$ are the ordinary harmonic 
oscillator eigenfuntions

\begin{eqnarray} \psi_{i}(\xi_{i}) &=& N_{v_i}e^{-(\alpha_{i}/2)\xi_{i}^{2}}
H_{v_i}(\sqrt{\alpha_{i}}\xi_{i}),\end{eqnarray}                          

\noindent
where $N_{v_i}$ is a normalization constant, $\alpha_{i} = \nu_{i}/{\hbar}$ and
$H_{v_i}(\sqrt{\alpha_{i}}\xi_{i})$ is a Hermite polynomial of the $v_i$th
degree. For large $\xi_{i}$ the eigenfunctions are governed by the exponential
functions which make the eigenfunctions go to zero very fast.  
\par The energy of each harmonic oscillator is

\begin{eqnarray} E_{i} &=& h\nu_{i}(v_{i} + \frac{1}{2}),\end{eqnarray}   

\noindent
where $v_{i} = 0,1,2,3,...$ and $\nu_i$ is the classical oscillation frequency
of the normal ``vibration'' $i$, and $v_i$ is the ``vibrational'' quantum 
number. The total energy of the system can assume only the values

\begin{eqnarray} E(v_{1},v_{2},v_{3}, ...v_{6}) &=& h\nu_{1}(v_{1} + 
\frac{1}{2}) + h\nu_{2}(v_{2} + \frac{1}{2}) + ... h\nu_{6}(v_{6} + 
\frac{1}{2}).\end{eqnarray}                                                

\par As was said above the three quarks in a baryon must always be in a plane. 
Therefore, each quark is composed of two oscillators and so we may rearrange 
the energy expression as 

\begin{eqnarray} E(n,m,k) &=& h\nu_{1}(n + 1) + h\nu_{2}(m + 1) + 
h\nu_{3}(k + 1),\end{eqnarray}                                             

\noindent
where $n=v_{1} + v_{2},m=v_{3} + v_{4},k=v_{5} + v_{6}$. Of course, $n,m,k$ can
assume the values, 0,1,2,3,... We may find the constants $h\nu$ from the 
ground states of some baryons. They are the known quark masses taken as 
$m_{u}=m_{d}= 0.31$Gev, $m_{s}= 0.5$Gev, $m_{c}=1.7$Gev,$m_{b}=5$Gev and
$m_{t}=174$GeV. 
\par Let us start the calculation with the states ddu(neutron), uud(proton) and
ddd($\Delta^{-}$), uuu ($\Delta^{++}$) and their resonances. All the energies 
below are given in Gev. The experimental values of baryon masses were taken 
from reference 67. Because $m_{u}=m_{d}$, we have that the energies calculated
by the formula  
\begin{eqnarray} E_{n,m,k} &=& 0.31(n+m+k + 3)\end{eqnarray}       
\noindent
correspond to many energy states. The calculated values are displayed in Table 
7. One observes in Table 7 that the particles that belong to it are $N$ and 
$\Delta$, which are particles that decay via the strong interaction either 
into $N$ or into $\Delta$(besides the electromagnetic decay, sometimes). For 
example 
\begin{itemize}
\item $\Delta(1232) {\rightarrow} N{\pi}$; 
\item $N(1440) {\rightarrow} N{\pi}$, $N{\pi}\pi$, $\Delta{\rho}$, $N{\rho}$; 
\item $\Delta(1600) {\rightarrow} N{\pi}$, $N{\pi}\pi$, $\Delta{\pi}$, 
$N{\rho}$, $N(1440)\pi$;
\item $N(1520) {\rightarrow} N{\pi}$, $N{\eta}$, $N{\pi}\pi$, $\Delta{\pi}$,
$N{\rho}$;
\end{itemize}
\noindent
Therefore, with the help of Table 7 we can easily understand the above decays. 
When a resonance decays into a particle of another table, then the decay is 
weak. For example, 
\begin{itemize}
\item $\Delta(1905) {\rightarrow} \Sigma{K}$;
\item $N(1650) {\rightarrow} \Lambda{K}$;
\end{itemize}
\par The other states appear to be composite states of a pion with each one of 
the states given by the above formulas. That is, they are sort of molecules.
For example, the $P_{11}$ state of $N$ is the composition of a pion with the 
state $P_{33}$, which corresponds in our notation to $(n,m,k)$(with $n+m+k=1$).
As a short notation we will just use $P_{33}\biguplus\pi$. 
Assuming that there is a weak
molecular type potential between a $(n,m,k)$ state and a pion we can say that
about the equilibrium position the potential is harmonic. The energies are then
described by(a system of two weakly interacting oscillators)
\begin{eqnarray} E_{(p;n,m,k;\pi)} &=& 2E_{n,m,k}(l + 0.5) + 
\pi_{j}\end{eqnarray}
\noindent							
in which $E_{n,m,k}$ is given by Eq. 81 above and $l,j=0,1,2,...$ and 
$\pi_{j}$ is calculated in section 21. The displayed values in Table 8
correspond to $l=j=0$ in the above formula. We may construct similar tables
for the other $l$ and $j$. In sections 20, 21 and 22 we will go further on this
subject. 
\par The energies of the particles $\Lambda$ and $\Sigma$, which are composed
of $uus$ and $uds$ are given by
\begin{eqnarray} E_{n,m,k} &=& 0.31(n+m+2) + 0.5(k+1).\end{eqnarray} 

\noindent
The results are displayed in Table 9. The agreement with the experimental 
values is excellent. As to the decay modes one observes the same as for
$N$ and $\Delta$, that is, decays via the strong interaction go as
$\Sigma {\rightarrow} \Lambda$ and $\Lambda {\rightarrow} \Sigma$. By means of
the weak interaction the two particles decay into $N$ and $\Delta$.
\par For the $\Xi$($ssd$) of ($ssu$) particle the energies are expressed by
\begin{eqnarray} E_{n,m,k} &=& 0.31(n+1) + 0.5(m+k+2).\end{eqnarray}
\noindent
See Table 10 to check the agreement with the experimental data. 
In this case we have just one kind of particle. The decays of it happen 
via the weak, the strong and the eletromagnetic interactions either into other
particles or into $\Xi$. For example,\newline
\noindent
$\Xi^{-} {\rightarrow} \Lambda{\pi^{-}}$, $\Sigma^{-}\gamma$, 
${\Lambda}e^{-}\bar{\nu}_{e}$, ${\Lambda}\mu^{-}\bar{\nu}_{\mu}$, 
${\Sigma^{0}}e^{-}\bar{\nu}_{e}$, ${\Sigma^{0}}\mu^{-}\bar{\nu}_{\mu}$,
${\Xi^{0}}e^{-}\bar{\nu}_{e}$, $n\pi^{-}$, $ne^{-}\bar{\nu}_{e}$, 
$n\mu^{-}\bar{\nu}_{\mu}$, \newline
$p\pi^{-}\pi^{-}$, $p\pi^{-}e^{-}\bar{\nu}_{e}$,
$p\pi^{-}\mu^{-}\bar{\nu}_{\mu}$; \newline
\noindent
$\Xi(1950) {\rightarrow} \Lambda\bar{K}$, $\Sigma\bar{K}$, $\Xi\pi$;

\newpage
\begin{center}
\begin{tabular}{||c||c|l|c|l||} \hline
& & & &  \\
$n,m,k$ & $E_{C}(Gev)$ & $E_{M}$(Gev) & Error(\%) & $L_{2I.2J}$ \\
& & & &  \\
\hline\hline 
0,0,0  & 0.93  & 0.938($N$) & 0.9 & $P_{11}$ \\ 
\hline 
$n+m+k=1$ & 1.24 & 1.232($\Delta$) & 0.6  & $P_{33}$ \\
\hline
$n+m+k=2$ & 1.55 & 1.52($N$) & 1.9 & $D_{13}$ \\
$n+m+k=2$ & 1.55 & 1.535($N$) & 1.0 & $S_{11}$ \\
$n+m+k=2$ & 1.55 & 1.6($\Delta$) & 3.1 & $P_{33}$ \\
$n+m+k=2$ & 1.55 & 1.62($\Delta$) & 4.5 & $S_{31}$ \\
\hline
$n+m+k=3$ & 1.86 & 1.90($N$) & 2.2 & $P_{13}$ \\
$n+m+k=3$ & 1.86 & 1.90($\Delta$) & 2.2 & $S_{31}$ \\
$n+m+k=3$ & 1.86 & 1.905($\Delta$) & 2.4 & $F_{35}$ \\
$n+m+k=3$ & 1.86 & 1.91($\Delta$) & 2.7 & $P_{31}$ \\
$n+m+k=3$ & 1.86 & 1.92($\Delta$) & 3.2 & $P_{33}$ \\
\hline
$n+m+k=4$ & 2.17 & 2.08($N$) & 4.1 & $D_{13}$ \\
$n+m+k=4$ & 2.17 & 2.09($N$) & 3.7 & $S_{11}$ \\
$n+m+k=4$ & 2.17 & 2.10($N$) & 3.2 & $P_{11}$ \\
$n+m+k=4$ & 2.17 & 2.15($\Delta$) & 0.9 & $S_{31}$ \\
$n+m+k=4$ & 2.17 & 2.19($N$) & 0.9 & $G_{17}$ \\
$n+m+k=4$ & 2.17 & 2.20($N$) & 1.4 & $D_{15}$ \\
$n+m+k=4$ & 2.17 & 2.20($\Delta$) & 1.4 & $G_{37}$ \\
$n+m+k=4$ & 2.17 & 2.22($N$) & 2.3 & $H_{19}$ \\
\hline
$n+m+k=5$ & 2.48 & 2.39($\Delta$) & 3.6 & $F_{37}$ \\
$n+m+k=5$ & 2.48 & 2.40($\Delta$) & 3.2 & $G_{39}$ \\
$n+m+k=5$ & 2.48 & 2.42($\Delta$) & 2.4 & $H_{3,11}$ \\
\hline
$n+m+k=6$ & 2.79 & 2.7($N$) & 3.2 & $K_{1,13}$ \\
$n+m+k=6$ & 2.79 & 2.75($\Delta$) & 1.4 & $I_{3,13}$ \\
\hline
$n+m+k=7$ & 3.10 & to be found & ? & ? \\
\hline
... & ... & ... & ... & ... \\
\hline\hline
\end{tabular}
\end{center}
\begin{center}
\parbox{4.5in}
{Table 7. Baryon states $N$ and $\Delta$. The energies $E_{C}$ were 
calculated according to the formula $E_{n,m,k} = 0.31(n+m+k+3)$ in which
$n,m,k$ are integers. $E_{M}$ is the measured energy.  The error means the 
absolute value of $(E_{C} - E_{M})/E_{C}$. We are able, of  course, to 
predict the energies of many other particles.}
\end{center}

\newpage

\begin{center}
\begin{tabular}{||c||c|l|c|l||} \hline
& & & &  \\
$n,m,k$ & $E_{C}(Gev)$ & $E_{M}$(Gev) & Error(\%) & $L_{2I.2J}$ \\
& & & &  \\
\hline\hline 
$(n,m,k)\biguplus\pi$, $n+m+k=1$ & 1.38 & 1.44($N$) & 4.3 & $P_{11}$ \\
\hline
$(n,m,k)\biguplus\pi$, $n+m+k=2$ & 1.69 & 1.65($N$) & 2.4 & $S_{11}$ \\
$(n,m,k)\biguplus\pi$, $n+m+k=2$  & 1.69 & 1.675($N$) & 0.9 & $D_{15}$ \\
$(n,m,k)\biguplus\pi$, $n+m+k=2$  & 1.69 & 1.68($N$) & 0.6 & $F_{15}$ \\
$(n,m,k)\biguplus\pi$, $n+m+k=2$  & 1.69 & 1.70($N$) & 0.6 & $D_{13}$ \\
$(n,m,k)\biguplus\pi$, $n+m+k=2$  & 1.69 & 1.70($\Delta$) & 0.6 & $D_{33}$ \\
$(n,m,k)\biguplus\pi$, $n+m+k=2$  & 1.69 & 1.71($N$) & 1.2 & $P_{11}$ \\
$(n,m,k)\biguplus\pi$, $n+m+k=2$  & 1.69 & 1.72($N$) & 1.8 & $P_{13}$ \\
\hline
$(n,m,k)\biguplus\pi$, $n+m+k=3$ & 2.00 & 1.93($\Delta$) & 3.5 & $D_{35}$ \\
$(n,m,k)\biguplus\pi$, $n+m+k=3$   & 2.00 & 1.94($\Delta$) & 3.0 & $D_{33}$ \\
$(n,m,k)\biguplus\pi$, $n+m+k=3$   & 2.00 & 1.95($\Delta$) & 2.5 & $F_{37}$ \\
$(n,m,k)\biguplus\pi$, $n+m+k=3$   & 2.00 & 1.99($N$) & 0.5 & $F_{17}$ \\
$(n,m,k)\biguplus\pi$, $n+m+k=3$   & 2.00 & 2.00($N$) & 0 & $F_{15}$ \\
$(n,m,k)\biguplus\pi$, $n+m+k=3$   & 2.00 & 2.00($\Delta$) & 0 & $F_{35}$ \\
\hline
$(n,m,k)\biguplus\pi$, $n+m+k=4$ & 2.31 & 2.25($N$) & 2.6 & $G_{19}$ \\
$(n,m,k)\biguplus\pi$, $n+m+k=4$  & 2.31 & 2.3($\Delta$) & 0.4 & $H_{39}$ \\
$(n,m,k)\biguplus\pi$, $n+m+k=4$  & 2.31 & 2.35($\Delta$) & 1.7 & $D_{35}$ \\
\hline
$(n,m,k)\biguplus\pi$, $n+m+k=5$ & 2.62 & 2.60($N$) & 0.8 & $I_{1,11}$ \\
\hline
$(n,m,k)\biguplus\pi$, $n+m+k=6$ & 2.93 & 2.95($\Delta$) & 0.7 & $K_{3,15}$ \\
\hline
$(n,m,k)\biguplus\pi$, $n+m+k=7$ & 3.24 & to be found & ? & ? \\
... & ... & ... & ... & ... \\
\hline\hline
\end{tabular}
\end{center}
\begin{center}
\parbox{4.5in}
{Table 8. Baryon states of $N$ and $\Delta$ which are composite states of a
pion and a regular $(n,m,k)$ state whose energy is calculated by Eq. 81. The 
calculated energies($E_{C}$) of the composite states above listed are expressed 
by Eq. 82 for $l=0$. $E_{M}$ is the measured energy.  The error means the 
absolute value of $(E_{C} - E_{M})/E_{C}$. We are able, of  course, to predict 
the energies of many other particles.}
\end{center}

\newpage

\begin{center}
\begin{tabular}{||c||c|l|c|c||} 
\hline\hline
& & & &  \\
State($n,m,k$) & $E_{C}(Gev)$ & $E_{M}$(Gev) & Error(\%) & $L_{I,2J}$ \\
& & & &  \\
\hline\hline 
0,0,0  & 1.12  & 1.116($\Lambda$) & 0.4 & $P_{01}$ \\ 
0,0,0  & 1.12  & 1.193($\Sigma$) & 6.5 & $P_{11}$ \\
\hline 
$n+m=1$, k=0 & 1.43 & 1.385($\Sigma$) & 3.2 & $P_{13}$ \\
$n+m=1$, k=0 & 1.43 & 1.405($\Lambda$) & 1.7 & $S_{01}$ \\
$n+m=1$, k=0 & 1.43 & 1.48($\Lambda$) & 3.5 & ? \\
\hline
0,0,1 & 1.62 & 1.52($\Lambda$) & 6.2 & $D_{03}$ \\
0,0,1 & 1.62 & 1.56($\Sigma$) & 3.7 & ? \\
0,0,1 & 1.62 & 1.58($\Sigma$) & 2.5 & $D_{13}$ \\
0,0,1 & 1.62 & 1.60($\Lambda$) & 1.2 & $P_{01}$ \\
0,0,1 & 1.62 & 1.62($\Sigma$) & 0 & $S_{11}$ \\
0,0,1 & 1.62 & 1.66($\Sigma$) & 2.5 & $P_{11}$ \\
0,0,1 & 1.62 & 1.67($\Sigma$) & 3.1 & $D_{13}$ \\
0,0,1 & 1.62 & 1.67($\Lambda$) & 3.1 & $S_{01}$ \\ 
\hline
$n+m=2$, k=0 & 1.74 & 1.69($\Lambda$) & 2.9 & $D_{03}$ \\
$n+m=2$, k=0 & 1.74 & 1.69($\Sigma$) & 2.9 & ? \\
$n+m=2$, k=0 & 1.74 & 1.75($\Sigma$) & 0.6 & $S_{11}$ \\
$n+m=2$, k=0 & 1.74 & 1.77($\Sigma$) & 1.7 & $P_{11}$ \\
$n+m=2$, k=0 & 1.74 & 1.775($\Sigma$) & 2.0 & $D_{15}$ \\
$n+m=2$, k=0 & 1.74 & 1.80($\Lambda$) & 3.4 & $S_{01}$ \\
$n+m=2$, k=0 & 1.74 & 1.81($\Lambda$) & 4.0 & $P_{01}$ \\
$n+m=2$, k=0 & 1.74 & 1.82($\Lambda$) & 4.6 & $F_{05}$ \\
$n+m=2$, k=0 & 1.74 & 1.83($\Lambda$) & 5.2 & $D_{05}$ \\
\hline\hline
\end{tabular}
\end{center}
\newpage
\begin{center}
\begin{tabular}{||c||c|l|c|l||}
\hline\hline
& & & & \\
State($n,m,k$) & $E_{C}$(Gev) & $E_{M}$(Gev) & Error(\%) & $L_{2I,2J}$ \\
& & & & \\
\hline\hline
$n+m=1$, k=1 & 1.93 & 1.84($\Sigma$) & 4.7 & $P_{13}$ \\
$n+m=1$, k=1 & 1.93 & 1.88($\Sigma$) & 2.6 & $P_{11}$ \\
$n+m=1$, k=1 & 1.93 & 1.89($\Lambda$) & 2.1 & $P_{03}$ \\
$n+m=1$, k=1 & 1.93 & 1.915($\Sigma$) & 0.8 & $F_{15}$ \\
$n+m=1$, k=1 & 1.93 & 1.94($\Sigma$) & 0.5 & $D_{13}$ \\
\hline
$n+m=3$, k=0 & 2.05 & 2.00($\Lambda$) & 2.5 & ? \\
$n+m=3$, k=0 & 2.05 & 2.00($\Sigma$) & 2.4 & $S_{11}$ \\
$n+m=3$, k=0 & 2.05 & 2.02($\Lambda$) & 1.5 & $F_{07}$ \\
$n+m=3$, k=0 & 2.05 & 2.03($\Sigma$) & 1.0 & $F_{17}$ \\
$n+m=3$, k=0 & 2.05 & 2.07($\Sigma$) & 1.0 & $F_{15}$ \\
$n+m=3$, k=0 & 2.05 & 2.08($\Sigma$) & 1.5 & $P_{13}$ \\
\hline
0,0,2 & 2.12 & 2.10($\Sigma$) & 0.9 & $G_{17}$ \\
0,0,2 & 2.12 & 2.10($\Lambda$) & 0.9 & $G_{07}$ \\
0,0,2 & 2.12 & 2.11($\Lambda$) & 0.5 & $F_{05}$ \\
\hline
$m+n=2$, k=1 & 2.24 & 2.25($\Sigma$) & 0.5 & ? \\
\hline
$n+m=4$, k=0 & 2.36 & 2.325($\Lambda$) & 1.5 & $D_{03}$ \\
$n+m=4$, k=0 & 2.36 & 2.35($\Lambda$) & 0.4 & ? \\
\hline
$n+m=1$, k=2 & 2.43 & 2.455($\Lambda$) & 1.0 & ? \\
\hline
$n+m=3$, k=1 & 2.55 & 2.585($\Lambda$) & 1.4 & ? \\
\hline
0,0,3 & 2.62 & 2.62($\Sigma$) & 0 & ? \\
\hline
$n+m=5$, k=0 & 2.67 & to be found & ? & ? \\
\hline
$n+m=2$, k=2 & 2.74 & to be found & ? & ? \\
\hline
$n+m=4$, k=1 & 2.86 & to be found & ? & ? \\
\hline
$n+m=1$, k=3 & 2.93 & to be found & ? & ? \\
\hline
$n+m=6$, k=0 & 2.98 & 3.00($\Sigma$) & 0.7 & ? \\
\hline
$n+m=3$, k=2 & 3.05 & to be found & ? & ? \\
\hline
$n=m=0$, k=4 & 3.12 & to be found & ? & ? \\
\hline
$n+m=5$, k=1 & 3.17 & 3.17($\Sigma$) & 0 & ? \\
\hline
$n+m=2$, k=3 & 3.24 & to be found & ? & ? \\
\hline
... & ... & ... & ...& ... \\
\hline\hline
\end{tabular}
\end{center}
\vskip .3in
\begin{center}
\parbox{4.5in}
{Table 9. Baryon states $\Sigma$ and $\Lambda$. The energies $E_{C}$ were
calculated according to the formula $E_{n,m,k}= 0.31(n+m+2) + 0.5(k+1)$.
$E_{M}$ is the measured energy. The error  means the absolute value of 
$(E_{C} - E_{M})/E_{C}$. We are able to predict the energy levels of many 
other particles.}
\end{center}

\newpage
 
\par In the same way the energies of $\Omega$($sss$) are obtained by
\begin{eqnarray} E_{n,m,k} &=& 0.5(n+m+k+3).\end{eqnarray}
\noindent
The energies are displayed in Table 11. The discrepancies are higher, of the 
order of 10\% and decreases as the energy increases. This is a tendency which 
is also observed for the other particles. This may mean that, at the bottom, 
the potential is less flat than the potential of a harmonic oscillator. The 
decays occur as with the $\Xi$, that is, one sees weak, electromagnetic and 
strong decays into other particles such as $\Xi$ and $\Lambda$. There are also
composite states whose energies are calculated according to Eq. 82. They are
shown in Table 12.
\par The energies of the charmed baryons($C=+1$) $\Lambda_{c}^{+}$, 
$\Sigma_{c}^{++}$, $\Sigma_{c}^{+}$ and $\Sigma_{c}^{0}$ are given by 
\begin{eqnarray} E_{n,m,k} &=& 0.31(n+m+2) + 1.7(k+1).\end{eqnarray}
\noindent
The levels are shown in Table 13. 
\par For the charmed baryons($C=+1$) $\Xi_{c}^{+}$ and $\Xi_{c}^{0}$ we have
\begin{eqnarray} E_{n,m,k} &=& 0.31(n+1) + 0.5(m+1) + 1.7(k+1).\end{eqnarray}
\noindent
The results are displayed in Table 14. 
\par As for the $\Omega_{c}^{0}$, its energies are
\begin{eqnarray} E_{n,m,k} &=& 0.5(n+m+2) + 1.7(k+1).\end{eqnarray}    

\noindent
Table 15 shows the results of the energy levels. The other charmed baryons 
should follow the same decay trends of the other baryons previously discussed.
\par We may predict the energies of many other baryons given by the formulas:

\begin{itemize} 
\item ucc and dcc, $E_{n,m,k} = 0.31(n+1) + 1.7(m+k+2)$;
\item scc, $E_{n,m,k} = 0.5(n+1) + 1.7(m+k+2)$;
\item ccc, $E_{n,m,k} = 1.7(n+m+k+3)$;
\item ccb, $E_{n,m,k} = 1.7(n+m+2) + 5(k+1)$;
\item cbb, $E_{n,m,k} = 1.7(n+1) + 5(m+k+2)$;
\item ubb and dbb, $E_{n,m,k} = 0.31(n+1) + 5(m+k+2)$ ;
\item uub, udb and ddb, $E_{n,m,k} = 0.31(n+m+2) + 5(k+1)$;
\item bbb, $E_{n,m,k} = 5(n+m+k+3)$;
\item usb and dsb, $E_{n,m,k} = 0.31(n+1) + 0.5(m+1) + 5(k+1)$;
\item sbb, $E_{n,m,k} = 0.5(n+1) + 5(m+k+2)$;
\item scb, $E_{n,m,k} = 0.5(n+1) + 1.7(m+1) + 5(k+1)$;
\item ucb, $E_{n,m,k} = 0.31(n+1) + 1.7(m+1) + 5(k+1)$;
\item ttt, $E_{n,m,k} = (174{\pm}17)(n+m+k+3)$;
\item and all combinations of t with u, d, c, s and b. 
\end{itemize}

\par The first state(0,0,0) of $udb$ which has an energy equal to 5.641GeV 
has been recently found. The above formula for this state yields the energy 
5.62GeV. The error is just 0.3\%. 
\par {\bf{All the recently found levels clearly shows that the naive model 
above described is quite good. They also confirm that the superstrong 
interaction has to exist. I am quite sure that more confirmation on this 
will come in the near future and I hope that this work and its previous 
versions will receive the proper credit for having demonstrated the need for
this new interaction.}}

\rule{0in}{.5in}

\begin{center}
\begin{tabular}{||c||l|l|l|c||} \hline
& & & & \\
State($n,m,k$) & $E_{C}$(Gev) & $E_{M}$(Gev) & Error(\%) & $L_{2I,2J}$ \\
& & & & \\
\hline\hline 
0,0,0  & 1.31  & 1.318 & 0.6 & $P_{11}$ \\ 
\hline 
1,0,0  & 1.62 & 1.53 & 5.6 & $P_{13}$ \\
1,0,0 & 1.62 & 1.62 & 0 & ? \\
1,0,0 & 1.62 & 1.69 & 4.3 & ? \\
\hline
n=0, $m+k=1$ & 1.81 & 1.82 & 0.6 & $D_{13}$ \\
\hline
2,0,0 & 1.93 & 1.95 & 1.0 & ? \\
\hline
n=1, $m+k=1$ & 2.12 & 2.03 & 4.2 & ? \\
n=1, $m+k=1$ & 2.12 & 2.12 & 0 & ? \\
\hline
n=3, $m=k=0$ & 2.24 & 2.25 & 0.5 & ? \\
\hline
n=0, $m+k=2$ & 2.31 & 2.37 & 2.6 & ? \\
\hline
n=2, $m+k=1$ & 2.43 & to be found & ? & ? \\
\hline
n=4, $m=k=0$ & 2.55 & 2.5 & 2.0 & ? \\
\hline
n=1, $m+k=2$ & 2.62 & to be found & ? & ? \\
... & ... & ... & ... & ... \\
\hline\hline
\end{tabular}
\end{center}
\vskip .2in

\begin{center}
\parbox{4.5in}
{Table 10. Baryon states $\Xi$. The energies $E_{C}$ were
calculated according to the formula $E_{n,m,k}= 0.31(n+1) + 0.5(m+k+2)$.
$E_{M}$ is the measured energy. The error means the absolute 
value of $(E_{C} - E_{M})/E_{C}$. We are able, of course, to 
predict the energies of many other particles. The state $\Xi(1530)P_{13}$
appears to be the lowest state of the composite $\Xi\biguplus\pi$. Its
decay is in fact $\Xi\pi$.}
\end{center}
\vskip .5in

\newpage

\begin{center}
\begin{tabular}{||c||l|c|l||} \hline
& & &  \\
State($n,m,k$) & $E_{C}$(Gev) & $E_{M}$(Gev) & Error(\%) \\
& & &  \\
\hline\hline 
0,0,0  & 1.5  & 1.672 & 11.7 \\ 
\hline 
$n+m+k=1$ & 2.0 & 2.25 & 12.5 \\
\hline
$2.0 + \pi$ & 2.14 & 2.38 & 11.2 \\
\hline
$n+m+k=2$ & 2.5 & 2.47 & 1.2 \\
\hline
$2.5 + \pi$ & 2.64 & to be found & ? \\
\hline
$n+m+k=3$ & 3.0 & to be found & ? \\
\hline
... & ... & ... & ... \\
\hline\hline
\end{tabular}
\end{center}
\vskip .2in

\begin{center}
\parbox{4.5in}
{Table 11. Baryon states $\Omega$. The energies $E_{C}$ were
calculated according to the formula $E_{n,m,k}= 0.5(n+m+k+3)$.
$E_{M}$ is the measured energy. The error means the absolute 
value of $(E_{C} - E_{M})/E_{C}$. We are able, of course, to 
predict the energies of many other particles.}
\end{center}

\rule{0in}{.5in}

\begin{center}
\begin{tabular}{||c||l|c|l||} \hline
& & &  \\
State($n,m,k$) & $E_{C}$(Gev) & $E_{M}$(Gev) & Error(\%) \\
& & &  \\
\hline\hline 
$(n,m,k)\biguplus\pi$, $n+m+k=1$ & 2.14 & 2.38 & 11.2 \\
\hline
$(n,m,k)\biguplus\pi$, $n+m+k=2$ & 2.64 & to be found & ? \\
\hline
$(n,m,k)\biguplus\pi$, $n+m+k=3$ & 3.14 & to be found & ? \\
\hline
... & ... & ... & ... \\
\hline\hline
\end{tabular}
\end{center}
\vskip .2in

\begin{center}
\parbox{4.5in}
{Table 12. Baryonic states which are composed of pion and a regular $\Omega$
state. The energies $E_{C}$ were calculated according to Eq. 82. $E_{M}$ is 
the measured energy. The error means the absolute value of 
$(E_{C} - E_{M})/E_{C}$. We are able, of course, to predict the energies of 
many other particles.}
\end{center}

\newpage

\begin{center}
\begin{tabular}{||c||l|c|l||} \hline
& & &  \\
State($n,m,k$) & $E_{C}$(Gev) & $E_{M}$(Gev) & Error(\%) \\
& & &  \\
\hline\hline 
0,0,0 & 2.32 & 2.285($\Lambda_{c}$) & 1.5 \\ 
$(0,0,0){\biguplus}\pi$ & 2.455 & 2.455($\Sigma_{c}$) & 0 \\
\hline 
$n+m=1$, k=0 & 2.63 & 2.594($\Lambda_{c}$) & 0.1 \\
$n+m=1$, k=0 & 2.63 & 2.627($\Lambda_{c}$) & 0.01 \\
\hline
$n+m=2$, k=0 & 2.94 & to be found & ? \\
\hline
... & ... & ... & ...\\
\hline\hline
\end{tabular}
\end{center}
\vskip .2in

\begin{center}
\parbox{4.5in}
{Table 13. Baryon states $\Lambda_{c}$ and $\Sigma_{c}$. The 
energies $E_{C}$ were calculated according to the formula 
$E_{n,m,k}= 0.31(n+m+2) + 1.7(k+1)$. $E_{M}$ is the measured 
energy. The error means the absolute value of $(E_{C} - E_{M})/E_{C}$.
We are able, of course, to predict the energies of many other 
particles. The state with energy 2.63 MeV had already been predicted
in another version of this work. The experimental levels 2.594 MeV
and 2.627 MeV have confirmed the theoretical values. It appears that
the level $\Sigma_{c}(2.455)$ is a composition of the level 
$(0,0,0)$(that is the 2.285 $Lambda_{c}$) with a pion as is also 
inferred from its decay.}
\end{center}

\rule{0in}{.3in}

\begin{center}
\begin{tabular}{||c||l|c|l||} \hline
& & &  \\
State($n,m,k$) & $E_{C}$(Gev) & $E_{M}$(Gev) & Error(\%) \\
& & &  \\
\hline\hline 
0,0,0  & 2.51  & 2.47($\Xi_{c}^{+}$ & 1.6 \\ 
\cline{2-3} 
       & 2.51  & 2.47($\Xi_{c}^{0}$ & 1.6 \\
\hline 
1,0,0 & 2.82 & to be found & ? \\
\hline
0,1,0 & 3.01 & to be found & ? \\
\hline
... & ... & ... & ...\\
\hline\hline
\end{tabular}
\end{center}
\vskip .3in
\begin{center}
\parbox{4.5in}
{Table 14. Baryon states $\Xi_{c}$. The energies $E_{C}$ were 
calculated according to the formula $E_{n,m,k}= 0.31(n+1) + 0.5(m+1) 
+ 1.7(k+1)$. $E_{M}$ is the measured energy. The error means the absolute 
value of $(E_{C} - E_{M})/E_{C}$. We are able, of course, to 
predict the energies of many other particles. The recently found level
$\Xi_{c}(2645)$ is just a composition of the regular level $\Xi_{c}^{+}$ with
a pion as its decay confirms.}
\end{center}

\newpage

\begin{center}
\begin{tabular}{||c||l|c|l||} \hline
& & &  \\
State($n,m,k$) & $E_{C}$(Gev) & $E_{M}$(Gev) & Error(\%) \\
& & &  \\
\hline\hline 
0,0,0  & 2.7  & 2.704($\Omega_{c}^{0}$ & 0 \\ 
\hline 
$n+m=1$, k=0 & 3.2 & to be found & ? \\
\hline
$n+m=2$, k=0 & 3.7 & to be found & ? \\
\hline
... & ... & ... & ...\\
\hline\hline
\end{tabular}
\end{center}
\vskip .2in
\begin{center}
\parbox{4.5in}
{Table 15. Baryon states $\Omega_{c}$. The energies $E_{C}$ were 
calculated according to the formula $E_{n,m,k}= 0.5(n+m+2) + 1.7(k+1)$.
$E_{M}$ is the measured energy. The error means the absolute 
value of $(E_{C} - E_{M})/E_{C}$. We are able, of course, to 
predict the energies of many other particles. The energy of the level 
$(0,0,0)$ above shown had been predicted in other versions of this work.}
\end{center}
\rule{0in}{.3in}

\par According to our considerations the total mass of a baryon must be given 
by 

\begin{eqnarray} M &=& M_{K} + V_{S,SS} + V_{e} + V_{spin} + V_{rot}
\end{eqnarray} 

\noindent
where $M_{K}$ is the total kinetic energy of the three constituent quarks,
$V_{S,SS}$ is the potential energy of the combination of the strong and 
superstrong interactions(our molecular type potential), $V_{e}$ is the 
electromagnetic interaction, and $V_{spin}$ is the spin dependent term of the 
mass.  We showed above that $V_{S,SS}$ is the leading term. Since quarks have 
charges the term $V_{e}$ must contribute in the splitting of the levels but
is a small effect. Also, rotation does not appear to play any significant role
as we can infer from the excellent agreement between the calculated levels and
the experimental values. The agreement also implies that quarks do not move at
relativistic speeds inside baryons. Actually, as is argued by 
Lichtenberg$^{68}$ 
and others it is hard to see how $SU(6)$ is a good approximate dynamical 
symmetry of baryons if quarks move at relativistic velocities inside baryons. 
\par We clearly see that the masses of baryons are expressed quite well by the 
simple model above described. It lends support to the general framework of 
having quarks as the basic building particles of baryons. Therefore, it agrees 
well with QCD. But, the model is also based on the idea of having a 
substructure for quarks.\newline
\vskip .3in

\noindent
18) GENERALIZATION OF THE GELL-MANN-OKUBO MASS FORMULA
\vskip .15in
\par The Gell-Mann-Okubo mass formula 
\begin{eqnarray} M &=& M_{0} + M_{1}Y + M_{2}\left(I(I+1) - \frac{Y^2}{4} 
\right)\end{eqnarray}                              
\noindent
where $M_{0}$, $M_{1}$ and $M_{2}$ are suitable constants, $I$ is the isospin,
and $Y$ is the hypercharge, has been widely used as a relation among the masses
of baryon states belonging either to an octet or to a decuplet.  This is
a phenomenological formula ``with no clear physical reasons for the 
assumptions on which it is based''$^{69}$. As we will show shortly the 
reason behind the above mass formula is the general formula for the mass of 
a baryon
\begin{eqnarray} E_{n,m,k} &=& \hbar{\nu_{1}}(n+1) + \hbar{\nu_{2}}(m+1) + 
\hbar{\nu_{3}}(k+1).\end{eqnarray} 
\noindent
For the decuplet of $SU_{3}$(u,d,s) Eq. (34) becomes
\begin{eqnarray} M = M_{0} + M_{1}Y\end{eqnarray} 
\noindent
where $Y$ is the hypercharge. The relation among the masses of  baryons
of the $SU_{3}$(u,d,s)decuplet is given by
\begin{eqnarray} M_{\Sigma} - M_{\Delta} &=& M_{\Xi} - M_{\Sigma} = 
M_{\Omega^{-}} - M_{\Xi}.\end{eqnarray} 
\noindent
According to Eq. (91) the equality of the first two terms of Eq. (37) is 
given by
\begin{eqnarray} 0.31(n+1) &+& 0.5(m+k+2) + 0.31(n+m+k+3) = \nonumber  \\
& & 2\left(0.31(n+m+2) + 0.5(k+1)\right)\end{eqnarray} 
\noindent
which is satisfied for any $n$, and $m=k$. Actually, instead of $\Delta$ we
may have either $\Delta$ or $N$. For example(see Tables 7, 9 and 10), 
\begin{itemize} 
\item $n=0$, $k=m=0$, $1.12 - 0.93 = 1.31 - 1.12 = 0.19$;
\item $n=0$, $k=m=1$, $1.93 - 1.55 = 2.31 - 1.93 = 0.38$;
\item $n=1$, $k=m=0$, $1.43 - 1.24 = 1.62 - 1.43 = 0.19$;
\item $n=1$, $k=m=1$, $2.24 - 1.86 = 2.62 - 2.24 = 0.38$;
\item $n=2$, $k=m=0$, $1.74 - 1.55 = 1.93 - 1.74 = 0.19$;
\item $n=3$, $k=m=0$, $2.05 - 1.86 = 2.24 - 2.05 = 0.19$;
\item ... ad infinitum
\end{itemize}
\noindent
The equality of the first term with the third term of Eq. (93) yields
\begin{eqnarray} 0.31(n+m+2) &+& 0.5(k+1) - 0.31(n+m+k+3) = \nonumber  \\
0.5(n+m+k+3) &-& 0.31(n+1) - 0.5(m+k+2)\end{eqnarray} 
\noindent
which is satisfied for any $n,m,k$. Again, instead of $\Delta$ we may have
$N$. For example(observe Tables 7, 9, 10 and 11),
\begin{itemize}
\item $n=m=k=0$, $1.12 - 0.93 = 1.5 - 1.31 = 0.19$;
\item $n=0$, $m+k=1$, $1.43 - 1.24 = 2.0 - 1.81 = 0.19$;
\item $n=k=0$, $m=2$, $1.74 - 1.55 = 2.5 - 2.31 = 0.19$;
\item ... ad infinitum
\end{itemize}
\noindent
Finally, equaling the second and third terms of Eq. (93) one obtains
\begin{eqnarray} 0.5(n+m+k+3) &+& 0.31(n+m+2) + 0.5(k+1) = \nonumber  \\
& & 2\left(0.31(n+1) + 0.5(m+k+2)\right)\end{eqnarray}
\noindent
which is satisfied if $n=m$ for any value of $k$. As examples one finds(see
Tables 9, 10 and 11)
\begin{itemize}
\item $n=m=k=0$, $1.31 - 1.12 = 1.5 - 1.31 = 0.19$;
\item $n=m=0$, $k=1$, $1.81 - 1.62 = 2 - 1.81 = 0.19$;
\item $n=m=1$, $k=0$, $2.12 - 1.74 = 2.5 - 2.12 = 0.38$;
\item ... ad infinitum
\end{itemize}
\par For an octet of $SU_{3}$(u,d,s) one obtains 
\begin{eqnarray} 3M_{\Lambda} + M_{\Sigma} &=& 2M_{N} - M_{\Xi}\end{eqnarray}
\noindent
which in terms of Eq. (91) becomes
\begin{eqnarray} 
2\left(0.31(n+m+2)\right. &{+}& \left. 0.5(k+1)\right) = \nonumber \\
0.31(n+m+k+3) &{-}& 0.31(n+1) - 0.5(m+k+2).\end{eqnarray}
\noindent
This equation is satisfied if $k=m$ for any $n$. For example, one has(see
Tables 7, 9 and 10)
\begin{itemize}
\item $n=m=k=0$, $2{\times}1.12 = 0.93 + 1.31$;
\item $n=1$, $m=k=0$, $2{\times}1.43 = 1.24 + 1.62$;
\item $n=2$, $m=k=0$, $2{\times}1.74 = 1.55 + 1.93$;
\item $n=3$, $m=k=0$, $2{\times}2.05 = 1.86 + 2.24$;
\item $n=0$, $m=k=1$, $2{\times}1.93 = 1.55 + 2.31$;
\item $n=k=m=1$, $2{\times}2.24 = 1.86 + 2.62$;
\item ... ad infinitum.
\end{itemize}
\par Let us now try to relate the constants $M_{0}$ and $M_{1}$ to the
quark masses. Let us consider, for example, the decuplet of $SU_{3}$(u,d,s).
In terms of the hypercharge the masses of the particles are described by
\begin{eqnarray} M_{\Omega^{-}} &=& M_{0} - 2M_{1};\end{eqnarray}  
\begin{eqnarray} M_{\Xi} &=& M_{0} - M_{1};\end{eqnarray}          
\begin{eqnarray} M_{\Sigma} &=& M_{0};\end{eqnarray}              
\begin{eqnarray} M_{\Delta} &=& M_{0} + M_{1}.\end{eqnarray}      
\noindent
As we calculated above from the masses of $\Xi$, $\Sigma$ and $\Delta$ one
finds that $m=k$(any $n$) and from the masses of $\Omega^{-}$, $\Xi$ and
$\Sigma$ one has $n=m$(any $k$). Therefore, in terms of $Y$ the masses of
$\Xi$, $\Sigma$ and $\Delta$ are given by
\begin{eqnarray} M_{n,m}(Y) &=& 0.31(n+m+2) + 0.5(m+1) - 
0.19(m+1)Y\end{eqnarray}
\noindent
and the mass of $\Omega^{-}$ is the above formula with $n=m$, that is,
\begin{eqnarray} M_{\Omega^{-}}(Y) &=& (1.12 - 0.19Y)(n+1).\end{eqnarray}
\noindent
It is easy to observe that the composite baryons(composed with pions) whose 
energies are given by $E_{n,m,k} + \pi$ do not obey the Gell-Mann-Okubo mass 
formula. 
\par From the $SU(4)$(Fig. 19) multiplets of baryons made of $u$, $d$, $s$, 
and $c$ quarks, and considering Eq. (91) one obtains, for example,
\begin{eqnarray} M_{\Omega_{ccc}} - M_{\Xi_{cc}} &=& M_{\Xi_{cc}} - 
M_{\Sigma_{c}} = M_{\Sigma_{c}} - M_{\Delta};\end{eqnarray}
\noindent
\begin{eqnarray} M_{\Omega_{ccc}} - M_{\Omega_{cc}} &=& M_{\Omega_{cc}} -
M_{\Omega_{c}} = M_{\Omega_{c}} - M_{\Omega}\end{eqnarray}
\noindent
and
\begin{eqnarray} 2M_{\Xi_{cc}} &=& M_{\Omega_{ccc}} + M_{\Sigma_{c}}
\end{eqnarray}
\noindent
or more generally, one obtains
\begin{eqnarray} M_{q_{1}q_{1}q_{1}} - M_{q_{2}q_{1}q_{1}} &=& 
M_{q_{2}q_{1}q_{1}} - M_{q_{2}q_{2}q_{1}} = M_{q_{2}q_{2}q_{1}} - 
M_{q_{2}q_{2}q_{2}}\end{eqnarray}

\noindent
and
\begin{eqnarray} 2M_{q_{1}q_{1}q_{2}} &=& M_{q_{1}q_{1}q_{1}} + 
M_{q_{1}q_{2}q_{2}}\end{eqnarray}
\noindent
in which we can consider $SU(6)$, that is, $q_{i}$ may be $u$, $d$, $c$, $s$
$b$, and $t$. In the case of considering $u$ and $d$, we may have the
combinations $ud$, $uu$, and $dd$ for $q_{i}q_{i}$. We also may have
\begin{eqnarray} M_{q_{1}q_{2}q_{3}} - M_{q_{4}q_{2}q_{3}} &=&
M_{q_{1}q_{i}q_{j}} - M_{q_{4}q_{i}q_{j}}.\end{eqnarray}
\noindent
\par We conclude this section saying that the Gell-Mann-Okubo mass
formula is a natural consequence of the pairwise interacting harmonic 
potential among quarks. 
\par We find in the literature several relations among the masses of baryons.
They are, actually, just special cases of the above formulas.
\newline
\vskip .3in
\noindent
19) THE EXCITED STATES OF QUARKS
\vskip .2in
\par As was shown at the end of section 1 the interaction between two 
primons brings forth the action of the strong force between them, and there is 
also the action of the superstrong interaction between them. Therefore, there 
is a sort of effective potential energy which can be approximated by a 
molecular type potential. This potential energy is harmonic around its minimum.
Considering that quarks do not rotate at relativistic speeds and doing in the 
same fashion as we did in section 7 we obtain that the masses of quarks 
should be given by 
\begin{eqnarray} E_{q} &=& 2\pi{\hbar}\nu(k + 1/2)\end{eqnarray} 
\noindent
in which $k = 0,1,2,3,...$ and $\bar{\hbar}$ has been previously defined.
\par Let us now determine the energy levels of the quarks u,d,s,c,t and b. The
ground state of the u quark is about 0.31GeV. Therefore, 
${\hbar}\nu=0.62$GeV. Thus, the energies of the $u$ and $d$ quarks are  
0.31GeV, 0.93GeV, 1.55GeV, 2.17GeV, ... Making the same for the other quarks 
we obtain the following results:
\begin{itemize}
\item $E_{u} = E_{d} = 0.62(k + 1/2)$;
\item $E_{s} = 1(k + 1/2)$;
\item $E_{c} = 3.4(k + 1/2)$;
\item $E_{b} = 10(k + 1/2)$;
\item $E_{t} = (348{\pm}35)(k + 1/2)$.
\end{itemize}
\par Since $k + 1/2$ is a half integer, we may also write $E_{q}$ as
\begin{eqnarray} E_{q} &=& E_{o}l\end{eqnarray}                         
\noindent
where $l$ is an odd integer and $E_{o}={\hbar}/2$. The calculated values 
are shown in Table 16. 
\rule{0in}{.2in}

\begin{center}
\begin{tabular}{||c||c|c|c|c|c||} \hline
& & & & & \\
State($n$) & $E_{u}=E_{d}$(GeV) & $E_{s}$(GeV) & $E_{c}$(GeV) & $E_{b}$(GeV) & 
$E_{t}$(GeV)\\
& & & & & \\
\hline\hline 
& & & & & \\
0 & 0.31 & 0.5 & 1.7 & 5 & $174{\pm}17$\\ 
& & & & & \\
\hline 
& & & & & \\
1 & 0.93 & 1.5 & 5.1 & 15 & $522{\pm}52$\\
& & & & & \\
\hline
& & & & & \\
2 & 1.55 & 2.5 & 8.5 & 25 & $870{\pm}87$\\
& & & & & \\
\hline
& & & & & \\
3 & 2.17 & 3.5 & 11.9 & 35 & $1218{\pm}122$\\
& & & & & \\
\hline
& & & & & \\
...& ... & ... & ... & ... &... \\
& & & & & \\
\hline\hline
\end{tabular}
\end{center}
\vskip .15in
\begin{center}
\parbox{4.5in}
{Table 16. Possible energy states of quarks u, d, c, s, b and t.}
\end{center}

\newpage
\par Because quarks are confined we could observe the
excited levels only by means of hadrons, but actually we can not. The reason 
is simple. Let us, for example, consider the baryons made of the first 
excited states of the u or d quarks. According to Eq.(80) such baryons have 
energies 
\begin{eqnarray} E^{*}_{111u} &=& E^{*}_{111d} = 0.93(n + 1) + 0.93(m + 1) +
0.93(k + 1) \nonumber \\
&=& 0.93(n + m + k + 3)\end{eqnarray}  
\noindent
in which the asterisk means excited state and the repeated ones mean that each
quark is in the first excited state.  But $0.93(n + m + k + 3) = 0.31(3(n + 
m + k) + 9)$ can be represented as $0.31(n' + m' + k' + 3)$, where $n', m'$, 
and $k'$ are also integers that satisfy the relation 
\begin{eqnarray} n' + m' + k' &=& 3(n + m + k + 2).\end{eqnarray} 
\noindent
\par Therefore, the baryons made of quark excited states cannot be observed
because their energy levels coincide with the energies of baryons made of 
quark ground states. Thus, we will not be able to know if the excited states 
of quarks exist. Actually, they may not exist for the potential well of quarks 
may allow just one level(ground state), depending on its depth and width.  
\vskip .2in
\noindent
20) HADRONIC MOLECULES
\vskip .15in
\par Having in mind what was considered in section 2 it is easy, then, to 
understand the formation of ``hadronic molecules''. The molecules are formed 
simply because we expect that in some cases there may exist a net binding 
molecular potential well between two hadrons or even among three or even four 
hadrons. Of course, this is only possible because the range of the superstrong 
interaction is smaller than that of the strong interaction. Thus, the existence
of such hadronic molecules is a direct proof of the existence of a 
superstrong(and repulsive) interaction. This binding may happen either among 
baryons or among mesons. This is quite in line with the existence of the 
deuteron, triton and alpha particle. 
\par As we know the energy of a system composed of N independent harmonic
oscillators is approximately equal to the sum of the energies of N harmonic
oscillators if they interact weakly and if we disregard rotation. That is how
we will calculate the energies of composite mesons.
\vskip .2in
\noindent
21) THE ENERGIES OF MESONS
\vskip .15in
a) $Q\bar{Q}$ Mesons
\vskip .3in
\par According to QCD a meson is a colorless state which transforms  under
$SU_{3}$ as
\begin{eqnarray} q^{in}q_{jn} &=& \bar{q}_{in}q_{jn}.\end{eqnarray}
\noindent
According to the theory presented above it is reasonable to admit that there is
also a harmonic effective potential energy in the interaction between a quark 
and an antiquark.  As we know they are also confined inside mesons, which, 
actually, supports the ideas above mentioned. Since we are assuming that 
rotation plays a minor role the energy levels are those of a simple harmonic 
oscillator because we reduce the two-body problem to one-body problem. In this 
fashion the energies of many mesons should be given by 
\begin{eqnarray} E_{n} &=& \hbar\nu(n + 1/2).\end{eqnarray}     
\noindent
Let us apply it to the mesons composed of $u\bar{d}$, $\bar{u}d$, $d\bar{d}$ 
and $u\bar{u}$. The ground state must correspond to the three pions $\pi^{+}$, 
$\pi^{-}$ and $\pi^{0}$. The splitting
comes from the electromagnetic interaction(and from the anharmonicity of the
potential energy). Let us choose for ${\hbar}\nu_{\pi}$ the value of 270MeV.
The values of the corresponding energies are shown in Table 17. The levels have
been labeled as $\pi_{n}$. There is a good agreement for almost all 
levels(error below 3\%). The levels with energies 405 and 675 MeV are 
forbidden, somehow. The experimental values were taken from reference 67. 
We may predict that there must exist mesons of the pion family with energies 
around 2565MeV(Table 17). Of course, we are considering that the quark and the 
antiquark may have their spins either parallel or antiparallel, so that, the 
levels will correspond to scalar or vector mesons.
\par For $K$ mesons let us take $\hbar\nu_{K}$=998MeV. The energy levels are
listed in Table 18. Tables 19, 20, 21, 22, 23 and 24 show the calculated and
experimental values for other $q\bar{q}$ mesons. 
\vskip .2in
\noindent
b) Composite Mesons
\par As we saw above there should exist hadronic molecules. As is well known
there are many non-$q\bar{q}$ candidates that have been proposed in the 
literature$^{67}$. The composition has been attributed to the gluon self
coupling. As we saw above it is actually due to primon bonding between two
different quarks. This happens when there is an effective molecular type
potential well between two mesons. In order to use the harmonic approximation
we will consider that the potential energy is quadratic in the small 
displacements of the individual oscillators. Of course we are assuming that
they interact weakly. This should be the case if the distance between the two
mesons is significant. In this case the energy of the system is just expressed
as 
\begin{eqnarray} E_{n_{1},n_{2}} &=& h\nu_{1}(n_{1} + 1/2) +
h\nu_{2}(n_{2} + 1/2)\end{eqnarray}     
\noindent
in which $h\nu_{i}$ is a constant that depends on the mass of each meson. 
It is approximately given by twice the lowest state mass of the given meson. 
\par Let us discuss some of these states. For example, the mass of the $\eta$ 
meson is 547.45MeV, which is very close to 405+135=540(MeV). Thus, we may 
suspect that it is the lowest state of the molecule 
${\pi}_{1}\biguplus{\pi}_{0}$, in which ${\pi}_{1}$ and $\pi_{0}$ are regular 
levels. The symbol $\biguplus$ between the two particles will indicate that it 
is a molecule. The other possible energy levels of such molecule are given by 
$E=810, 1080, 1350, 1620, 1890(MeV),... $ They are probably the particles 
$\omega(782)$, $\rho(770)$, $\phi(1020)$, $a_{2}(1318)$, $\omega(1394)$,
$f_{0}(1400)$, $f_{0}(1587)$, $\omega(1594)$, $\omega_{3}(1668)$, 
$\pi_{2}(1670)$, $\phi(1680)$, $\rho_{3}(1690)$, $\phi_{3}(1854)$. From now on 
we will omit the MeV since this unit will be used for all energies.  The meson 
${\eta}'(958)$ may be either the state $\pi_{3}$ or the lowest energy level of 
the composite system 
$(\pi_{2}\biguplus\pi_{0})\pi_{0}=(\rho(770)\biguplus\pi_{0})$ or 
$(\pi_{2}\biguplus\pi_{0})\pi_{0}=(\omega(782)\biguplus\pi_{0})$. The meson 
$f_{0}(980)$ may be either the $\pi_{3}$ state or the lowest state of the 
$K\biguplus\bar{K}$ molecule whose energy levels are aproximately 996, 1992,
2988, etc. This is quite in line with the decays of this meson into 
$\pi\pi$ and into $K\bar{K}$. The meson $a_{0}(980)$ is either the $\pi_{3}$ 
level or the lowest level of the molecule $K\biguplus\bar{K}$. It may also be 
the lowest level of the molecule $(\pi_{1}\biguplus\pi_{0})\biguplus\pi_{1}$. 
The decays of this meson into $K\bar{K}$ and into ${\eta}\pi$ agrees well with 
these considerations. The $b_{1}(1235)$ meson may be the $\pi_{4}$ state and 
also the lowest state of the molecules $\omega(782)\biguplus\pi_{1}$ and 
$\eta\biguplus\rho$ whose energies are around 1187 and 1222, respectively. The 
meson $a_{1}(1260)$ appears to be the molecule $\rho(770)\biguplus\pi_{1}$ 
with quite a lot of kinetic rotational energy(about 100). In this case our 
naive model is not good. The meson $f_{2}(1270)$ is either $\pi_{4}$ or the 
molecules $K\biguplus\bar{K}$ plus energy and $\eta\biguplus\eta$ plus energy. 
The energies of the lowest states of the molecules 
$\pi_{n}\biguplus\pi_{0}$ are listed in Table 25.
\par There are many strange mesons that are molecules. The first ones are the
mesons $K^{*}(892)^{\pm}$ and $K^{*}(896)^{0}$. They are probably the lowest 
state of the molecule $K\biguplus\pi_{1}$ which has an energy of about 
494+405=899. This agrees well with its decay into $K\pi$. The meson 
$K_{1}(1270)$ is the lowest state of the following molecules: $K\biguplus\rho$ 
which has an energy of approximately 494+770=1264; $K\biguplus\omega$ which 
has an energy of about 494+782=1276; $K^{*}(892)\biguplus\pi_{1}$. Doing in 
the same way with the other levels we obtain Table 26 below. The meson 
$\phi(1680)$ may also be the lowest level of the molecule 
$s\bar{s}_{0}\biguplus\pi_{2}$ whose energy is 1019+675=1694. The meson 
$f_{2}(2010)$ is clearly the lowest state of the molecule 
$({s\bar{s}}_{0})\biguplus({s\bar{s}}_{0})$ which has an energy of about 
1019+1019=2038.

\rule{0in}{.3in}

\rule{0in}{.3in}
\begin{center}
\begin{tabular}{||c||c|l||} \hline
& & \\
State & $E_{n}$(MeV) & Particles \\
& & \\
\hline\hline 
$\pi_{0}$ & 135 & $\pi^{\pm}(140)$, $\pi^{0}(135)$ \\ 
\hline 
$\pi_{1}$ & 405 & ? \\
\hline
$\pi_{2}$ & 675 & ? \\
\hline
$\pi_{3}$ & 945 & ${\eta}'(958)$, $f_{0}(980)$, $a_{0}(980)$ \\
\hline
$\pi_{4}$ & 1215 & $h_{1}(1170)$, $b_{1}(1235)$, $a_{1}(1260)$, \\
& & $f_{2}(1270)$, $f_{1}(1285)$, $\eta(1295)$, \\
& & $\pi(1300)$ \\
\hline
$\pi_{5}$ & 1485 & $\rho(1465)$, $f_{1}(1512)$, $f_{2}'(1525)$ \\
\hline
$\pi_{6}$ & 1755 & $\pi(1770)$, $\phi_{3}(1854)$ \\
\hline
$\pi_{7}$ & 2025 & $f_{2}(2011)$, $f_{4}(2049)$ \\
\hline
$\pi_{8}$ & 2295 & $f_{2}(2297)$, $f_{2}(2339)$ \\
\hline
$\pi_{9}$ & 2565 & to be found \\
\hline
$\pi_{10}$ & 2835 & to be found \\
\hline
... & ... & ... \\
\hline
\hline
\end{tabular}
\end{center}
\vskip .2in
\begin{center}
\parbox{6in}
{Table 17. The energies of states $q\bar{q}$ in which $q$ is $u$ or $d$. 
The states are named $\pi_{n}$. $E_{n}$ was calculated by the formula 
${\hbar}\nu(n + 0.5)$ taking for $\hbar\nu$ the values 270MeV. The error 
between experimental and calculated values is in general below 3\%, 
for every particle.}
\end{center}

\newpage

\begin{center}
\begin{tabular}{||c||c|l||} \hline
& & \\
State($K_{n}$) & $E_{n}$(MeV) & Particles \\
& & \\
\hline\hline 
$K_{0}$ & 494 & $K^{\pm}(494)$, $K^{0}(498)$ \\ 
\hline 
$K_{1}$ & 1482 & $f_{2}'(1525)$, $f_{1}(1512)$ \\
\hline
$K_{2}$ & 2470 & to be found \\
\hline
... & ... & ... \\
\hline\hline
\end{tabular}
\end{center}
\vskip .2in
\begin{center}
\parbox{5in}
{Table 18. The energy states of kaons. $E_{n}$ was calculated by the formula 
${\hbar}\nu(n + 0.5)$ taking for $\hbar\nu$ the value of 998MeV. The error 
between experimental and calculated values is in general below 2\%, for every 
particle.}
\end{center}

\rule{0in}{.2in}

\begin{center}
\begin{tabular}{||c||c|l||} \hline
& & \\
State & $E_{n}$(MeV) & Particles \\
& & \\
\hline\hline 
$s\bar{s}_{0}$ & 1019 & $\phi(1020)$ \\ 
\hline 
$s\bar{s}_{1}$ & 3057 & to be found \\
\hline
$s\bar{s}_{2}$ & 5095 & to be found \\
\hline
... & ... & ... \\
\hline\hline
\end{tabular}
\end{center}
\vskip .2in
\begin{center}
\parbox{4.5in}
{Table 19. The energy states of $s\bar{s}$ mesons. $E_{n}$ was calculated
according to the formula $E_{n}={\hbar}\nu(n+0.5)$ with $\hbar\nu=2038$MeV.}
\end{center}

\vskip .4in

\begin{center}
\begin{tabular}{||c||c|l||} \hline
& & \\
State & $E_{n}$(MeV) & Particles \\
& & \\
\hline\hline 
$D_{0}$ & 1864 & $D^{\pm}(1869)$, $D^{0}(1864)$ \\ 
\hline 
$D_{1}$ & 5595 & to be found \\
\hline
$D_{2}$ & 9325 & to be found \\
\hline
... & ... & ... \\
\hline\hline
\end{tabular}
\end{center}
\vskip .2in
\begin{center}
\parbox{4.5in}
{Table 20. The energy states of $D$ mesons. $E_{n}$ was calculated
according to the formula $E_{n} = \hbar\nu(n + 0.5)$ with 
$\hbar\nu = 3730$MeV.}\end{center}

\newpage

\begin{center}
\begin{tabular}{||c||c|l||} \hline
& & \\
State & $E_{n}$(MeV) & Particles \\
& & \\
\hline\hline 
$(D_{S})_{0}$ & 1969 & $D_{S}^{\pm}(1969)$ \\ 
\hline 
$(D_{S})_{1}$ & 5907 & to be found \\
\hline
$(D_{S})_{2}$ & 9845 & to be found \\
\hline
... & ... & ... \\
\hline\hline
\end{tabular}
\end{center}
\vskip .2in
\begin{center}
\parbox{4.5in}
{Table 21. The energy states of $D_{S}$ mesons. $E_{n}$ was calculated
according to the formula $E_{n} = \hbar\nu(n + 0.5)$ with 
$\hbar\nu = 3938$MeV.}\end{center}

\vskip .2in

\begin{center}
\begin{tabular}{||c||c|l||} \hline
& & \\
State & $E_{n}$(MeV) & Particles \\
& & \\
\hline\hline 
$B_{0}$ & 5279 & $B^{0}(5279)$, $B^{\pm}(5279)$ \\ 
\hline 
$B_{1}$ & 15837 & to be found \\
\hline
$B_{2}$ & 26395 & to be found \\
\hline
... & ... & ... \\
\hline\hline
\end{tabular}
\end{center}
\vskip .2in
\begin{center}
\parbox{4.5in}
{Table 22. The energy states of $B$ mesons. $E_{n}$ was calculated
according to the formula $E_{n} = \hbar\nu(n + 0.5)$ with 
$\hbar\nu = 10558$MeV.}\end{center}

\vskip .4in

\begin{center}
\begin{tabular}{||c||c|l||} \hline
& & \\
State & $E_{n}$(MeV) & Particles \\
& & \\
\hline\hline 
$c\bar{c}_{0}$ & 2978 & $\eta_{c}(2978)$ \\ 
\hline 
$c\bar{c}_{1}$ & 8967 & to be found \\
\hline
$c\bar{c}_{2}$ & 14890 & to be found \\
\hline
... & ... & ... \\
\hline\hline
\end{tabular}
\end{center}
\vskip .3in
\begin{center}
\parbox{4.5in}
{Table 23. The energy states of $c\bar{c}$ mesons. $E_{n}$ was calculated
according to the formula $E_{n} = \hbar\nu(n + 0.5)$ with 
$\hbar\nu = 5956$MeV.}\end{center}

\newpage

\begin{center}
\begin{tabular}{||c||c|l||} \hline
& & \\
State & $E_{n}$(MeV) & Particles \\
& & \\
\hline\hline 
$b\bar{b}_{0}$ & 9460 & $\Upsilon(9460)$ \\ 
\hline 
$b\bar{b}_{1}$ & 28380 & to be found \\
\hline
$b\bar{b}_{2}$ & 47300 & to be found \\
\hline
... & ... & ... \\
\hline\hline
\end{tabular}
\end{center}
\vskip .3in
\begin{center}
\parbox{4.5in}
{Table 24. The energy states of $b\bar{b}$ mesons. $E_{n}$ was calculated
according to the formula $E_{n} = \hbar\nu(n + 0.5)$ with 
$\hbar\nu = 18920$MeV.}\end{center}

\vskip .3in
\begin{center}
\begin{tabular}{||c||c|l||} \hline
& & \\
Particle & Possible Composition & Calculated Energy(MeV) \\
& & \\
\hline\hline 
$\eta(548)$ & $\pi_{1}\biguplus\pi_{0}$ & 405+135=540 \\ 
\hline 
$\rho(768)$, $\omega(782)$ & $\pi_{2}\biguplus\pi_{0}$ & 675+135=810 \\
\hline
$\phi(1019)$ & $\pi_{3}\biguplus\pi_{0}$ & 945+135=1080 \\
\hline
$a_{2}(1320)$, $f_{1}(1420)$ & $\pi_{4}\biguplus\pi_{0}$ & 1215+135=1350 \\
\hline
$\omega(1600)$, $\omega_{3}(1670)$ & $\pi_{5}\biguplus\pi_{0}$ & 
1485+135=1620 \\
\hline
$\phi_{3}(1850)$ & $\pi_{6}\biguplus\pi_{0}$ & 1755+135=1890 \\
\hline
? & $\pi_{7}\biguplus\pi_{0}$ & 2025+135=2160 \\
\hline
? & $\pi_{8}\biguplus\pi_{0}$ & 2295+135=2430 \\
... & ... & ... \\
\hline
\hline
\end{tabular}
\end{center}
\vskip .2in
\begin{center}
\parbox{5in}
{Table 25. The lowest energy states of mesons which are the molecules 
$\pi_{n}\biguplus\pi_{0}$. The errors are in general below 5\%, for every 
particle.}\end{center} 

\newpage

\begin{center}
\begin{tabular}{||l||l|l||} \hline
& & \\
Particle & Possible Composition & Calculated Energy(MeV) \\
& & \\
\hline\hline 
$K^{*}(892)^{\pm}$, $K^{*}(892)^{0}$ & $K\biguplus\pi_{1}$ & 494+405=899 \\ 
\hline 
$K_{1}(1270)$ & $K\biguplus\rho$ & 494+770=1264 \\ \cline{2-3}
              & $K\biguplus\omega$ & 494+782=1276 \\ \cline{2-3}
              & $K^{*}(892)\biguplus\pi_{1}$ & 892+405=1297 \\ 
\hline
$K_{1}(1400)$ & $K^{*}(892)\biguplus\pi_{1} + energy$ & 
892+405=1297 plus 103 of energy \\ \cline{2-3}
              & $K\biguplus\rho + energy$ & 
494+770=1264 plus 136 of energy \\ \cline{2-3}
              & $K\biguplus\omega + energy$ & 
494+782=1276 plus 124 of energy \\
\hline
$K^{*}(1410)$ & $K^{*}(892)\biguplus\pi_{1} + energy$ & 
892+405=1297 plus 113 of energy \\ \cline{2-3}
              & $K\biguplus\rho + energy$ & 
494+770=1264 plus 146 of energy \\ \cline{2-3}
              & $K\biguplus\pi_{3}$ & 494+945=1439 \\
\hline
$K^{*}_{0}(1430)$ & $K\biguplus\pi_{3}$ & 494+945=1439 \\
\hline
$K^{*}_{2}(1430)$ & $K\biguplus\pi_{3}$ & 494+945=1439 \\ \cline{2-3}
                  & $K^{*}(892)\biguplus\pi_{1} + energy$ & 
892+405=1297 plus 133 of energy \\ \cline{2-3}
              & $K\biguplus\rho + energy$ & 
494+770=1264 plus 166 of energy \\ \cline{2-3}
              & $K\biguplus\omega + energy$ & 
494+782=1276 plus 154 of energy \\
\hline
$K^{*}(1680)$ & $K\biguplus\pi_{4}$ & 494+1215=1709 \\ \cline{2-3}
              & $K^{*}(892)\biguplus(\pi_{2}\biguplus\pi_{0})$ & 
892+810=1702 \\ \cline{2-3} 
              & ($K_{0}{\biguplus}K_{0}){\biguplus}\rho$ & 988+770=1758 \\
\hline
$K_{2}(1770)$ & $K^{*}_{2}(1430)\biguplus\pi_{0} + energy$ & 
1430+135=1565 plus 205 of energy \\ 
\hline\hline
\end{tabular}
\end{center}
\vskip .2in
\begin{center}
\parbox{5in}
{Table 26. The energies of mesons which are molecules composed of kaons 
with $\pi_{n}$. The first column refers to the experimental data and
the second column to the possible composition.}\end{center}

\vskip .3in

\begin{center}
\begin{tabular}{||l||l|l||} \hline
& & \\
Particle & Possible Composition & Energy(MeV) \\
& & \\
\hline\hline 
$D^{*}(2007)^{0}$ & $D^{0}\biguplus\pi^{0}$ & 1864+135=1999 \\ 
\hline 
$D^{*}(2010)^{\pm}$ & $D^{0}\biguplus\pi^{+}$ & 1864+140=2004 \\ \cline{2-3}
                    & $D^{+}\biguplus\pi^{0}$ & 1869+135=2004 \\
\hline
$D_{1}(2420)^{0}$ & $D^{*}(2010)^{+}\biguplus{(\pi_{1})^{-}}$ & 
2010+405=2415 \\
\hline
$D_{2}^{*}(2460)^{0}$ & $D^{+}\biguplus{\pi_{1}\biguplus(\pi_{0})^{-}}$ & 
1869+(405+135)=2409 \\ \cline{2-3}
                    & $D^{*}(2010)^{+}\biguplus{(\pi_{1})^{-}}$ & 
2010+405=2415 \\
\hline
$D_{2}^{*}(2460)^{\pm}$ & $D^{0}\biguplus{\pi_{1}\biguplus(\pi_{0})^{+}}$ & 
1864+(405+135)=2409 \\
\hline
$D_{sJ}(2573)^{\pm}$ & $D^{0}{\biguplus}K^{+} + energy$ & 1864+494=2358 plus 
215 of energy \\
... & ... & ... \\
\hline\hline
\end{tabular}
\end{center}
\vskip .2in
\begin{center}
\parbox{4.5in}
{Table 27. Some possible molecules of $D$ mesons.}
\end{center}

\newpage

\begin{center}
\begin{tabular}{||c||c|l||} \hline
& & \\
Particle & Possible Composition & Energy(MeV) \\
& & \\
\hline\hline 
$D_{s}^{*\pm}$ & $D_{s}^{+}\biguplus\pi^{0}$ & 1969+135=2104 \\ 
\hline 
$D_{s1}(2536)^{\pm}$ & $D^{*}(2010)^{+}{\biguplus}K^{0}$ & 
2010+494=2504 \\ \cline{2-3}
                     & $D^{*}(2007)^{0}{\biguplus}K^{+}$ & 2010+498=2508 \\
\hline
... & ... & ... \\
\hline\hline
\end{tabular}
\end{center}
\vskip .2in
\begin{center}
\parbox{4.5in}
{Table 28. Some mesons that probably are molecules of $D_{s}$.}\end{center}

\vskip .3in

\begin{center}
\begin{tabular}{||c||c|l||} \hline
& & \\
Particle & Possible Composition & Energy(MeV) \\
& & \\
\hline\hline 
$J/{\Psi}(3097)$ & $\eta_{c}(1S)\biguplus\pi_{0}$ & 2978+135=3113 \\ 
\hline 
$\chi_{c0}(3415)$ & $\eta_{c}(1S)\biguplus\pi_{1}$ & 2978+405=3383 \\
\hline
$\chi_{c1}(3510)$, $\chi_{c2}(3556)$, $h_{c}(1P)(3526)$  & 
$(J/{\Psi(1S)})\biguplus\pi_{1}$ & 3097+405=3502 \\
\hline
$\Psi(2S)(3686)$ & $(J/{\Psi(1S)})\biguplus\eta(548)$ & 
3097+548=3645 \\ \cline{2-3}
                 & $\chi_{c0}(3415)\biguplus(\pi^{0}\biguplus\pi^{0})$ &
3415+(135+135)=3685 \\ \cline{2-3}
                 & $\chi_{c0}(3415)\biguplus(\pi^{+}\biguplus\pi^{-})$ &
3415+(135+135)=3685 \\ \cline{2-3}
                 & $\chi_{c1}(3510)\biguplus\pi_{0}$ & 
3510+135=3645 \\ \cline{2-3}
                 & $\chi_{c2}(3556)\biguplus\pi_{0}$ & 
3556+135=3691 \\
\hline
$\Psi(3770)$ & $D\biguplus\bar{D}$ & 3725 \\
\hline
$\Psi(4040)$ & $D^{*}(2007)\biguplus\bar{D^{*}}$ & 4014 \\
\hline
$\Psi(4160)$ & $\Psi(4040)\biguplus\pi_{0}$ & 4040+135=4175 \\
\hline
... & ... & ... \\
\hline\hline
\end{tabular}
\end{center}
\vskip .2in
\begin{center}
\parbox{5in}
{Table 29. Possible composition of many particles of the $c\bar{c}$ 
family.}\end{center}

\newpage

\par Many other mesons that have been found experimentally are composite
systems, composed of regular states. That is, they are hadronic molecules.
Tables 25, 26, 27, 28, and 29 display the possible composition of many
mesons in terms of the regular(basic) particles.\newline
\vskip .2in
\noindent
22) THE NUCLEAR POTENTIAL AND THE STABILITY OF THE DEUTERON
\vskip .15in
\par The most accurate empirical nuclear potential to date is the Paris 
potential$^{70}$. It has two expressions: one for the antisymmetric 
states(with respect to spin), allowed for two protons, two neutrons, as well 
as a proton and a neutron, and one for the symmetric states(with respect to 
spin), accessible only for the n-p system. In any case, when $S=0$, there is 
only a central potential between any two nucleons($V_{C0}$). According to our 
model this situation
corresponds to repulsion among the four ${p_{2}}'s$ of the outer layers and 
to attraction between the two inner shells. Although there is an equilibrium
position around 1fm, the potential well is not sufficiently deep to produce 
bound states(see Figs. 20a, 20b and 20c). Of course, it is easy to see that
the $p-p$, $n-n$ and $n-p$ interactions have about the same strength in this 
case. Why the only possibility is to have ${p_{2}}'s$ of different outer shells
repelling each other? Let us answer this question considering the diproton. 
In this system the ${p_{1}}'s$ of the outer shell will tend to stay away from 
each other. The $p_{1}$ of an outer shell is not attracted by a $p_{2}$ of the 
other outer shell simply because $p_{1}$ is bound with $p_{2}$ forming the u 
quark. Therefore, by means of the strong and superstrong interactions(that 
change $p_{1}$ into $p_{2}$ and vice versa) one would force the binding of
$p_{2}$ to $p_{2}$ which does not happen. That is, the situation shown in Fig. 
21 does not take place. 
\par Let us now see how the total spin turns out to be zero. Let us consider, 
for example, the diproton(Fig. 20a). As we saw in section 1 primons  have 
their net spins parallel in order to form a quark.  Therefore, the primons 
$p_{1}$(of an inner shell) and $p_{3}$(of the other inner shell) tend to be
parallel, and the same also happens in each quark. Thus, there are only two 
possible arrangements(of course, in the other one all spins are inverted). 
\par For $S=1$(symmetric states), the situation is more 
complicated. The Paris group has found that the potential has four different 
terms and is described by $^{70,71)}$
\begin{eqnarray} V(r) &=& V_{C1}(r) + V_{T}(r)\Omega_{T} + 
V_{S0}(r)\Omega_{S0} + V_{S02}(r)\Omega_{S02}\end{eqnarray}
\noindent
where 
\begin{eqnarray*} \Omega_{T} &=& 
3\frac{(\vec{\sigma_{1}}.\vec{r})(\vec{\sigma_{2}}.\vec{r})}{r^2} 
- \vec{\sigma_{1}}.\vec{\sigma_{2}},\end{eqnarray*}
\begin{eqnarray*} \hbar\Omega_{S0} &=& 
(\vec{\sigma_{1}} + \vec{\sigma_{2}}).\vec{L},\end{eqnarray*}
\begin{eqnarray*} \hbar^{2}\Omega_{S02} &=& 
(\vec{\sigma_{1}}.\vec{L})(\vec{\sigma_{2}}.\vec{L}) +
(\vec{\sigma_{2}}.\vec{L})(\vec{\sigma_{1}}.\vec{L}).\end{eqnarray*} 
\noindent
In these equations $\vec{L}$ is the total orbital angular momentum of the 
nucleons, ${\hbar/2}\sigma$ is the spin operator of each nucleon, the 
subscripts 1 and 2 in $\sigma$ refer to the two nucleons, and the subscript
1 in the first term refers to $S=1$. The first three terms  are responsible
for binding the deuteron. The term $V_{T}(r)$ is associated also with the large
electric quadrupole moment of the deuteron$^{71}$. As we saw in section 2
the strong binding between p and n occurs in the primon configuration described
below(Fig. 22). Also, as was discussed in that section this arrangement of
primons produces a strong electric quadrupole moment. We clearly see that the
spatial part of the wavefunction must be antisymmetric. Of course, the spin
wavefunctions $|S, S_{z}>$($|1,-1>$, $|1,0>$ and $|1,1>$) are symmetric under
particle exchange. It is expected that the inner shells will play a minor role
in this case. That is why $V_{C1}(r)$ is very shallow and most of the binding
is due to the $V_{T}(r)$ term. It is interesting to notice that the strong
binding between $p{1}$ and $p_{3}$ generates a quasi quark(charm in this 
case). 
\par Let us now see why the configuration shown in Fig. 22 has $S=1$. The
primons $p_{1}$ and $p_{3}$(of the outer shells) will tend to have parallel
spins and the $p_{2}$(of the inner shell of p) which is bound to this $p_{1}$ 
will also have its spin parallel. The same happens to the spin of the $p_{2}$
of the inner shell of n which is bound to $p_{3}$. And since each nucleon has 
$S=1/2$, the total spin is one. It is worth mentioning that it also explains 
the stability of the deuteron 
which is not explained at all with pointlike quarks. Actually, without 
considering the above model, since the deuteron has three d quarks we would 
expect it to be a very unstable system that would decay very fast. Fig. 23 
displays the behavior with r of the different terms of the Paris potential.
\par In the light of what was discussed above we can understand the large
decay constant of triton. We know that the spins of the two neutrons cancel
each other so that the spin of triton comes from the proton. The configuration
of primons(and quarks) of the system is describe below in Fig. 24. There is a
net binding between $p_{1}$ and the two ${p_{3}}'s$. Actually, it must be an
alternate binding between $p_{1}$ and each $p_{3}$. This binding makes $p_{3}$
more stable so that instead  of decaying in 920s it decays in about 
$3.87{\times}10^{8}$s. The addition of another $p_{1}$ would make the system
completely stable. Therefore, the alfa particle primon configuration should be
given by Fig. 25 and {\bf{ it is a planar configuration and not piramidal}}.
Due to the attraction of the four inner shells the system is
very tightly bound and, of course, very stable. The eight ${p_{2}}'s$ of the
outer layers will tend to stay away from each other. We infer, thus, that
the system has the following electric charge distribution: the center(the
region where the two ${p_{1}}'s$ and the two ${p_{3}}'s$ are) has a net charge
of about $2{\times}(+5/6) - 2{\times}(-1/6)=+4/3$; a middle 
region(corresponding to the position of the four inner shells) with a charge
of about $4{\times}(+1/2)=+2$; and an outer region(corresponding to the 
positions of the eight ${p_{2}}'s$) with a charge of $8{\times}(-1/6)=-4/3$.
The system, of course, as we see, has no quadrupole moment. It is interesting
to notice that an alpha particle is not, therefore, a system of two deuterons.
In this way we explain that the saturation of the nuclear force is quite 
similar to the saturation of chemical bonds. We can also understand the
reason behind the tensorial character of the nuclear force. It arises simply
due to the spatial arrangement of primons. \newline
\vskip .2in
\noindent
23) THE ABSENCE OF NUCLIDES WITH 5 AND THE INSTABILITY OF ${4}_Be^{8}$
\vskip .2in
\par It is well known that there is no nuclide with A=5. It simply does not
form, even for a brief time. Why is it so? Taking a look at the primon
configuration of the alpha particle we can understand why. As we saw above
the binding happens in the middle among the four primons: the two pairs of
$p_{1}-p_{3}$. Besides it is a planar structure. Thus, there is no room for
another nucleon, that is, there is no bond left. We have a strong binding
if we put a neutron on one side and a proton on the other side because in 
this case there will be another bond $p_{1}-p{3}$. That is why ${3}_Li^{6}$ is
stable. 
\par We can also see that it is impossible to bind to alpha particles
since there is no bond left in any of them. Actually, the bonding could occur
only by means of the $p_{2}$'s of the outer layers, but there is no bonding
between equal primons and, therefore, the binding does not take place. 
We know that ${4}_Be^{8}$ is formed only for an extremely brief time(about
$10^{-23}$s) and breaks up into two alpha particles.\newline
\vskip .2in
\noindent
25) THE DESIRED UNITY
\vskip .15in
\par The energy levels of hadrons and the hadronic molecules shows us clearly
that Nature repeats itself in different scales. That is, the so-called 
autosimilarity is an intrinsic and very important property of Nature. Thus,
a quark, which has a variable size is the smallest kind of ``medium''.  It has
a similarity with the atomic nucleus which is also a medium. Both are part of a
generalized state of matter called the structured state. This also shows that
Nature has an intrinsic fractality in space and in time.
\par In the same way a nucleon is quite similar to an atom and to a primon.
They are units. When we consider Nature in larger scales we expect that it
will have other units and other manifestations of the structured state. 
While the Universe expands its units, the galaxies, also expand and 
aggregate. In the two tables below we observe the great 
design of Nature and its most important general principle: 
{\bf{Space is filled with different units. When it is small it is filled with 
small units and when it is large it is filled with large units. The 
aggregation of the different units form the different types of structured 
states by means of the different kinds of fundamental forces.}} We clearly see 
that there is a close association between each unit, the size of the Universe 
and each fundamental force(except the weak force). {\bf{THE GREAT DESIGN OF 
NATURE IS THEREFORE VERY SIMPLE.}}

\rule{0in}{.5in}

\begin{center}
\begin{tabular}{c c c c c c c} \hline\hline\\ 
& primon & & quark & & nucleon & \\
\\
& nucleon & & nucleus & & atom & \\
\\
& atom & & gas & & galaxy & \\
& & & liquid & & & \\
& & & solid & & & \\
\\
& galaxy & & galactic liquid & & Universe \\
\\
\hline\hline\\
\end{tabular}
\end{center}
\vskip .2in

\begin{center}
\parbox{4in}
{Table 30. The units of Nature(left and right columns) and the different 
kinds of structered states(middle column). The table is arranged in such a 
way as to show the stepwise aggregation of matter in larger units.}
\end{center}

\newpage

\begin{center}
\begin{tabular}{c c c} 
\hline\hline\\ 
superstrong force & superstrong force &  \\
& strong force & strong force \\
\\
strong force & strong force & \\
& electromagnetic force & electromagnetic force \\
\\
electromagnetic force & electromagnetic force & \\
& gravitational force & gravitational force \\
\\
gravitational force & gravitational force & \\
& superweak force & superweak force \\
\\
\hline\hline\\
\end{tabular}
\end{center}
\vskip .2in

\begin{center}
\parbox{5in}
{Table 31. Five of the six fundamental forces of nature. Each force 
appears twice and is linked to another force by means of a structured 
state. Compare with Table 28 above. The weak force does not appear 
because it is not linked to the formation of any structered state.
It is rather related to instability and is quite different from the 
other forces. It violates parity and has no static potential associated
to it.}
\end{center}

\vskip .3in

\noindent
REFERENCES
\newline
\noindent
1. M.E. de Souza, in {\it{Proceedings of the XII Brazilian National Meeting
of the Physics of Particles and Fields}}, Caxambu, Minas Gerais, Brazil,
September 18-22, 1991.
\newline
\noindent
2. M.E. de Souza, {\it{IX Meeting of Physicists of the North and Northeast}}, 
Macei\'{o}, Alagoas, Brazil, November 07 and 08, 1991.
\newline
\noindent
3. M.E. de Souza, {\it{13th Interantional Conference on General Relativity and
Gravitation}}, Huerta Grande, Cordoba, Argentina, June 28-July 4, 1992.
\newline
\noindent
4. M.E. de Souza, in {\it{Proceedings of the XIII Brazilian National Meeting
of the Physics of Particles and Fields}}, Caxambu, Minas Gerais, Brazil,
September 16-20, 1992. 
\newline
\noindent
5. M.E. de Souza, {\it{X Meeting of Physicists of the North/Northeast}}, 
Recife, Pernambuco, Brazil, December 2-4, 1992.
\newline
\noindent
6. M.E. de Souza, in {\it{Proceedings of the XIV Brazilian National Meeting
of the Physics of Particles and Fields}}, Caxambu, Minas Gerais, Brazil,
September 29-October 3, 1993.
\newline 
\noindent
7. M.E. de Souza, {\it{XI Meeting of Physicists of the North/Northeast}}, 
Jo\~{a}o Pessoa, Para\'{\i}ba, Brazil, November 17-19, 1993.
\newline
\noindent
8. M.E. de Souza, in {\it{The Six Fundamental Forces of Nature}}, Universidade
Federal de Sergipe, S\~{a}o Crist\'{o}v\~{a}o, Sergipe, Brazil, February 1994.
\newline
\noindent
9. M.E. de Souza, {\it{International Symposium Physics Doesn't Stop: Recent 
Developments in Phenomenology}}, University of Wisconsin, Madison(Wisconsin), 
USA, April 11-13, 1994.
\newline
\noindent
10. M.E. de Souza, in {\it{Proceedings of the XV Brazilian National Meeting
of the Physics of Particles and Fields}}, Angra dos Reis, Rio de Janeiro, 
Brazil, October 4-8, 1994.
\newline
\noindent
11. M.E. de Souza, {\it{XVI Brazilian National Meeting of the Physics of 
Particles and Fields}}, Caxambu, Minas Gerais, Brazil, October 24-28, 1995.
\newline
\noindent
12. M.E. de Souza, {\it{XVII Brazilian National Meeting of the Physics of 
Particles and Fields}}, Serra Negra, Minas Gerais, Brazil, October 24-28, 1996.
\newline
\noindent
13. E.J. Eichten, K.D. Lane, and M.E. Peskin, {\it{Phys. Rev. Lett.}} 
{\bf{50}}, 811 (1983).
\newline
\noindent
14. K. Hagiwara, S. Komamiya, and D. Zeppenfeld, {\it{Z. Phys.}} {\bf{C29}}, 
115 (1985).
\newline
\noindent
15. N. Cabibbo, L. Maiani, and Y. Srivastava, {\it{Phys. Lett.}} {\bf{139}},
459 (1984).
\newline
\noindent
16. H. Fritzsch, in {\it{Proceedings of the twenty-second Course of the 
International School of Subnuclear Physics, 1984}}, ed. by A. Zichichi
(Plenum Press, New York, 1988).
\newline
\noindent
17. G. 'tHooft, in {\it{Recent Developments in Gauge Theories}}, eds. G. 
'tHooft et al., Plenum Press, New York, 1980.
\newline
\noindent
18. E. E. Chambers and R. Hofstadter, {\it{Phys. Rev.}} {\bf{103}}, 
1454 (1956).
\newline
\noindent
19. C. Kittel, W. D. Knight and M. A. Ruderman, in {\it{Mechanics, Berkeley
Physics Course}}, Vol. 1, pg. 451, McGraw-Hill Book Company, New York(1965).
\newline
\noindent
20. H. G. Kolsky, T. E. Phipps, Jr., N. F. Tamsey, and H. B. Silsbee, 
{\it{Phys. Rev.}} {\bf{87}}, 395(1952); recalculated by E. P. Auffray, 
{\it{Phys. Rev. Letters}} {\bf{6}}, 120 (1961).
\newline
\noindent
21. P. Renton, in {\it{Electroweak Interactions}}, p. 315, Cambridge 
University Press, Cambridge(1990).
\newline
\noindent
22. Particle Data Group, {\it{Review of Particle Properties}}, Phys. Rev. D, 
{\bf{54}}, Part II, No. 1 (1996).
\newline
\noindent
23. P. Amaudruz et al. (CERN NMC), {\it{Phys. Rev. Lett.}} {\bf{66}}, 
2712(1991); N. Arneodo et al. {\it{Phys. Rev. D}} {\bf{50}}, R1 (1994).
\newline
\noindent
24. A. Baldi et al. {\it{Phys. Lett. B}} {\bf{332}}, 244(1994).
\newline
\noindent
25. E.A. Hawker et al. (Fermilab E866/NuSea Collaboration), 
{\it{Phys.Rev.Lett.}} {\bf{80}}, 3715(1998).
\newline
\noindent
26. G. Arnison, {\it{et al., Phys. Lett.}} {\bf{136B}}, 294(1984).
\newline
\noindent
27. P. Bagnaia, {\it{et al., Phys. Lett.}} {\bf{138B}}, 430(1984).
\newline
\noindent
28. S.L. Shapiro and S.A. Teukolsky, in {\it{Black Holes, White Dwarfs, and 
Neutron Stars}}, John Wiley \& Sons, New York(1983).
\newline
\noindent
29. J.D. Walecka, {\it{Annals of Phys.}} {\bf{83}}, 491(1974).
\newline
\noindent
30. Yu.A.Baurov and A.V.Kopajev, hep-ph/9701369.
\newline
\noindent
31. M.J.McCaughrean and M.-M. Mac Low, astro-ph/9611058.
\newline
\noindent
32. S. Edwards, T.P. Ray, and R. Mundt, in {\it{Protostars and Planets III}},
eds. E.H. Levy and J.I.Lunine (Tucson: University of Arizona Press), p. 567,
1993.
\newline
\noindent
33. G. Mellema and A. Frank, astro-ph/9710255.
\newline
\noindent
34. C.F. Prosser, J.R. Stauffer, L. Hartmann, D.R. Soderblom, B.F. Jones, M.W.
Werner, and M.J. McCaughrean, {\it{Ap.J.}} {\bf{421}}, 517(1994).
\newline
\noindent
35. I. M\'{a}rquez, F. Durret, and P. Petitjean, astro-ph/9810012.
\newline
36. A. Yahil, astro-ph/9803052.
\newline
\noindent
37. P.A. Shaver, L.M. Hook, C.A. Jackson, J.V. Wall, and K.I. Kellermann,
astro-ph9801211.
\newline
\noindent
38. M. Pettini, C.C. Steidel, M. Dickinson, M. Kellogg, M. Giavalisco, and
K.L. Adelberger, in {\it{The Ultraviolet Universe at Low and High Redshift:
Probing the Progress of Galaxy Evolution}} (Eds. W.H. Waller et al.) AIP,
1997.
\newline
\noindent
39. H. di Nella et G. Paturel, {\it{C.R.Acad.Sci. Paris}}, {\bf{t.319}}, 
S\'{e}rie II, p. 57-62, 1994.
\newline
\noindent
40. G. K. Miley and A.P. Hartsuijker, {\it{A \& AS}} {\bf{291}}, 29 (1978).
\newline
\noindent
41. T.S. Slatler, I.R. King, P. Krane, and R.I. Jedrzejwski, astro-ph/9810264.
\newline
\noindent
42. E.J. M. Colbert, A. S. Wilson, and J. Bland-Hawthorn, {\it{The Radio
Emission from the Ultra-Luminous Far-Infrared Galaxy NGC 6240}}, preprint 
network astro-ph; astro-ph/9405046, May 1994. 
\newline
\noindent
43. W. J. Kaufmann,III, in {\it{Galaxies and Quasars}}(W.H.Freeman and Company,
San Francisco, 1979).
\newline
\noindent
44. P.A. Shaver, L.M. Hook, C.A. Jackson, J.V. Wall, and K.I. Kellermann,
astro-ph9801211.
\newline
\noindent
45. S.C. Chapman, G.A.H. Walker and S.L. Morris, astro-ph/9810250.
\newline
\noindent
46. H. Falcke and P.L. Biermann, astro-ph/9810226.
\newline
\noindent
47. T. M. Heckman, K. C. Chambers and M. Postman, {\it{Ap.J.}}, 
{\bf{391}}, 39(1992).
\newline
\noindent
48. S. Baum and T. M. Heckman, {\it{Astrophys. J}} {\bf{336}}, 702(1989).
\newline
\noindent
49. N. Jackson and I. Browne, {\it{Nature}}, {\bf{343}}, 43(1990).
\newline
\noindent
50. A. Lawrence, {\it{Mon. Not. R. Astr. Soc.}}, 1992, in press.
\newline
\noindent
51. R. W. Goodrich and M. H. Cohen, {\it{Astrophys. J.}} {\it{391}}, 623(1992).
\newline
\noindent
52. Y. Sofue, {\it{Astro. Lett. Comm.}} {\bf{28}}, 1(1990).
\newline
\noindent
53. N. Nakai, M. Hayashi, T. Handa, Y. Sofue, T. Hasegawa and M. Sasaki, 
{\it{Pub. Astr. Soc. Japan}} {\bf{39}}, 685(1987).
\newline
\noindent
54. K. L. Visnovsky, C. D. Impey, C. B. Foltz, P. C. Hewett, R. J. Weymann and
S. L. Morris, {\it{Astrophys. J.}} {\bf{391}}, 560(1992).
\newline
\noindent
55. T. A. Boroson and S. E. Persson, {\it{Astrophys. J.}} {\bf{293}}, 
120(1985).
\newline
\noindent
56. M. V. Berry, in {\it{Principles of Cosmology and Gravitation}}(Adam Hilger,
Bristol, 1991).
\newline
\noindent
57. O. Y. Gnedin, J. Goodman, and Z. Frei, Princeton University Observatory
preprint {\it{Measuring Spiral Arm Torques: Results for M100}}, preprint 
network Astro-ph/9501112.
\newline
\noindent
58. E. Fischbach, in {\it{Proceedings of the NATO Advanced Study Institute on
Gravitational Measurements, Fundamental Metrology and Constants, 1987}},
ed. by V. de Sabbata and V. N. Melnikov(D. Reidel Publishing Company, Dordrecht,
Holland, 1988).
\newline
\noindent
59. E. G. Adelberger, B. R. Heckel, C. W. Stubbs and W. F. Rogers, {\it{Annu. 
Rev. Nucl. Part. Sci.}} {\bf{41}}, 269(1991).
\newline
\noindent
60. H. El-Ad, T. Piran, and L.N. da Costa, {\it{Mon. Not. R. Astron. Soc.}}
{\bf{287}}, 790 (1997).
\newline
\noindent
61. M.E. de Souza, in {\it{The Six Fundamental Forces of Nature}}, p. 6,
Universidade Federal de Sergipe, 1994.
\newline
\noindent
62. J. E. Lennard-Jones, {\it{Proc. Roy. Soc.}} {\bf{A106}}, 463(1924).
\newline
\noindent
63. D. L. Goodstein, in {\it{States of Matter}}(Prentice-Hall, Englewood Cliffs,
1975).
\newline
\noindent
64. P. Ring and P. Schuck, in {\it{The Nuclear Many-Body Problem}}, 
Springer-Verlag, New York(1980).
\newline
\noindent
65. A. Hosaka, H. Toki, and M. Tokayama, {\it{Mod. Phys. Lett.}} {\bf{13}},
1699 (1998).
\newline
\noindent
66. L. Pauling and E. B. Wilson Jr., {\it{Introduction to Quantum Mechanics}},
McGraw-Hill, New York(1935).
\newline
\noindent
67. Particle Data Group, {\it{Review of Particle Properties}}, Phys. Rev. D, 
{\bf{45}}, Part II, No. 11 (1992).
\newline
\noindent
68. Lichtenberg, D. B., {\it{Unitary Symmetry and Elementary Particles}},  
Academic Press, New York, N.Y(1970).
\newline
\noindent
69. D. H. Perkins, {\it{Introduction to High Energy Physics}}(Addison-Wesley
Publishing Company, Inc., Menlo Park, California) 1987.
\newline
\noindent
70. M. Lacombe, {\it{et al., Phys. Rev.}} {\bf{C21}}, 861 (1980).
\newline
\noindent
71. W. N. Cottingham and D. A. Greenwood, {\it{An Introduction to Nuclear
Physics}}, Cambridge University Press, Cambridge(1992).
\newpage
\centerline{BRIEF VITA}
\vskip .2in
\noindent
Bachelor in Physics: Universidade Federal de Pernambuco, Recife, Pernambuco,
Brazil.\newline
Master Degree(in Physics): Universidade Federal de Pernambuco, Recife, 
Pernambuco, Brazil.\newline
Doctor of Philosophy(in Physics): University of Illinois at Chicago, Chicago,
Illinois, USA.

\newpage

\large
\noindent
Complete address:
\vskip .15in
\noindent
M\'{a}rio Everaldo de Souza,
\newline     
\noindent
Universidade Federal de Sergipe
\newline
\noindent
Departamento de F\'{\i}sica - CCET,
\newline
\noindent
49000 Aracaju, Sergipe, Brazil
\newline
\noindent
Phone nos. (55)(79)212-6630, (55)(79)212-6634
\newline
\noindent
e-mail mdesouza@sergipe.ufs.br

\newpage

\pagestyle{empty}

\bf
%Figure captions:
\parbox{5.5in}
{Fig. 1. Configuration of the spins of primons and quarks in the proton. For
simplifying the visualization let us consider that all primons are in the
XY plane and that the +Z direction points upwards perpendicularly to the
plane. The spin of each primon is $\hbar/2$. The angle between the spins of
the two primons of a quark is always equal to $2\pi/3$ and the angle of each
primon spin with the Z axis(that is, with the baryon spin) is always equal to
$\pi/3$. Hence, each primon contributes with $\hbar/4$ to the spin of its
corresponding quark. The XY components of the spins of the two primons of a
particular quark cancel out so that the total spin of a quark is $\hbar/2$.
Therefore, primons have to move in such a way as to maintain such angles. Due
to the exchange of vector bosons the primon spins become inverted eventually
but the overall baryon spin does not change. With respect to spin, a baryon is,
of course, a highly ordered system. In the specific case shown above(for L=0)
the total spin points in the +Z direction. For other values of L the total
spin will be equal to J= L + S. It is important to notice that the spins of
each pair of primons may precess about the direction of each quark spin.
In the above figure $\theta=\pi/3$.}

\vskip .3in

\parbox{5.5in}
{Fig. 2. Distribution of electric charge (a) in the proton and (b) in the 
neutron. The area under the curve is equal to each particle's charge.}

\vskip .3in
\parbox{6in}
{Fig. 3. The arrangement of primons in the proton. The two mean radii are
approximately to scale, according to the two peaks seen in Fig. 2a. The 
small white circle is $p_{1}$, the large white circle is $p_{2}$, and the
black circle is $p_{3}$. The sizes of these circles do not mean anything,
since primons are supposed to be pointlike. The supercolors are $\alpha$,
$\beta$ and $\gamma$. The primons in each layer have different supercolors. 
The thick black lines mean the strong bonds that link primons, that is, they
mean the quarks, and the thin black lines mean the weak bonds between any two
primons. Due to the exchange of gluons the weak bonds change all the time. The
large circles represent the mean radii of the two shells.}

\vskip .3in
\parbox{6in}
{Fig. 4. The arrangement of primons in the neutron. The two mean radii are
approximately to scale, according to the two peaks seen in Fig. 2b. The 
small white circle is $p_{1}$, the large white circle is $p_{2}$, and the
black circle is $p_{3}$. The sizes of these circles do not mean anything,
since primons are supposed to be pointlike. The supercolors are $\alpha$,
$\beta$ and $\gamma$. The primons in each layer have different supercolors.}

\vskip .3in
\parbox{5.5in}
{Fig. 5. The arrangement of primons in the deuteron corresponding to $S=1$. 
The exchange of the charged pions happen between $p_{1}$ and $p_{3}$. Probably
one nucleon disturbs completely the spherical character of the shells. The
supercolors have been omitted and just a few weak bonds are shown.}

\vskip .3in
\parbox{5.5in}
{Figs. 6a and 6b. Experimental data on the structure functions of the nucleons, as 
measured in deep inelastic electron scattering at the Stanford Linear 
Accelerator. }

\vskip .3in
\parbox{5.5in}
{Fig. 7. A possible arrangement of primons in $\Delta^{++}$. This arrangement
generates the charge distribution shown in Fig. 9.}

\vskip .3in
\parbox{5.5in}
{Fig. 8. The other possible configuration of primons in $\Delta^{++}$. It
produces the charge distribution shown in Fig. 10.}

\vskip .3in
\parbox{4.5in}
{Fig. 9. Charge distribution of $\Delta^{++}$ in the configuration 
$(p_{1}p_{2}p_{2})^{1}(p_{2}p_{1}p_{1})^{2}$.}

\vskip .3in
\parbox{5.5in}
{Fig. 10. Charge distribution of $\Delta^{++}$ which corresponds to the
configuration $(p_{1}p_{2}p_{1})^{1}(p_{2}p_{1}p_{2})^2$.}

\vskip .3in
\parbox{4.5in}
{Fig. 11. Charge distribution of $\Delta^{-}$ in the configuration
$(p_{2}p_{3}p_{2})^{1}(p_{3}p_{2}p_{3})^2$.}

\vskip .3in
\parbox{4.5in}
{Fig. 12. Charge distribution of $\Sigma_{0}$ and $\Lambda$ in the 
configuration $(p_{2}p_{3}p_{4})^{1}(p_{1}p_{2}p_{2})^2$.}

\vskip .3in
\parbox{5.5in}
{Fig. 13. Distribution of charge in $\Sigma_{c}^{++}$, which has the
primon arrangement $(p_{1}p_{2}p_{3})^{1}(p_{2}p_{1}p_{1})^{2}$.}

\vskip .3in
\parbox{5.5in}
{Fig. 14. The effective potential between two primons which results from the
actions of the strong and superstrong interactions. The ranges of the
superstrong and strong forces are $(\mu_{1})^{-1}$ and $(\mu_{2})^{-1}$,
respectively.}

\vskip .3in
\parbox{5.5in}
{Fig. 15. Graphical solution of Eq. 06.  The solution $r=r_{0}$ exists 
only if $\alpha>\beta$.}
\vskip .3in

\parbox{5.5in}
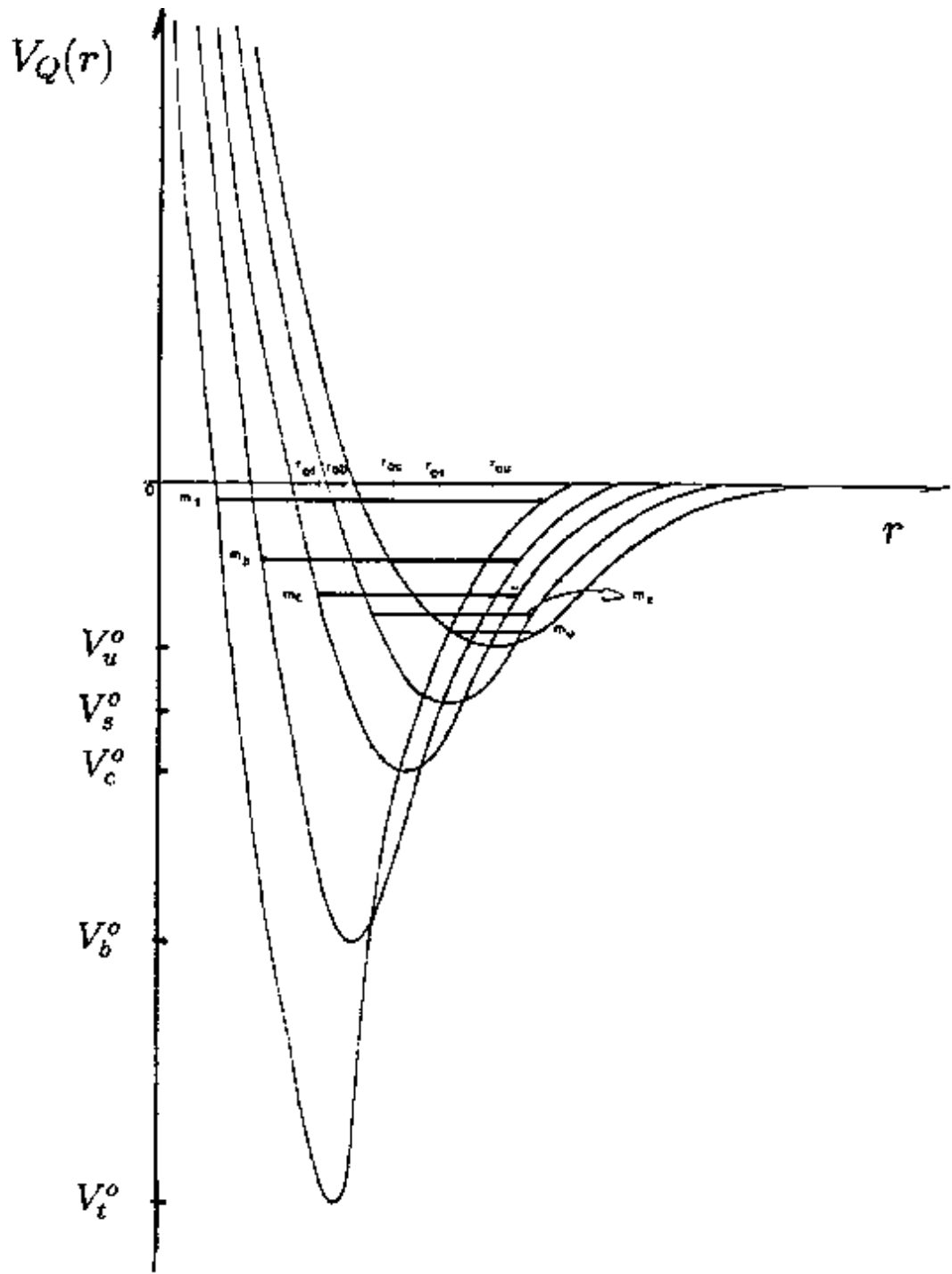
{Fig. 16. Potentials of quarks u, s, c, b, and t. The ground state levels are
given by $m_{u}$, $m_{s}$, $m_{c}$, $m_{b}$, and $m_{t}$. The larger the 
ground state is, the narrower and deeper should be the corresponding potential
well.  The level of the $d$ quark is not shown. The equilibrium distance
between any two primons is $r_{o}$.}
\vskip .3in

\parbox{5.5in}
{Fig. 17. The usual QCD potential $V = 
- \frac{4}{3}\frac{\alpha_{s}}{r} + {\beta}r(\beta=1$GeV$fm^{-1}$) and  a 
Yukawa potential of the type
$V = - \frac{(g_{s}^{Q})^{2}e^{-\mu_{s}r}}{r}$ with $\mu_{s}=1$ $fm^{-1}$, and
$(g_{s}^{Q})^{2}= \alpha_{s}$.}
\vskip .3in

\parbox{6in}
{Fig. 18. The data points were found at the CERN $p\bar{p}$ collider, at
$q^{2}{\simeq}2000$ GeV$^{2}$. One clearly sees that the first point at the
left is off the straight line and shows some sort of saturation, indicating
a Yukawa type of potential.} 
\vskip .4in

\parbox{5.5in}
{Fig. 19. The 20-plet of $SU_{4}$ with an $SU_{3}$ decuplet.}
\vskip .4in

\parbox{6in}
{Figs. 20. The probable arrangement of primons in the nucleons when they 
form the systems p-p(a), n-n(b) and n-p(c) for $S=0$. The small and large 
white circles are $p_{1}$ and $p_{2}$, respectively, and the black circle is
$p_{3}$. The arrows indicate the spin directions. One can easily see that 
in these arrangements the attraction is about the same in the three cases.
Just some weak bonds are shown. Each arrow means $\hbar/4$ as discussed in the
first section. this also holds for Figs. 22, 23 and 24.}
\vskip .4in

\parbox{5.5in}
{Fig. 21. One of the arrangements of primons in the diproton that does not 
occur because if $p_{1}$ and $p_{2}$ are exchanged(the broken line) it would 
force the binding of $p_{2}$ with $p_{2}$ and of $p_{1}$ with $p_{1}$. Only 
some weak bonds are shown.}
\vskip .4in

\parbox{6in}
{Fig. 22. The probable configuration of primons in the deuteron. The arrows
indicate the spin directions. The total spin is one. Most of the binding
is due to the interaction between $p_{1}$ and $p_{3}$ of the outer layers.
The interaction causes the change of $p_{3}$ into $p_{1}$ and vice versa, so
that the rest of the nucleons remains essentially unaltered. In this way the
decay of $p_{3}$ is avoided and the system is highly stable. We expect that
the shells are not spherical anymore. Just a few weak bonds are shown.}
\vskip .4in

\parbox{5.5in}
{Fig. 23. The most important terms of the Paris nuclear potential.} 
\vskip .4in

\parbox{5.5in}
{Fig. 24. The configuration of primons(and quarks) in the triton. The arros
indicate the spin directions. The total spin is 1/2. Most of the binding
happens between $p_{1}$ and the two ${p_{3}}'s$ and slows down the decay of
$p_{3}$. Probably all shells are no longer spherical, so that the above
circles may be substituted by other more realistic curves. Of course, it is 
not an easy task since we are dealing with a very complex system. The figure
shows just a few weak bonds.} 
\vskip .4in

\parbox{5.5in}
{Fig. 25. The arrangement of primons in the alpha particle. Probably the 
$p-p$ bond is perpendicular to the $n-n$ bond. The arrows indicate the
spin directions. The total spin is zero. We easily see that the system has no
quadrupole moment. We expect that there should exist attraction among the
inner shells. Probably all shells are no longer spherical, so that the above
circles may be substituted by other more realistic curves. Of course, it is 
not an easy task since we are dealing with a very complex system. Only four
weak bonds are shown.}

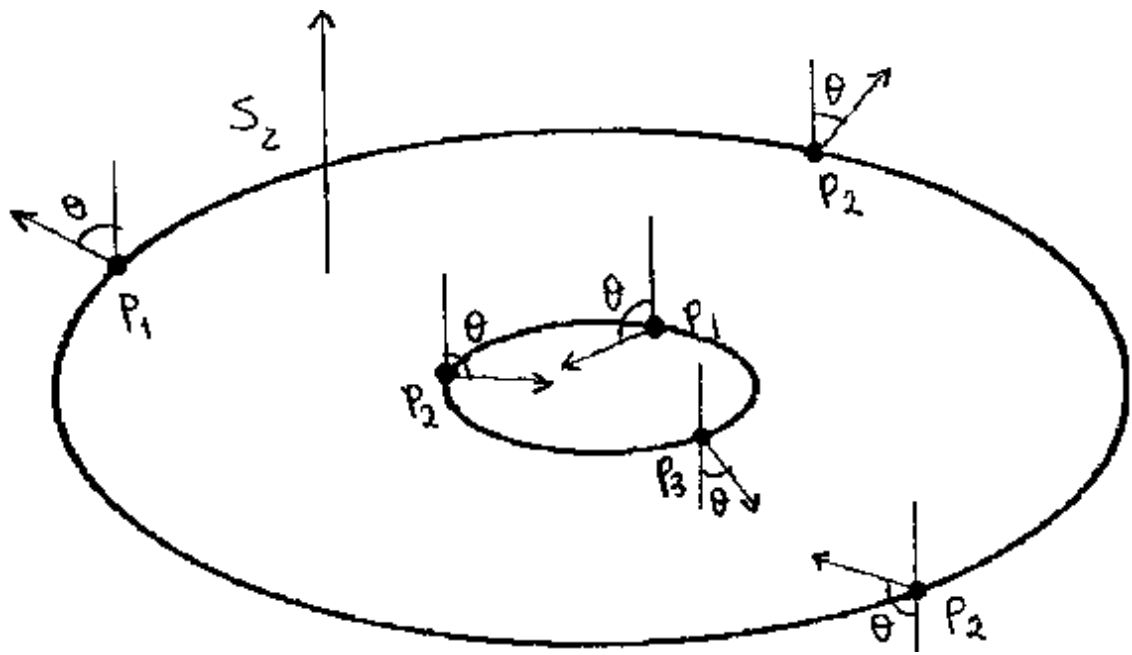
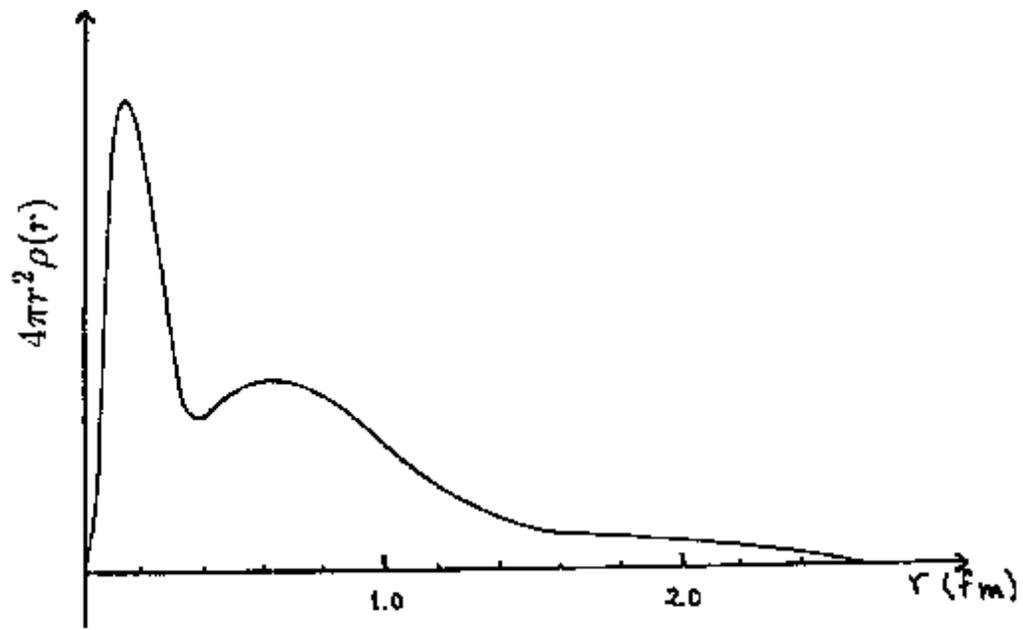
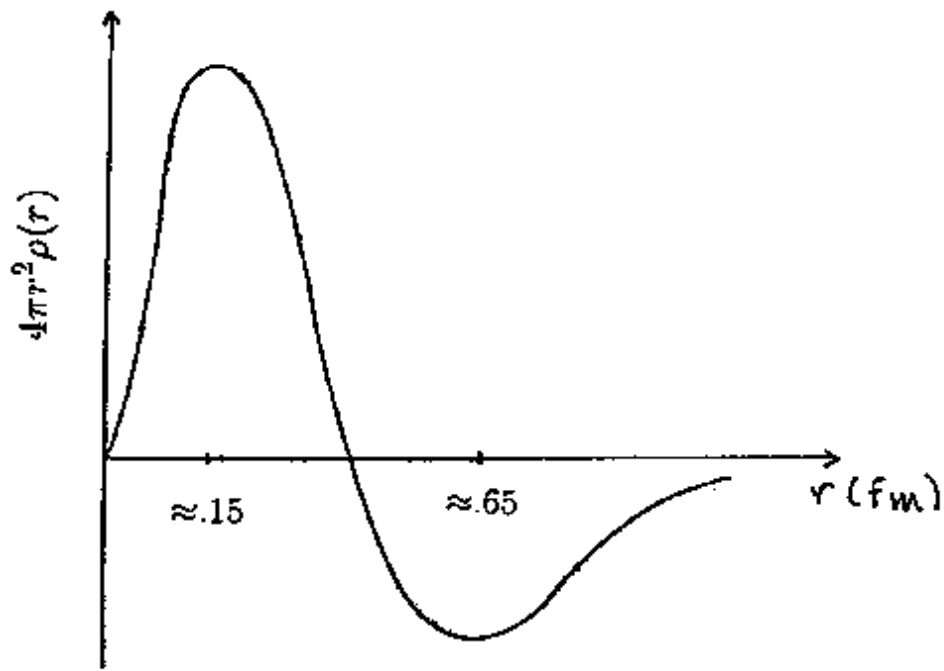
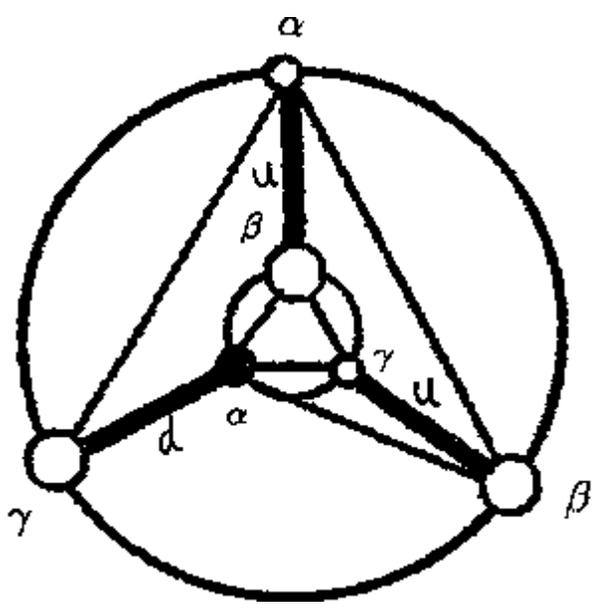
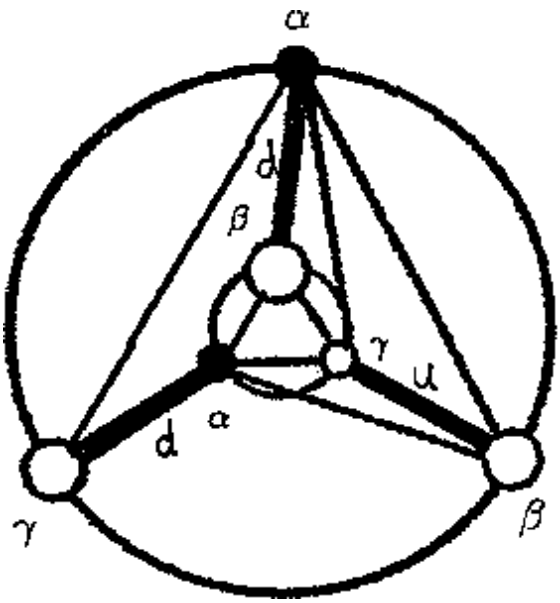
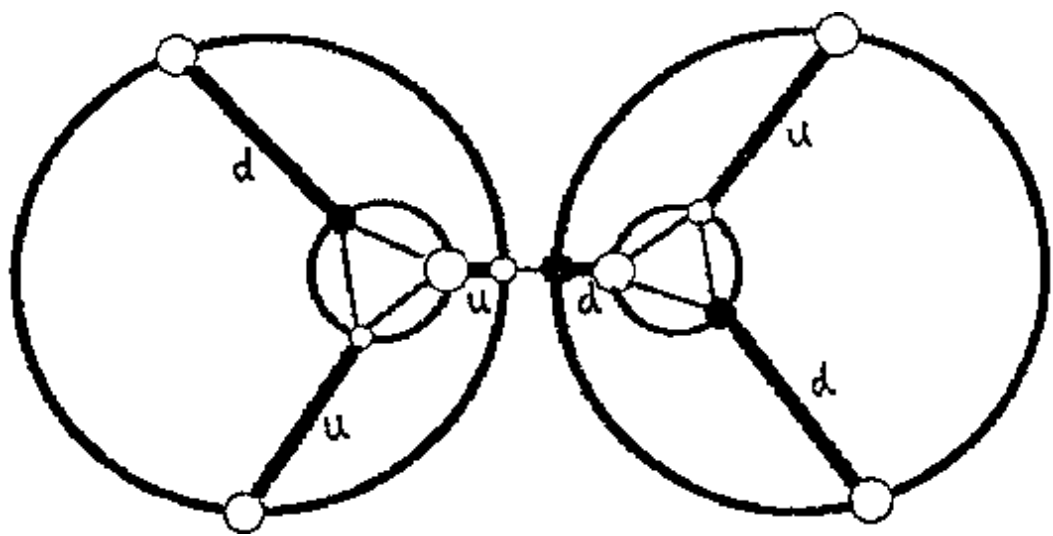
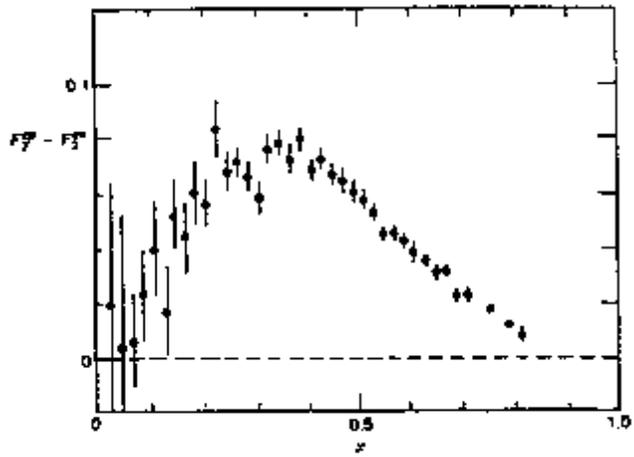
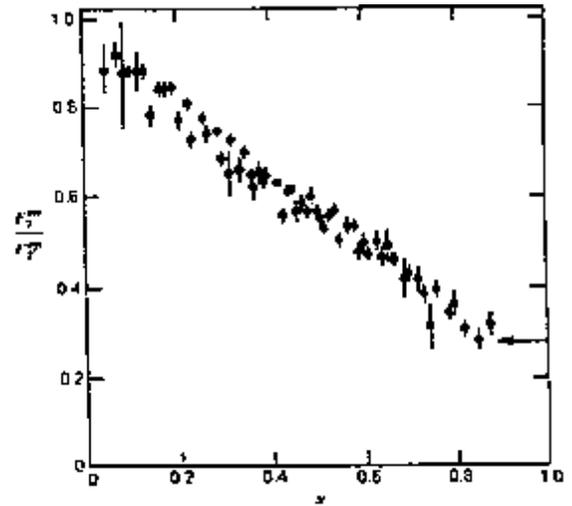
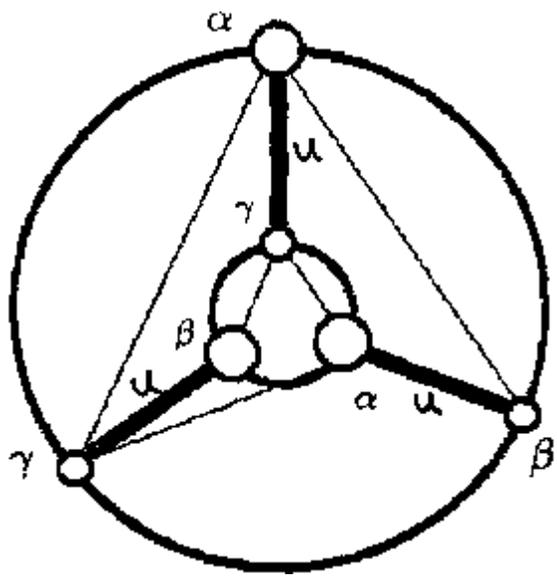
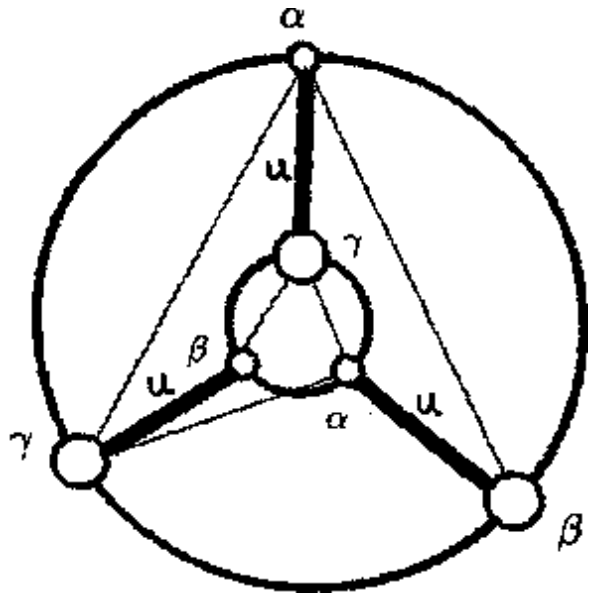
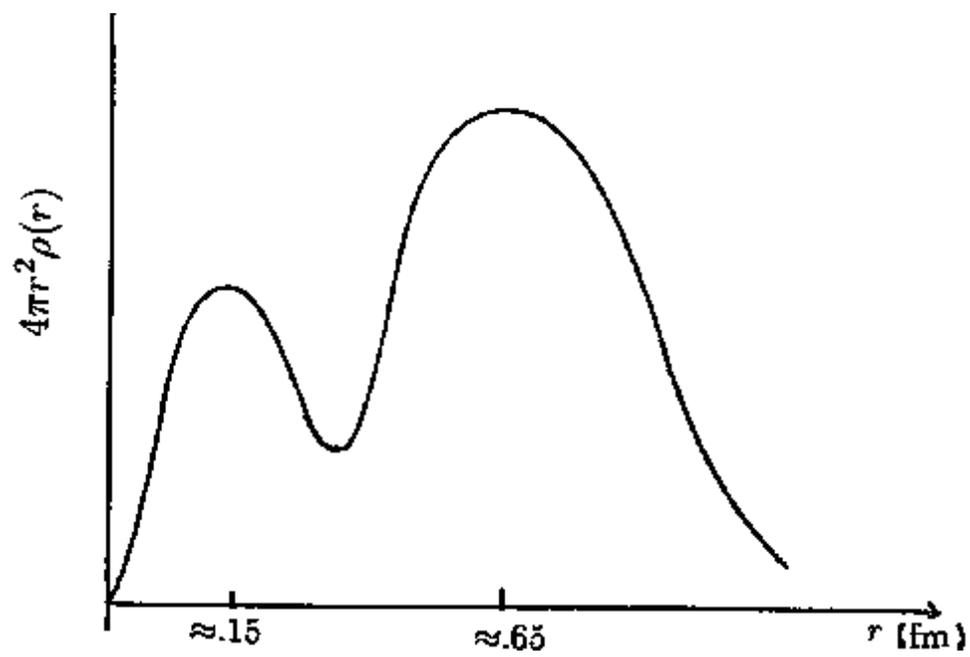
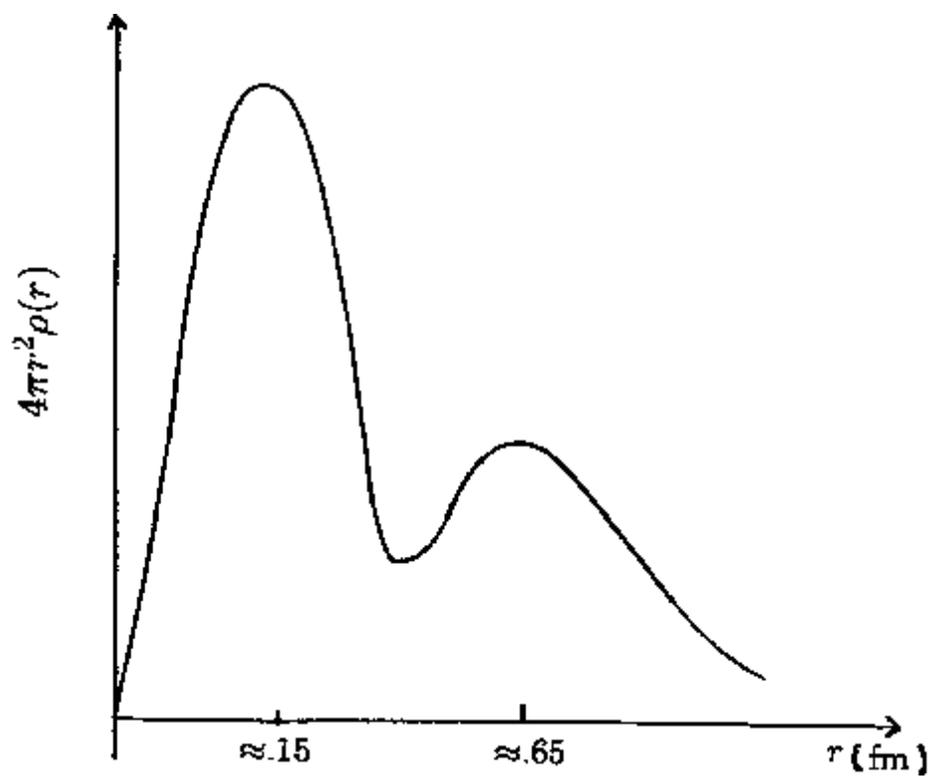
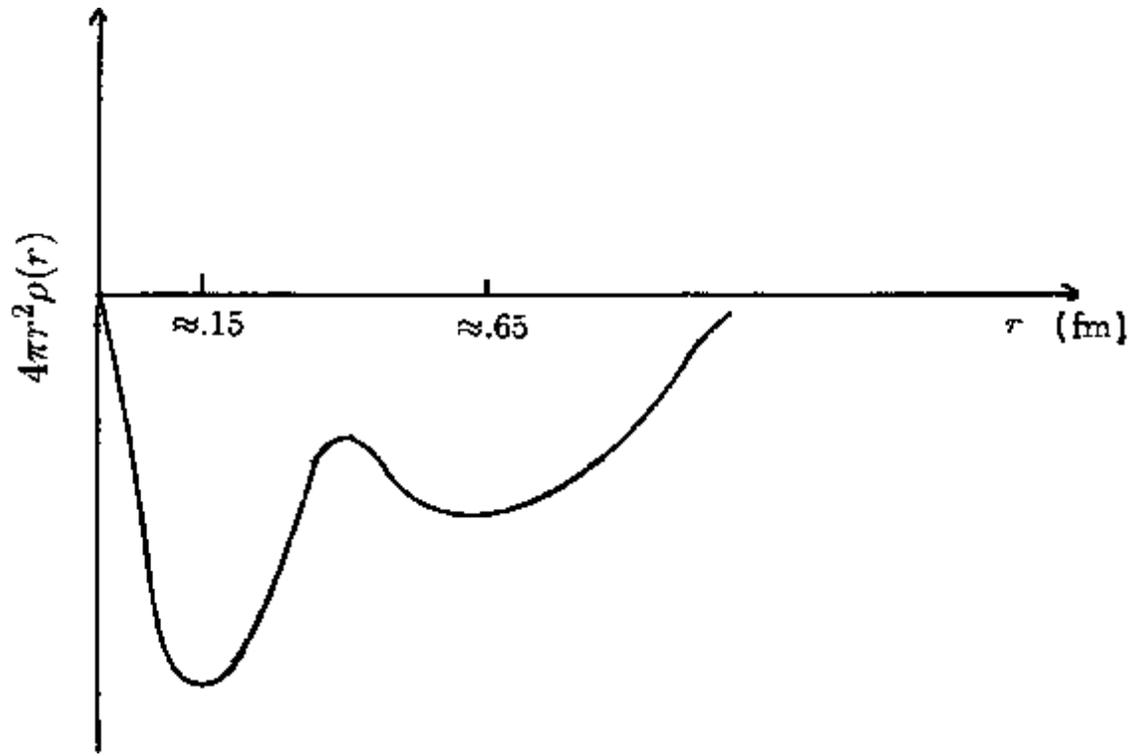
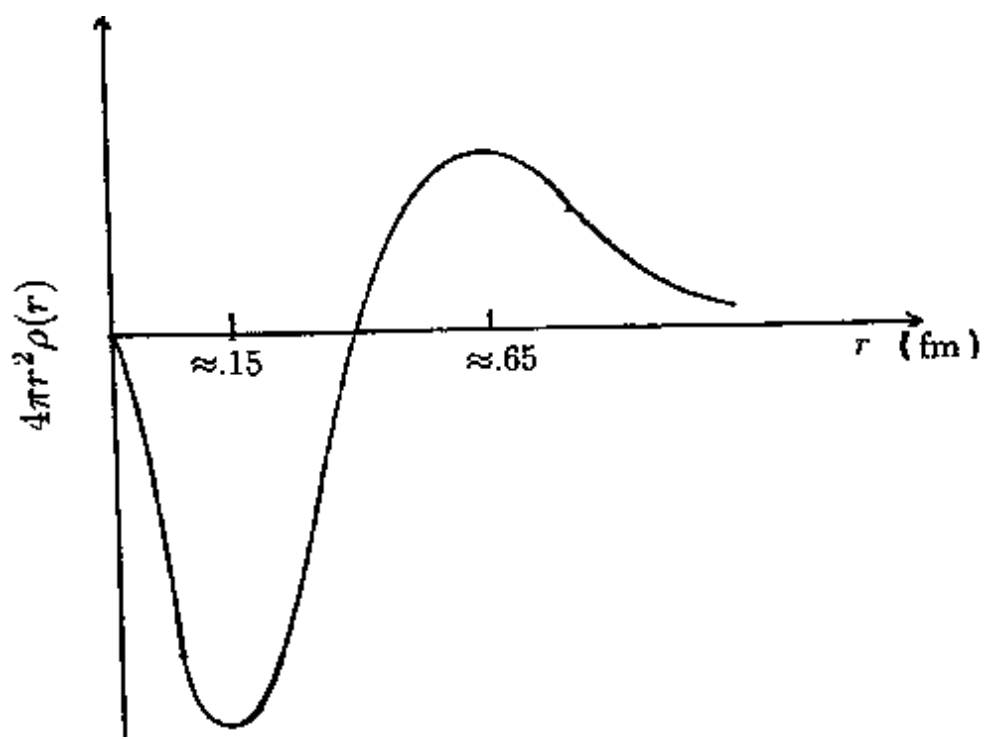
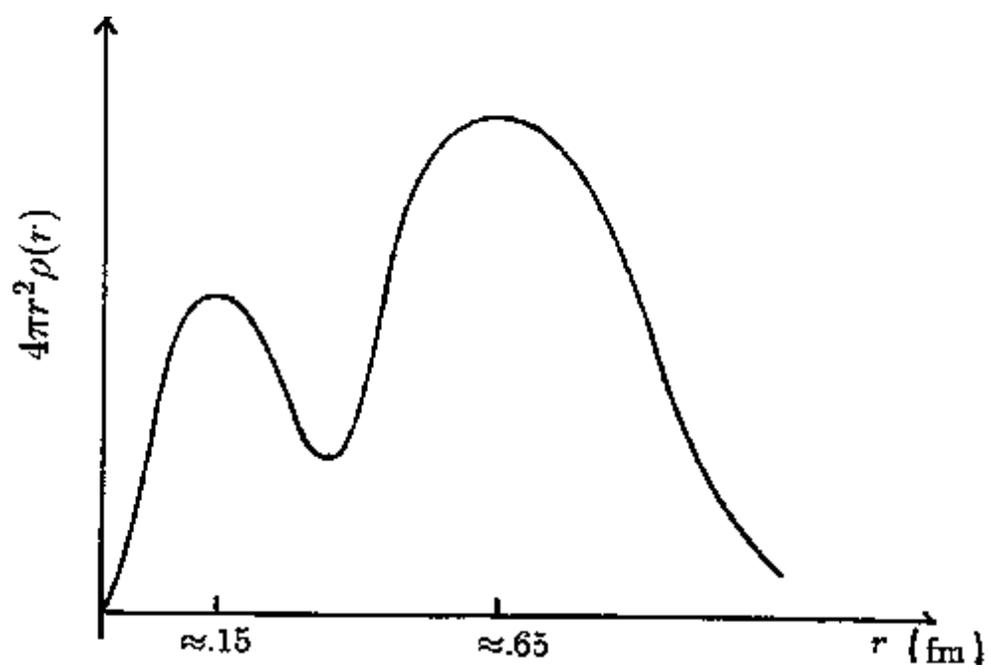
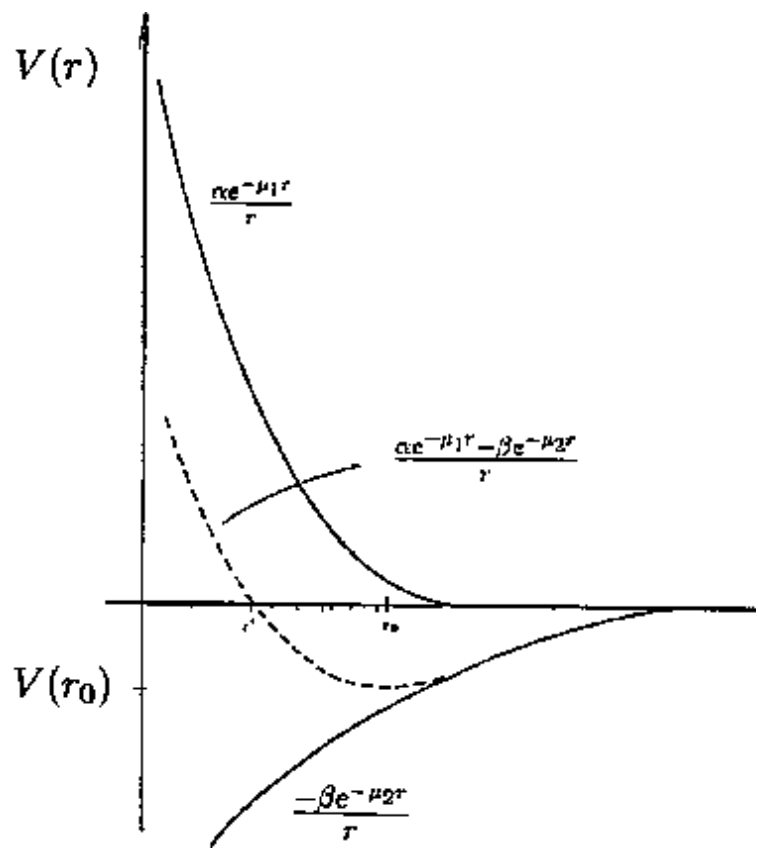
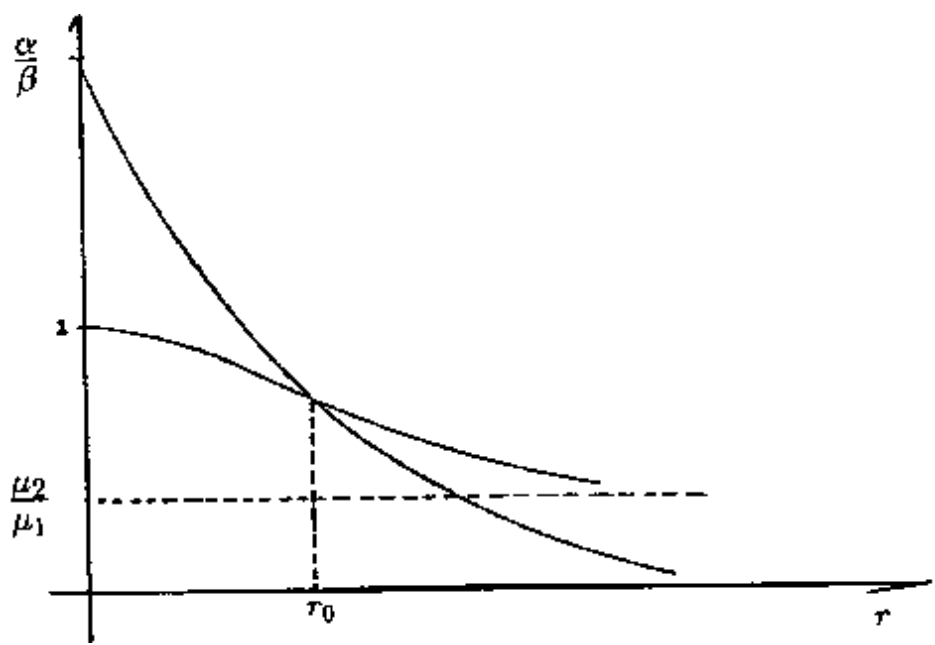
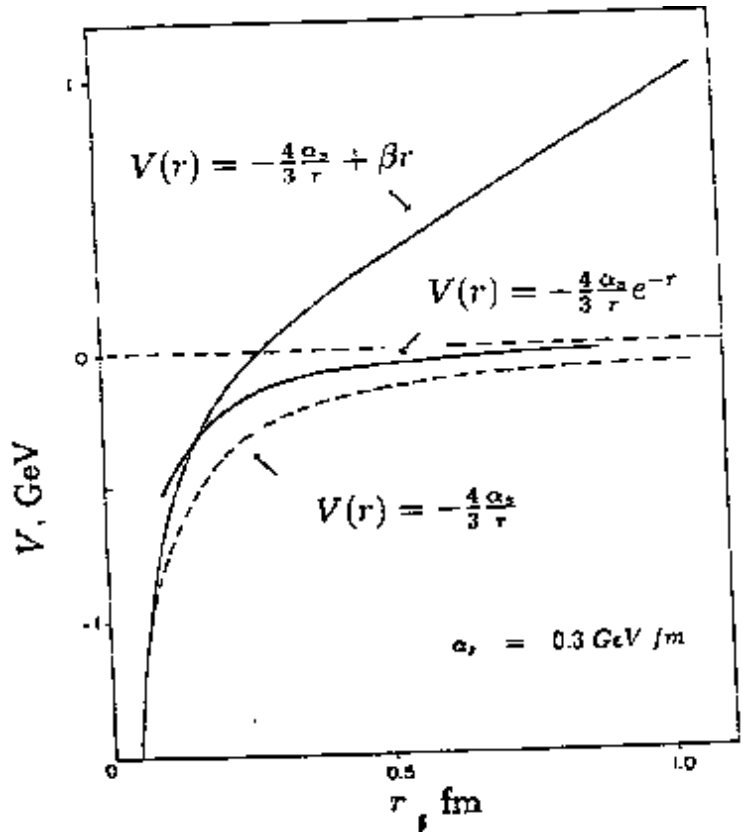
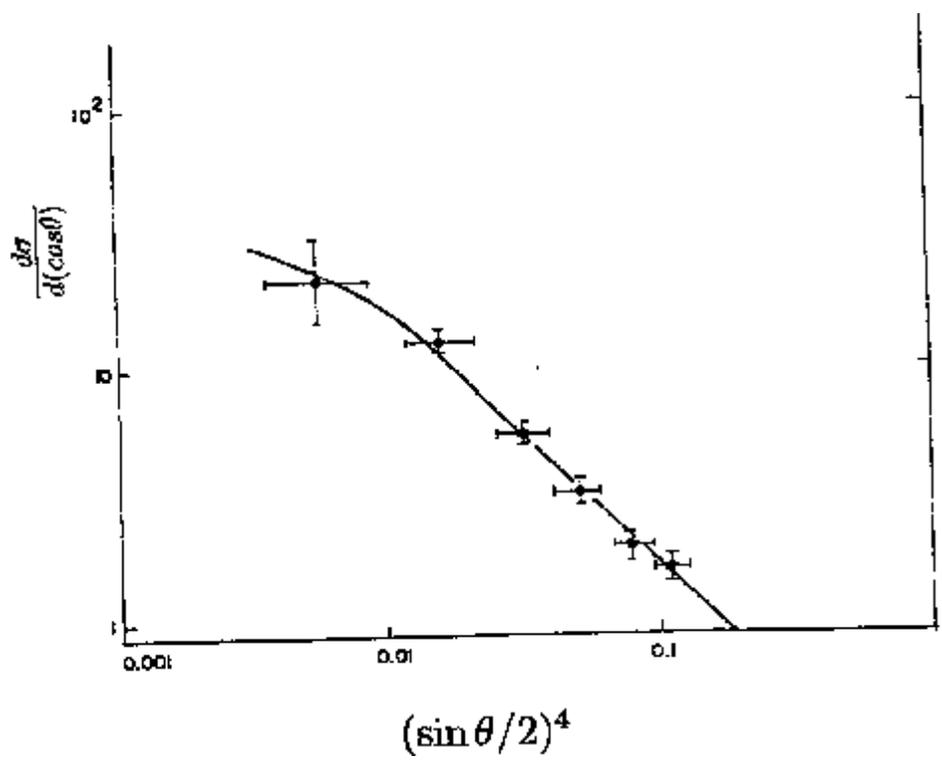
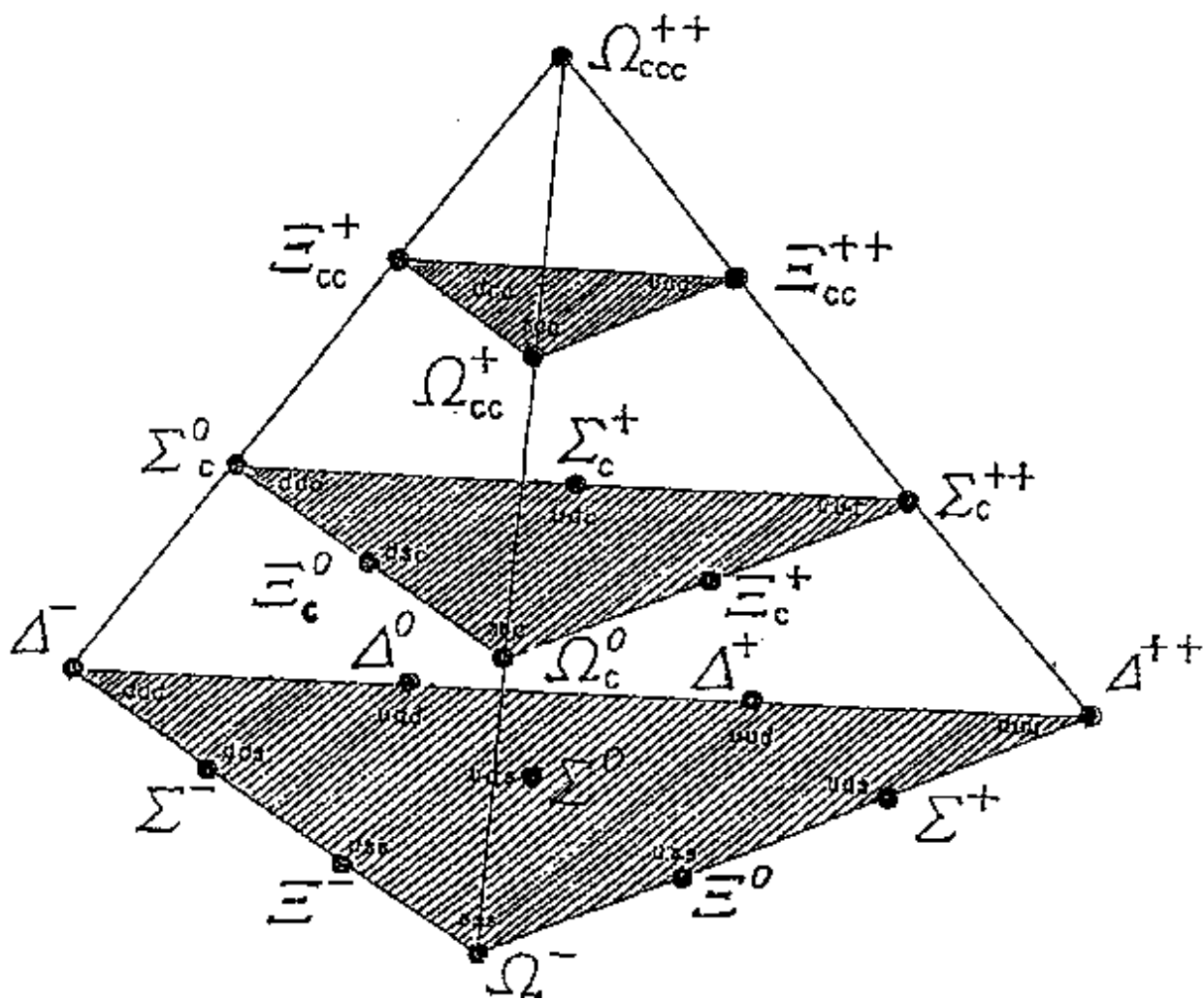
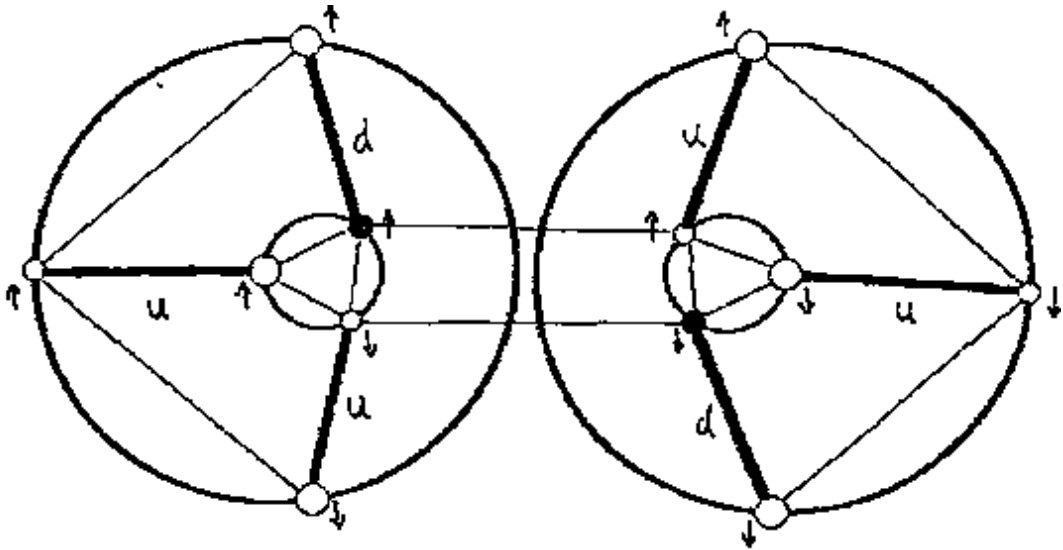
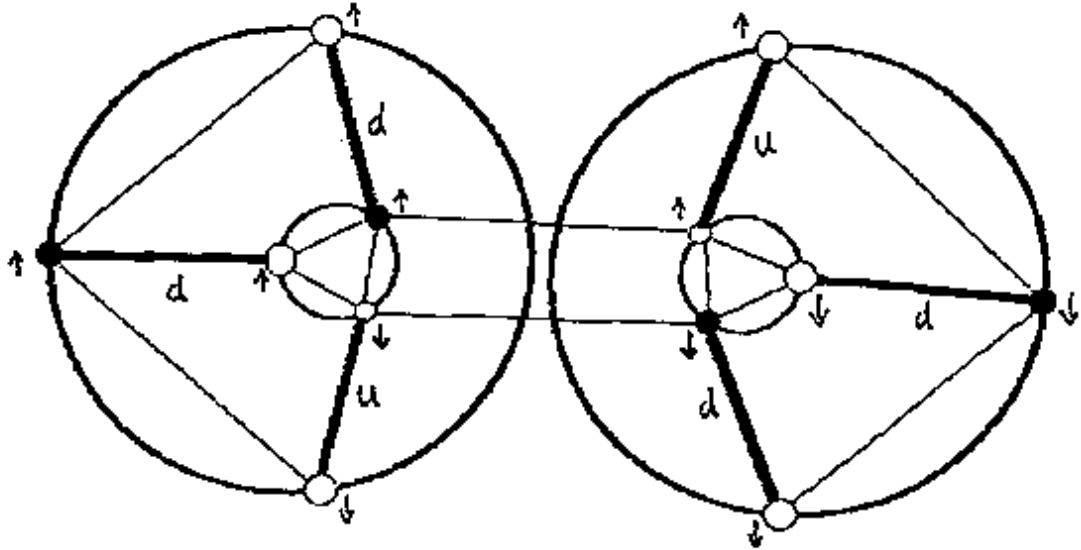
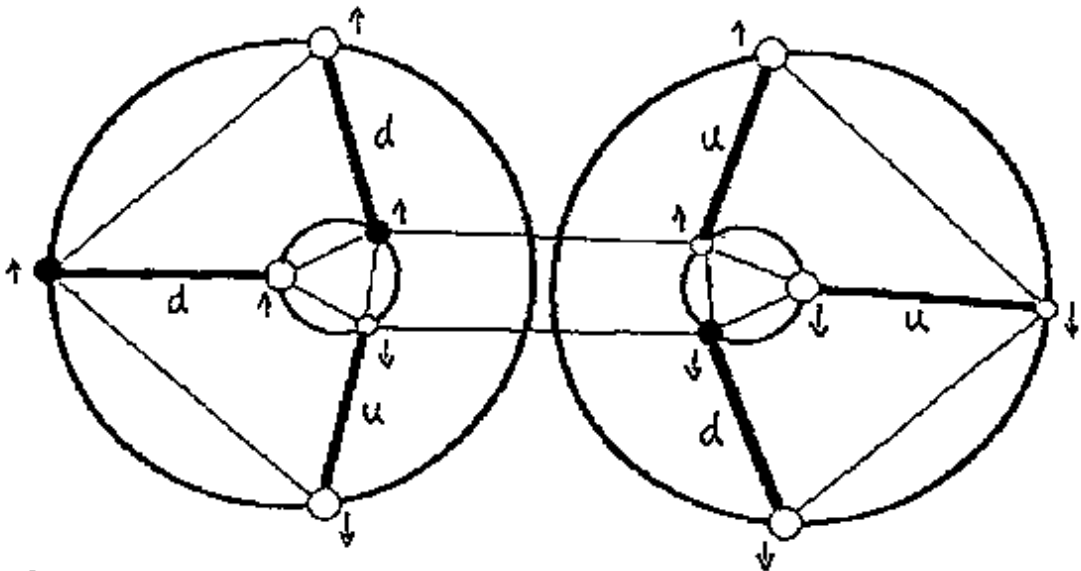
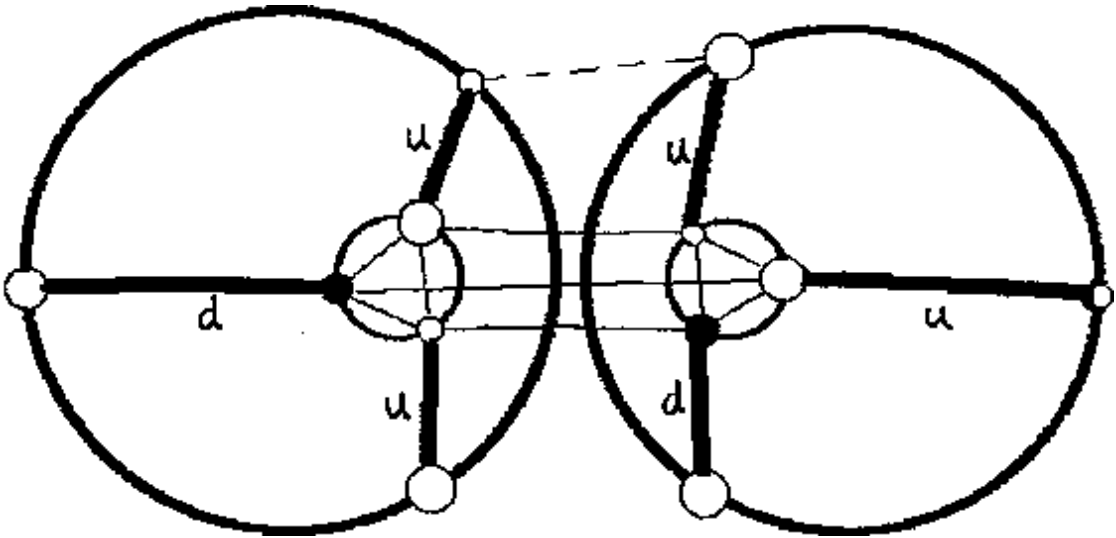
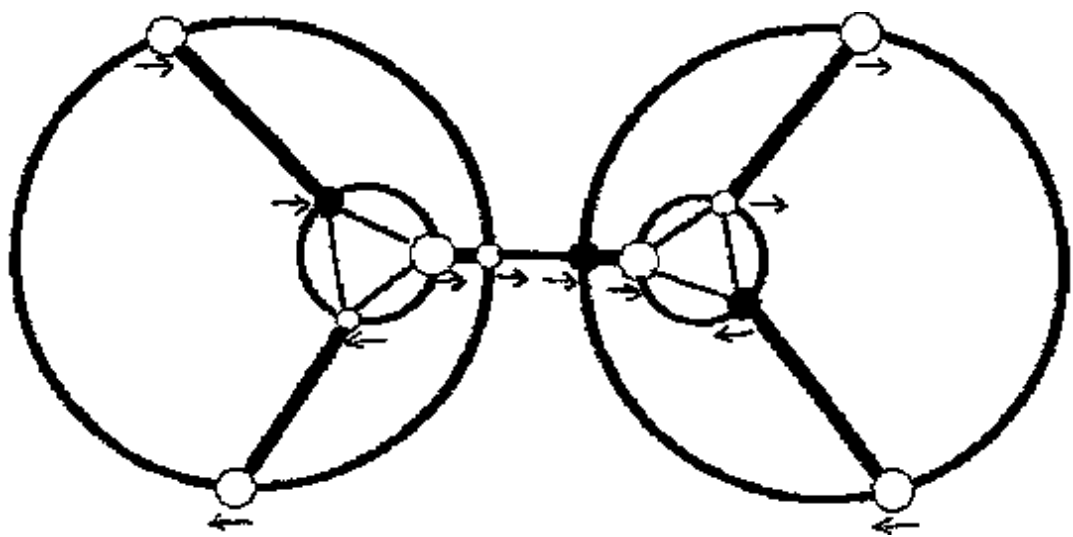
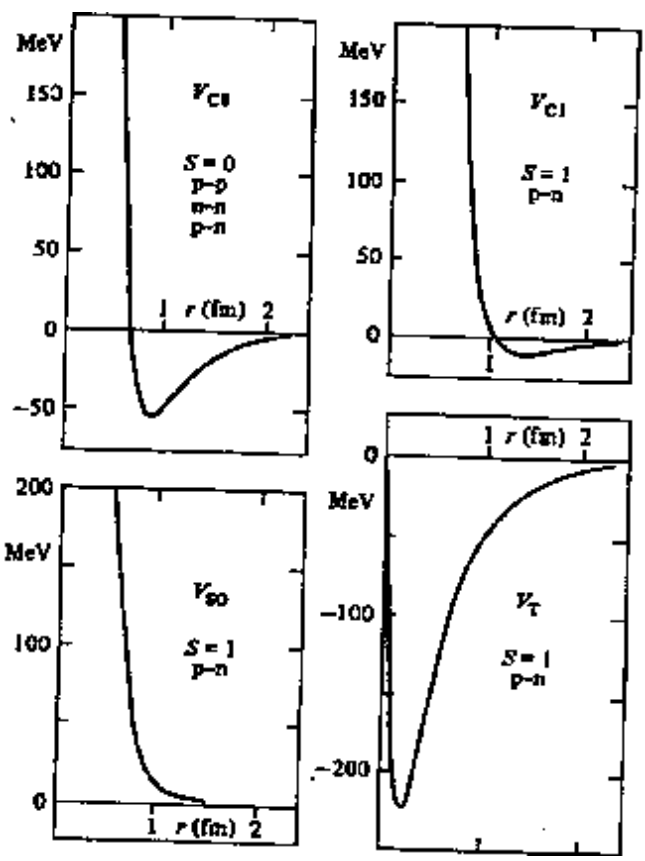
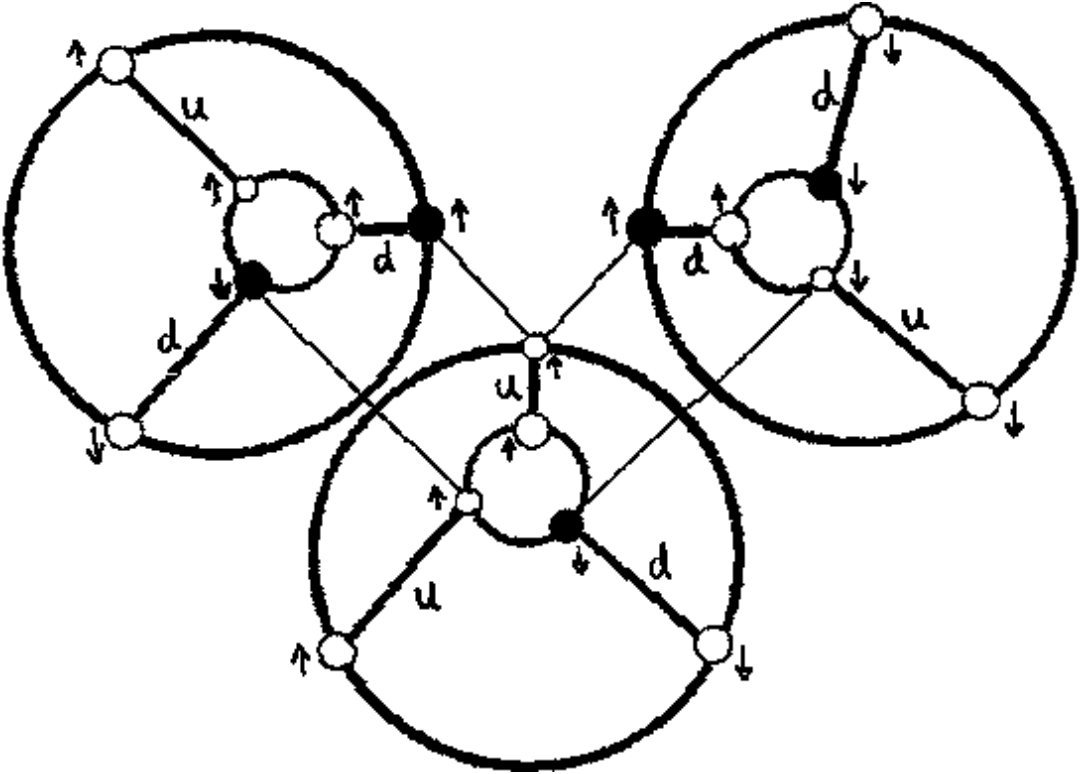
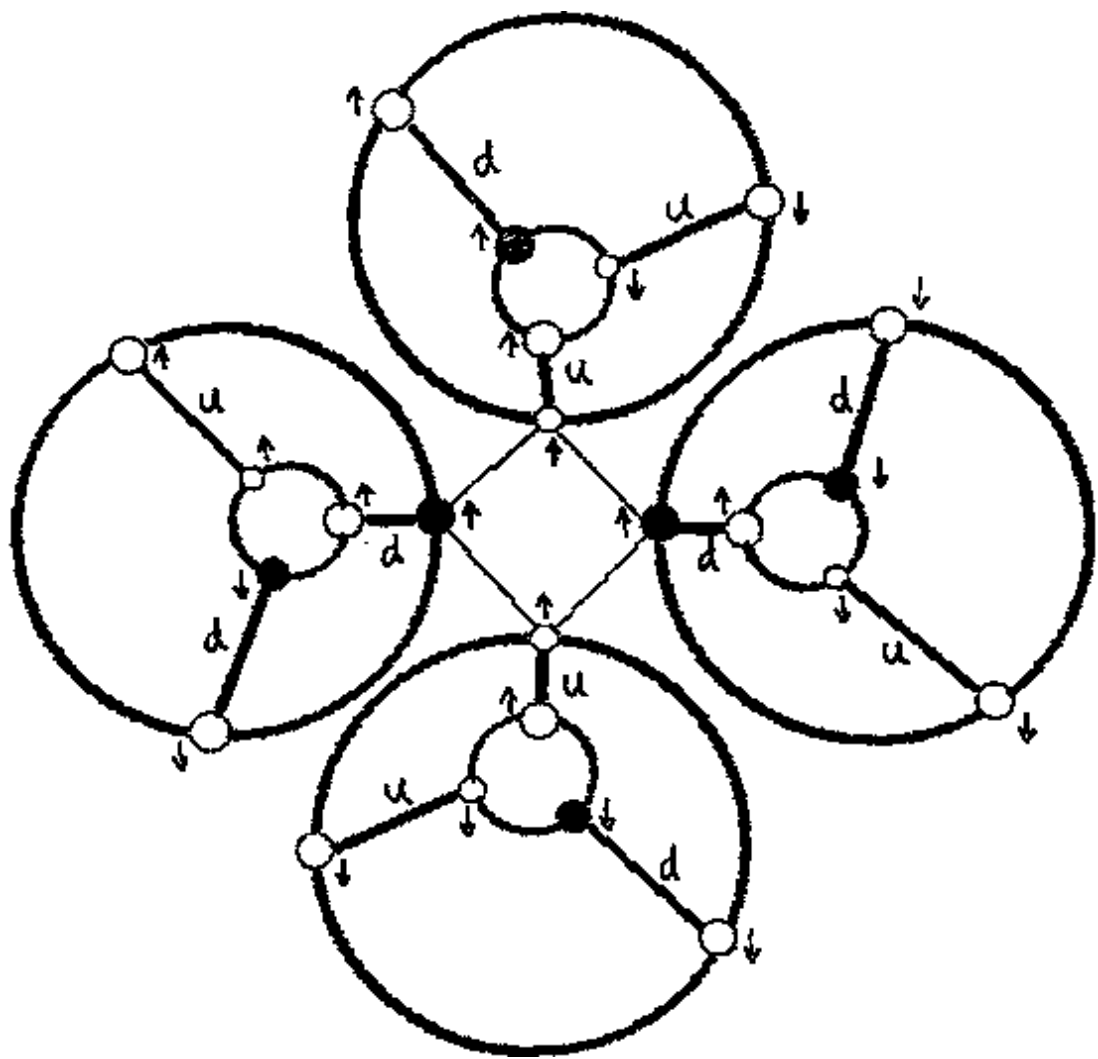
\end{document}